%% file: tritandb.tex
\documentclass[format=acmsmall, review=false, screen=true]{acmart}
\usepackage{booktabs} 
\usepackage{multirow}
\usepackage{paralist} 
\usepackage[utf8]{inputenc}
\usepackage{pgfplots}
\usepackage[ruled]{algorithm2e} 
\usepackage{newfloat,caption}
\usepackage{subcaption}
\usepackage{qtree}
\usepackage{listings}

\SetAlFnt{\small}
\SetAlCapFnt{\small}
\SetAlCapNameFnt{\small}
\SetAlCapHSkip{0pt}
\IncMargin{-\parindent}

\definecolor{gray1}{gray}{0.6}
\definecolor{gray2}{gray}{0.8}
\definecolor{Dark2-6-1}{RGB}{27,158,119}
\definecolor{Dark2-6-A}{RGB}{27,158,119}
\definecolor{Dark2-6-2}{RGB}{217,95,2}
\definecolor{Dark2-6-B}{RGB}{217,95,2}
\definecolor{Dark2-6-3}{RGB}{117,112,179}
\definecolor{Dark2-6-C}{RGB}{117,112,179}
\definecolor{Dark2-6-4}{RGB}{231,41,138}
\definecolor{Dark2-6-D}{RGB}{231,41,138}
\definecolor{Dark2-6-5}{RGB}{102,166,30}
\definecolor{Dark2-6-E}{RGB}{102,166,30}
\definecolor{Dark2-6-6}{RGB}{230,171,2}
\definecolor{Dark2-6-F}{RGB}{230,171,2}
\pgfplotscreateplotcyclelist{mycolorlist}{%
	Dark2-6-1,every mark/.append style={fill=Dark2-6-1},mark=*\\
	Dark2-6-2,every mark/.append style={fill=Dark2-6-2},mark=square*\\%
	Dark2-6-3,every mark/.append style={fill=Dark2-6-3},mark=otimes\\%
	Dark2-6-4,every mark/.append style={fill=Dark2-6-4},mark=star\\%
	Dark2-6-5,every mark/.append style={fill=Dark2-6-5},mark=diamond\\%
	Dark2-6-6,densely dashed,every mark/.append style={fill=Dark2-6-6},mark=*\\%
	brown!60!black,densely dashed,every mark/.append style={
	solid,fill=brown!80!black},mark=square*\\%
	black,densely dashed,every mark/.append style={solid,fill=gray},mark=otimes*\\%
	blue,densely dashed,mark=star,every mark/.append style=solid\\%
	red,densely dashed,every mark/.append style={solid,fill=red!80!black},mark=diamond*\\%
}

\DeclareFloatingEnvironment[fileext=frm,placement={!ht},name=Listing,within=section]{listing}

\lstdefinelanguage{Kotlin}{
  keywords={package, as, typealias, this, super, val, var, fun, for, null, true, false, is, in, throw, return, break, continue, object, if, try, else, while, do, when, yield, typeof, yield, typeof, class, interface, enum, object, override, public, private, import, abstract, synchronized},
  ndkeywords={@Deprecated, Iterable, Int, Integer, Float, Double, String, Runnable, dynamic},
  emph={println, return@, forEach,},
  identifierstyle=\color{black},
  sensitive=true,
  commentstyle=\color{gray}\ttfamily,
  comment=[l]{//},
  morecomment=[s]{/*}{*/},
  morestring=[b]",
  morestring=[s]{"""*}{*"""},
}

\pgfplotsset{compat=1.10}

\lstdefinestyle{wind} {morekeywords={name,hasValue,hasTime,isA,measures,type,description,\$schema,title,properties,locatedAt,\$ref,has,units,format,latitude,longitude}}

\lstdefinestyle{sparql} {morekeywords={filter,hours}}

\acmDOI{123456}

\received{September 2017}
\received[revised]{NA}
\received[accepted]{NA}

\begin{document}
\title{TritanDB: Time-series Rapid Internet of Things Analytics} 
\author{Eugene Siow}
\author{Thanassis Tiropanis}
\author{Xin Wang}
\author{Wendy Hall}
\affiliation{%
  \institution{University of Southampton}}
  
\begin{abstract}
The efficient management of data is an important prerequisite for realising the potential of the Internet of Things (IoT). Two issues given the large volume of structured time-series IoT data are, addressing the difficulties of data integration between heterogeneous Things and improving ingestion and query performance across databases on both resource-constrained Things and in the cloud. In this paper, we examine the structure of public IoT data and discover that the majority exhibit unique flat, wide and  numerical characteristics with a mix of evenly and unevenly-spaced time-series. We investigate the advances in time-series databases for telemetry data and combine these findings with microbenchmarks to determine the best compression techniques and storage data structures to inform the design of a novel solution optimised for IoT data. A query translation method with low overhead even on resource-constrained Things allows us to utilise rich data models like the Resource Description Framework (RDF) for interoperability and data integration on top of the optimised storage. Our solution, TritanDB, shows an order of magnitude performance improvement across both Things and cloud hardware on many state-of-the-art databases within IoT scenarios. Finally, we describe how TritanDB supports various analyses of IoT time-series data like forecasting.
\end{abstract}

%
%


%
%


%

\maketitle


\input{body}

\end{document}

%% file: body.tex
\section{Introduction}

The rise of the Internet of Things (IoT) brings with it new requirements for data management systems. Large volumes of sensor data form streams of time-series input to IoT platforms that need to be integrated and stored. IoT applications that seek to provide value in real-time across a variety of domains need to retrieve, process and analyse this data quickly. Hence, data management systems for the IoT should support the collection, integration and analysis of time-series data.

Performance and interoperability for such systems are two pressing issues explored in this paper. Given the large volume of streaming IoT data coupled with the emergence of Edge and Fog Computing networks \cite{Chiang2017} that distribute computing and storage functions along a cloud-to-thing continuum in the IoT, there is a case for investigating the specific characteristics of IoT data to optimise databases, both on resource-constrained Things as well as dynamically-provisioned, elastically-scalable cloud instances, to better store and query IoT data. The difficulties in data integration between heterogeneous IoT Things, possibly from different vendors, different industries and conforming to specifications from different standard bodies also drives our search for a rich data model, that encourages interoperability, to describe and integrate IoT data, which can then be applied to databases with minimal impact on performance.

The Big Data era has driven advances in data management and processing technology with new databases emerging for many specialised use cases. Telemetry data from DevOps performance monitoring scenarios of web-scale systems has pushed time-series databases to the forefront again. IoT data is a new frontier, a potentially larger source of time-series data given it ubiquitous nature, with data that exhibits its own unique set of characteristics. Hence, it follows that by investigating the characteristics of IoT time-series data and the compression techniques, data structures and indexing used in state-of-the-art time-series database design, we can design solutions for IoT time-series data optimised for performance on both Things and the cloud.

Data integration is another challenge in the IoT due to fragmentation across platforms, ``a bewildering variety of standards" \cite{Things2016}, and multiple independent vendors producing Things which act as data silos that store personal data in the vendor's proprietary, cloud-based databases. There is a strong case for a common data model and there are proposals to use YANG \cite{schonwalder2010network}, JSON Schema \cite{galiegue2013json}, CBOR/CDDL \cite{bormann2013concise} and JSON Content Rules \cite{cordell2016language} amongst others. However, the Resource Description Framework (RDF) data model, the foundation of publishing, integrating and sharing data across different applications and organisations on the Semantic Web \cite{Bizer2009} has demonstrated its feasibility as a means of connecting and integrating rich and heterogeneous web data using current infrastructure \cite{Heath2011}. Barnaghi \emph{et~al.} \cite{Barnaghi2012} support the view that this can translate to interoperability for cyber-physical IoT systems with ontologies for describing sensors and observations from the W3C \cite{Compton2012} already present.

RDF is formed from statements consisting triples with a subject, predicate and object. For example, in the statement: \texttt{`sensor1 has windSpeedObservation1'}, the subject is \texttt{`sensor1'}, the predicate is \texttt{`has'} and the object is \texttt{`weatherObservation1'}. The union of the four triples in Listing \ref{lst:triples} with our original triple forms an RDF graph telling us of a weather observation at 3pm on the 1st of June 2017 from a wind sensor that measures wind speed with a value of 30 knots. 

\begin{lstlisting}[style=wind,caption={Four RDF triple statements describing a wind speed observation},label={lst:triples},basicstyle=\scriptsize\ttfamily]
weatherObservation1 hasValue "30.0knots"
weatherObservation1 hasTime "2017-06-01 15:46:08"
sensor1 isA windSensor
windSensor measures windSpeed
\end{lstlisting}

This data representation, though flexible (almost any type of data can be expressed in this format), has the potential for serious performance issues with almost any interesting query requiring several self-joins on the underlying triples when the triples are stored as a table. State-of-the-art RDF stores get around this by extensive indexing \cite{Neumann2009} \cite{Bishop2011} and partitioning the triples for query performance \cite{Abadi2009}. However, Buil-Aranda \emph{et~al.} \cite{Buil-Aranda2013} have examined traditional RDF store endpoints on the web and shown that performance for generic queries can vary by up to 3-4 orders of magnitude and stores generally limit or have worsened reliability when issued with a series of non-trivial queries. 

By investigating the characteristics of IoT time-series data, how it is can be more optimally stored, indexed and retrieved, how it is modelled in RDF and the structure of analytical queries, we design an IoT-specific solution, TritanDB, that provides both performance improvements in terms of writes, reads and storage space over other state-of-the-art time-series, NoSQL and relational databases and supports rich data models like RDF that encourage semantic interoperability.

Specifically, the main contributions of this paper are that:
\begin{enumerate}
\item We identify the unique structure and characteristics of both public IoT data and RDF sensor data modelled according to existing ontologies from a database optimisation perspective.
\item We also investigate, with microbenchmarks on real-world IoT data, how to exploit the characteristics of IoT data and advances in time-series compression, data structures and indexing to optimally store and retrieve IoT data, this leads to a novel design for an IoT time-series database, using a re-ordering buffer and an immutable, time-partitioned store.
\item We also define a specialised query translation method with low overhead, even on resource-constrained Things, that allows us to utilise the Resource Description Framework (RDF) as a data model for interoperability and integration. We compare the performance of our solution with other state-of-the-art databases within IoT scenarios on both Things and the cloud, showing an order of magnitude improvement.
\item  Finally, we build support for various analyses of IoT time-series data like forecasting on our TritanDB engine.
\end{enumerate}

The structure of the rest of the paper is as follows, Section \ref{sec:related_work} details the related work, Section \ref{sec:examine} covers our examination of the shape and characteristics of IoT data and rich, RDF-modelled IoT data while Section \ref{sec:microbenchmarks} discusses appropriate microbenchmarks for the design of an IoT database taking into account the characteristics of time-series IoT data. Section \ref{sec:translation} introduces our query translation method for minimising the overhead of rich data models like RDF, Section \ref{sec:design} presents our design for a time-series database engine using the microbenchmark results and query translation technique while Section \ref{sec:experiments} compares benchmark results against other time-series databases in terms of ingestion, storage size and query performance. Finally, Section \ref{sec:analytics} describes analytics like forecasting for time-series data built on our engine that supports resampling and moving average conversion to evenly-spaced time-series and Section \ref{sec:conclusion} concludes and discusses future work.

\section{Related Work}
\label{sec:related_work}

On the one hand, the large volume of Internet of Things (IoT) data is formed from the observation space of Things - sensors observe the environment and build context for connected applications. Streams of sensor readings are recorded as time-series data.

On the other hand, large amounts of telemetry data, monitoring various metrics of operational web systems, has provided a strong use case for recent research involving time-series databases. Facebook calls this particular use case an Operational Data Store (ODS) and Gorilla \cite{Pelkonen2015} was designed as fast, in-memory time-series database to this end. Gorilla introduces a time-series compression method which we adopt and build upon in this paper, specifically adapting it for Internet of Things (IoT) data. 

Spotify has Heroic \cite{Spotify} that builds a monitoring system on top of Cassandra \cite{LakshamAvinash2010} for time-series storage and Elasticsearch \cite{Elastic} for meta data indexing. Hawkular \cite{Redhat} by RedHat, KairosDB \cite{kairosdb} and Blueflood from Rackspace \cite{Rackspace2013} are all built on top of Cassandra, while OpenTSDB \cite{OpenTSDB} is similarly built on top of a distributed store, HBase \cite{Wlodarczyk2012}. InfluxDB \cite{InfluxData2017} is a native time series database with a rich data model allowing meta data for each event. It is utilises a Log-structured merge-tree \cite{ONeil1996}, Gorilla compression for time-series storage and has an SQL-like query language. In our experiments, we compare our solution against these time-series databases: InfluxDB and OpenTSDB while also benchmarking against Cassandra and Elastic Search engines (which underly the other systems). We not only evaluate their performance across IoT datasets but also on resource-constrained Things.

Anderson \emph{et~al.} introduce a time-partitioned, version-annotated, copy-on-write tree data structure in BTrDb \cite{Andersen2016} to support high throughput time-series data from microsynchophasors deployed within an electrical grid. Akumuli \cite{EugeneLazin2017} is a similar time-series database built for high throughput writes using a Numeric B+ tree (a log-structured, append-only B+ tree) as its storage data structure. Timestamps within both databases use delta-of-delta compression similar to Gorilla's compression algorithm and Akumuli uses FPC \cite{Burtscher2009} for values while BTrDb uses delta-of-delta compression on the mantissa and exponent from each floating point number within the sequence. Both databases also utilise the tree structures to store aggregate data to speed up specific aggregate queries. We investigate each of their compression methods in detail and a tree data structure for our time-series storage engine while also benchmarking against Akumuli in our experiments.

\begin{table}%
\caption{A Summary of Time-series Databases Storage Engines and Querying Support}
\label{tab:tsdbs}
\begin{minipage}{\columnwidth}
\begin{center}
\begin{tabular}{rl|llllll}
  \toprule
Database & \multicolumn{2}{c}{Storage Engine} & Read & \multicolumn{4}{c}{Query Support\footnote{$\Pi$ = Projection, $\sigma$ = Selection, where $t$ is selection by time only, $\bowtie$ = Joins, $\Gamma$ = Aggregation functions}}\\
  	\midrule
{} & \multicolumn{2}{l}{} & {} & $\Pi$ & $\sigma$ & $\bowtie$ & $\Gamma$ \\	  	
Heroic \cite{Spotify}					& \multicolumn{2}{c}{\multirow{4}{*}{Cassandra \cite{LakshamAvinash2010}}} & HQL & $\checkmark$ & $\checkmark$ & $\times$ & $\checkmark$  \\
KairosDB \cite{kairosdb}			 	& \multicolumn{2}{l}{} & JSON & $\times$ & $t$ & $\times$ & $\checkmark$\\
Hawkular \cite{Redhat}					& \multicolumn{2}{l}{} & \multirow{2}{*}{REST} & $\times$ & $t$ & $\times$ & $\checkmark$\\
Blueflood \cite{Rackspace2013}								& \multicolumn{2}{l}{} & {} & $\times$ & $t$ & $\times$ & $\times$\\
\midrule
OpenTSDB \cite{OpenTSDB}				& \multicolumn{2}{c}{HBase \cite{Wlodarczyk2012}} & REST & $\times$ & $\checkmark$ & $\times$ & $\checkmark$ \\
\midrule
Cube \cite{Square2012}				& \multicolumn{2}{c}{MongoDB \cite{banker2011mongodb}} & REST & $\times$ & $t$ & $\times$ & $\checkmark$ \\
\midrule
InfluxDB \cite{InfluxData2017}			& \multirow{5}{*}{Native}						& LSM-based  & InfluxQL & $\checkmark$ & $\checkmark$ & $\times$ & $\checkmark$\\
Vulcan/Prometheus \cite{PrometheusAuthors2016} & {}						& Chunks-on-FS   & PromQL & $\checkmark$ & $\checkmark$ & $\times$ & $\checkmark$\\
Gorilla/Beringei \cite{Pelkonen2015}	&  {}	& In-memory &  \multirow{3}{*}{REST} & $\times$ & $t$ & $\times$ & $\times$ \\

BTrDb \cite{Andersen2016}				& {}						& COW-tree &  {} & $\times$ & $t$ & $\times$ & $\checkmark$ \\
Akumuli \cite{EugeneLazin2017}			& {}						& Numeric-B+-tree & {} & $\times$ & $t$ & $\times$ & $\checkmark$\\
\midrule
DalmatinerDB \cite{FiFo2014}			& \multicolumn{2}{c}{\multirow{2}{*}{Riak}}  &  DQL & $\checkmark$ & $\checkmark$ & $\times$ & $\checkmark$\\
Riak-TS \cite{Basho2016}				& \multicolumn{2}{l}{}						  & SQL & $\checkmark$ & $\checkmark$ &  $\times$ & $\checkmark$\\
\midrule
Timescale \cite{Timescale2017}		& \multicolumn{2}{c}{\multirow{2}{*}{Postgres}}	& \multirow{2}{*}{SQL} & $\checkmark$ & $\checkmark$ & $\checkmark$ & $\checkmark$\\
Tgres \cite{Trubetskoy2017}				& \multicolumn{2}{l}{}  & {} & $\checkmark$ & $\checkmark$ & $\checkmark$ & $\checkmark$\\
    \bottomrule
\end{tabular}
\end{center}
\bigskip\centering
\end{minipage}
\end{table}%

Other time-series databases include DalmatinerDB \cite{FiFo2014} and Riak-TS \cite{Basho2016} which are built on top of Riak, Vulcan from DigitalOcean which adds scalability to the Prometheus time-series database \cite{PrometheusAuthors2016}, Cube from Square \cite{Square2012} which uses MongoDB \cite{banker2011mongodb} and relational database solutions like Timescale \cite{Timescale2017} and Tgres \cite{Trubetskoy2017} which are built on PostgresSQL. 

Table \ref{tab:tsdbs} summarises the storage engines, method of reading data from the discussed time-series databases and query support for basic relational algebra with projections ($\Pi$), selections ($\sigma$) and joins ($\bowtie$) and also aggregate functions ($\Gamma$) essential for time-series data. InfluxDB, DalmatinerDB and Riak-TS implement SQL-like syntaxes while Timescale and Tgres have full SQL support. KairosDB provides JSON-based querying while Prometheus and Heroic have functional querying languages HQL and PromQL respectively. Gorilla, BTrDb, Akumuli, Hawkular, Blueflood and OpenTSDB have REST interfaces allowing query parameters. It can be seen that expressive SQL and SQL-like query languages provide the most query support and in this work we seek to build on this expressiveness with the rich data model of RDF and its associated query language, SPARQL \cite{harris2013sparql}.

Efficient SPARQL-to-SQL translation that improves performance and builds on previous literature has been investigated by Rodriguez-Muro \emph{et~al.} \cite{Rodriguez-muro2014}, Priyatna \emph{et~al.} \cite{Priyatna2014} and Siow \emph{et~al.} \cite{Siow2016b}. None of the translation methods supports time-series databases at the time of writing though and we build on previous work in efficient query translation to create a general abstraction for graph models and query translation to work on time-series IoT databases.

\section{Examining the characteristics of IoT Time-series and RDF for IoT}
\label{sec:examine}

\subsection{The shape of IoT data}
\label{subsec:shape}

To investigate the shape of data from IoT sensors, we collected the public schemata of 11,560 unique IoT Things from data streams on Dweet.io\footnote{http://dweet.io/see} for a month in 2016. These were from a larger collected set of 22,662 schemata of which 1,541 empty and 9,561 non-IoT schemata were filtered away. The non-IoT schemata were largely from the use of Dweet.io for rooms in a relay chat stream.

Dweet.io is a cloud-based platform that supports the publishing of sensor data from any IoT Things in JavaScript Object Notation (JSON). It was observed from the schemata that 11,468 (99.2\%) were flat, row-like with a single level of data, while only 92 (0.8\%) were complex, tree-like/hierarchical with multiple nested levels of data. Furthermore, we discovered that the IoT data was mostly wide. A schema is considered wide if there are 2 or more fields beside the timestamp. We found that 80.0\% of the Things sampled had a schema that was wide while the majority (57.3\%) had 5 or more fields related to each timestamp. Only about 6\% had more than 8 fields though, which is considerably less than those in performance-monitoring telemetry use cases (MySQL by default measures 350 metrics\footnote{https://dev.mysql.com/doc/refman/5.7/en/server-status-variables.html}). The most common set of fields was intertial sensor (\emph{tilt\_x, tilt\_y, tilt\_z}) at 31.3\% and metrics (\emph{memfree, avgrtt, cpu, hsdisabled, users, ploss, uptime}) at 9.8\%. 122 unique Thing schemata were environment sensors with (\emph{temperature, humidity}) that occupied 1.1\%.

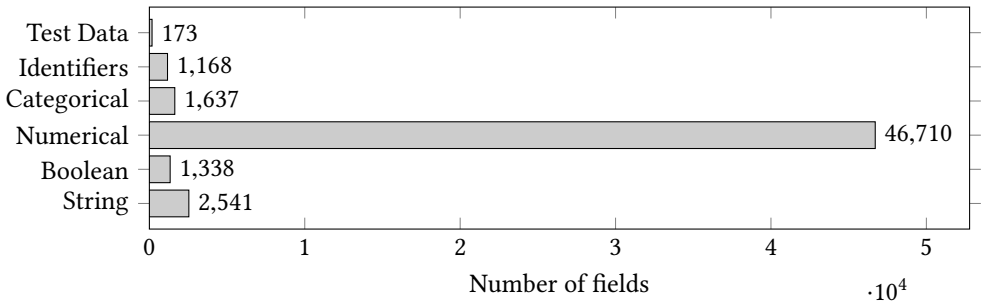
\begin{figure}
\centering
\begin{tikzpicture}
\begin{axis}[
    xbar,
    xmin=0.0,
    xtick={0,10000,20000,30000,40000,50000},
    width=12.5cm,
    height=4.5cm,
    enlarge x limits={rel=0.13,upper},
    ytick={1,2,3,4,5,6},
    yticklabels={{String},{Boolean},{Numerical},{Categorical},{Identifiers},{Test Data}},
    enlarge y limits=0.15,
    xlabel={Number of fields},
    ytick=data,
    nodes near coords,
    nodes near coords align=horizontal
]
\addplot [draw=black, fill=gray2] coordinates {
    (2541,1)
    (1338,2)
    (46710,3)
    (1637,4)
    (1168,5)
    (173,6)
};
\end{axis}
\end{tikzpicture}
\caption{Number of fields of varying types in a sample of over 11k IoT Schemata}
\label{fig:fields}
\end{figure}

Finally, we observed that the majority of fields (87.2\%) beside the timestamp were numerical as shown in Fig. \ref{fig:fields}. Numerical fields include integers, floating point numbers and time values. Identifiers (2.2\%), categorical fields (3.1\%) that take on only a limited number of possible values, e.g. a country field, and Boolean fields (2.5\%) occupied a small percentage each. Some test data (0.3\%) like \emph{`hello world'} and \emph{`foo bar'} was also discovered and separated from String fields. String fields occupied 4.7\% with 738 unique keys of 2,541 keys in total, the most common being `\emph{name}' with 13.7\%, `\emph{message}' with 8.1\% and `\emph{raw}' with 3.2\%.

We also obtained a smaller alternative sample of 614 unique Things (over the same period) from Sparkfun\footnote{https://data.sparkfun.com/streams}, that only supports flat schemata, which confirmed that most IoT Things sampled have wide (76.3\%) schema and 93.5\% of the fields were numerical while only 4.5\% were string fields.

Hence, to summarise the observations of the sampled public schemata, the shape of IoT data is largely flat, wide and numerical in content. All schemata are available from a public repository \footnote{http://dx.doi.org/10.5258/SOTON/D0076}.

These characteristics were verified by a series of surveys of public IoT schemata from different application domains and multiple independent sources as shown in Table \ref{tab:cross_iot_studies}. These include 614 schemata from SparkFun \footnote{https://data.sparkfun.com/streams} which records public streams from Arduino devices, 18 schemata from the Array of Things (AoT) \footnote{https://arrayofthings.github.io/} which is a smart city deployment in Chicago, 4,702 weather station schemata from across the United States in Linked Sensor Data \cite{Patni2010}, 9,033 schemata from OpenEnergy Monitor's \footnote{https://emoncms.org/} open-hardware meters measuring home energy consumption, 9,007 schemata from ThingSpeak \footnote{https://thingspeak.com/channels/public} which is a cloud-based, MatLab-connected IoT analytics platform. 

All the studies consisted of flat schemata with a majority of numerical-typed data. The majority of schemata were also wide accept for the AoT and OpenEnergy Monitor study where only about half the schemata were. This was because in both cases, a mix of sensor modules were deployed where some only measured a single quantity and resulted in narrow schemata. The schemata analysed are available from a repository \footnote{http://dx.doi.org/10.5258/SOTON/D0202}.

\begin{table}%
\caption{Summary of the Cross-IoT Study on Characteristics of IoT Data}\bigskip
\label{tab:cross_iot_studies}
\begin{minipage}{\columnwidth}
\begin{center}
\begin{tabular}{lrrrrrr}
  \toprule
 \multicolumn{2}{c}{Study Details} & \multicolumn{5}{c}{Characteristics (\%)}  \\
\midrule
{} 					& \#\footnote{Number of unique schemata} 		& Flat 	& Wide & Num 	& Periodic & $0_{\emph{MAD}}$\footnote{Percentage with Median Absolute Deviation (MAD) of zero (approximately evenly-spaced time-series) } \\
  	
SparkFun 			& 614 		& 100.0 & 76.3 & 93.5	& 0.0 & 27.6\\
Array of Things		& 18			& 100.0 & 50.0 & 100.0	& 0.0 & 100.0 \\
LSD Blizzard		& 4702		& 100.0 & 98.8 & 97.0	& 0.004 & 91.8 \\
OpenEnergy Monitor	& 9033		& 100.0 & 52.5 & 100.0	& - & - \\
ThingSpeak			& 9007		& 100.0 & 84.1 & 83.2	& 0.004	& 46.9 \\
    \bottomrule
\end{tabular}
\end{center}
\end{minipage}
\end{table}%

\subsection{Evenly-spaced VS Unevenly-spaced IoT data}
\label{subsec:spacing}

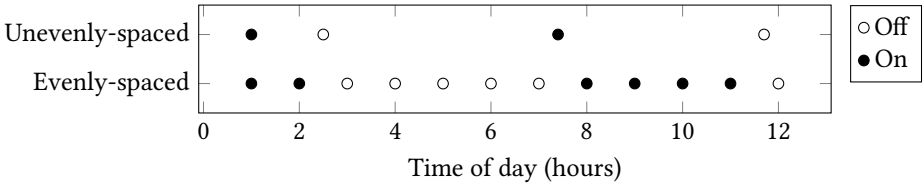
\begin{figure}
\centering
\begin{tikzpicture}
\begin{axis}[%
scatter/classes={%
    a={mark=o,draw=black},
    b={mark=*,draw=black}},
    ytick={1,2},
    width=10cm,
    yticklabels={{Evenly-spaced},{Unevenly-spaced}},
    enlarge y limits=0.6,
    xlabel={Time of day (hours)},
    height=3.0cm,
    legend pos=outer north east]
\addplot[scatter,only marks,%
    scatter src=explicit symbolic]%
table[meta=label] {
x y label
1 1 b
1 2 b
2 1 b
2.5 2 a
3 1 a
4 1 a
5 1 a
6 1 a
7 1 a
7.4 2 b
8 1 b
9 1 b
10 1 b
11 1 b
11.7 2 a
12 1 a
    };
    \legend{Off,On}
\end{axis}
\end{tikzpicture}
\caption{Unevenly-spaced event-triggered VS evenly-spaced smart light bulb time-series}
\label{fig:spacing}
\end{figure}

One of the differences between the Internet of Things and traditional wireless sensor networks is the advent of  an increasing amount of event-triggered sensors within smart Things instead of sensors that record measurements at regular time intervals. For example, a smart light bulb that measures when a light is on can either send the signal only when the light changes state, i.e. is switched on or off, or send its state regularly every second. The former type of event-triggered sensor gives rise to an unevenly-spaced time series as shown in Fig. \ref{fig:spacing}.

Event-triggered sensing has the advantages of
\begin{inparaenum}[1)]
\item more efficient energy usage preserving the battery as long as events occur less often than regular updates, 
\item better time resolution as timestamps of precisely when the state changed are known without needing to implement buffering logic on the sensor itself between regular signals and
\item less redundancy in sensor data storage.
\end{inparaenum}
However, there is the potential disadvantage that missing a signal can cause large errors although this can be addressed by an infrequent `heartbeat' signal to avoid large measurement errors.

We retrieved the timestamps of the last 5 `dweets' from the sample of IoT Things in Section \ref{subsec:shape} over a 24 hour period and observed that, of those available, 62.1\% are unevenly-spaced while 37.9\% are evenly-spaced. This tells us that IoT databases should be able to support both evenly and unevenly-spacing data. Hence, time-series databases that use fixed-sized, fixed-interval, circular buffers like the Round Robin Database Tool (RRDTool) \cite{oetiker2005rrdtool} or the Whisper database which was designed as a storage backend for the Graphite stack \cite{Sharma2016} are less suitable for handling the IoT's variable spacing time-series data. Both even and unevenly-spaced data were also present across the studies shown in Table \ref{tab:cross_iot_studies}. To take into account slight fluctuations in the period that could be a result of transmission delays caused by the communication medium, processing delays or highly precise timestamps, a statistical measure, the Median Absolute Deviation (MAD), the median of the absolute deviations from the data's median, was used to measure the spacing within time-series. Given the set of differences between subsequent timestamps, $X$, the equation \ref{eq:mad} defines its MAD.

\begin{equation}\label{eq:mad}
\text{MAD} = \text{median}(| X - \text{median}(X)|)
\end{equation}

A zero value of MAD reflects an approximately evenly-spaced time-series. It was not possible to determine the MAD of OpenEnergy Monitor streams as the format of data retrieved had to have a fixed, user-specified period as this was the main use case for energy monitoring dashboards.

\subsection{The characteristics of RDF IoT data}
\label{subsec:rdf_iot_data}

We observe that RDF sensor data from IoT datasets can be divided into 3 categories
\begin{inparaenum}[1)]
	\item \emph{device metadata} like the location and specifications of sensors,
	\item \emph{observation metadata} like the units of measure and types of observation
	\item \emph{observation data} like timestamps and actual readings.
\end{inparaenum}
Table \ref{tab:lsd_rdf} shows a sample of RDF triples divided into the 3 categories from weather observations of rainfall in the Linked Sensor Data (LSD) \cite{Patni2010} dataset.

\begin{table}%
\caption{Sample of RDF Triples according to IoT Data Categories from rainfall observations in LSD}
\label{tab:lsd_rdf}
\begin{minipage}{\columnwidth}
\begin{center}
\begin{tabular}{ll}
  \toprule
 IoT Data Category & Sample RDF Triples\\
  	\midrule
\multirow{2}{*}{Device metadata} 		& \texttt{sensor1 ssw:processLocation point1} \\
{}										& \texttt{point1 wgs:lat "40.82944"}  \\
  	\midrule
\multirow{4}{*}{Observation metadata} 	& \texttt{obs1 a weather:RainfallObservation} \\
{} 										& \texttt{obs1 ssw:result data1} \\
{} 										& \texttt{obs1 ssw:samplingTime time1} \\
{}										& \texttt{data1  ssw:uom weather:degree} \\
  	\midrule
\multirow{2}{*}{Observation data}		& \texttt{data1 ssw:floatValue "0.1"} \\
{} 										& \texttt{time1 time:inXSDDateTime "2017-06-01T15:46:08"} \\
    \bottomrule
\end{tabular}
\end{center}
\bigskip\centering
\end{minipage}
\end{table}%

For the LSD Blizzard dataset with 108,830,518 triples and the LSD Hurricane Ike dataset with 534,469,024 triples, only 12.5\% is observation data, 0.17\% is device metadata, while 87.3\% is observation metadata. In the Smart Home Analytics dataset \cite{Siow2016b} based on a different ontology, a similarly large 81.7\% of 11,157,281 triples are observation metadata.

Observation metadata which connects observations, time and measurement data together, consists of identifiers like \texttt{obs1}, \texttt{data1} and \texttt{time1}, which might not be returned in queries. In practice, the majority of time-series data, 97.8\% of fields, does not contain identifiers (Section \ref{subsec:shape}). As such, publishers of RDF observation metadata often generate long 128-bit universally unique identifiers (UUIDs) to serve as observation, time and data identifiers. In the 17 queries proposed for the streaming RDF/SPARQL benchmark, SRBench \cite{Zhang2012a}, and the 4 queries in the Smart Home Analytics Benchmark \cite{Siow2016b}, none of the queries project any these identifiers from observation metadata. 


\section{Microbenchmarks}
\label{sec:microbenchmarks}

\subsection{Internet of Things Datasets}
\label{subsec:iot_datasets}

To evaluate the performance of various algorithms and system designs with microbenchmarks, we collated a set of publicly available Internet of Things datasets. The use of public, published data, as opposed to proprietary data, enables reproducible evaluations and a base for new systems and techniques to make fair comparisons.

Table \ref{tab:iot_datasets} summarises the set of datasets collated, describing the precision of timestamps, Median Absolute Deviation (MAD) of deltas, $\delta_{\emph{MAD}}$, Interquartile Range (IQR) of deltas, $\delta_{\emph{IQR}}$, and the types of fields for each dataset. 

\subsubsection{SRBench}
SRBench \cite{Zhang2012a} is a benchmark based on the established Linked Sensor Data \cite{Patni2010} dataset that describes sensor data from weather stations across the United States with recorded observations from periods of bad weather. In particular, we used the Nevada Blizzard period of data from 1st to 6th April 2003 which included more than 647 thousand rows with over 4.3 million fields of data from 4702 of the stations. Stations have timestamp precision in seconds with the median $\delta_{\emph{MAD}}$ and $\delta_{\emph{IQR}}$ across stations both zero, showing regular, periodic intervals of measurement. The main field type was small floating point numbers mostly up to a decimal place in accuracy.

\subsubsection{Shelburne}
Shelburne is an agriculture dataset aggregating data from a network of wireless sensors obtained from a vineyard planting site in Charlotte, Vermont. Each reading includes a timestamp and fields like solar radiation, soil moisture, leaf wetness, etc. The dataset is available on SensorCloud \footnote{https://sensorcloud.microstrain.com/SensorCloud/data/FFFF0015C9281040/} and is collected from April 2010 to July 2014 with 12.4 million rows and 74.7 million fields. Timestamps are recorded up to nanosecond precision. The $\delta_{\emph{MAD}}$ is zero as the aggregator records at regular intervals (median of 10s), however, due to the high precision timestamps and outliers, there is a $\delta_{\emph{IQR}}$ of 293k (in microsecond range). All fields are floating point numbers recorded with a high decimal count/accuracy.

\subsubsection{GreenTaxi} This dataset includes trip records from green taxis in New York City from January to December 2016. Data is provided by the Taxi and Limousine Commission \footnote{http://www.nyc.gov/html/tlc/html/about/trip\_record\_data.shtml} and consists of 4.4 million rows with 88.9 million fields of data. Timestamp precision is in seconds and is unevenly-spaced as expected from a series of taxi pick-up times within a big city with a $\delta_{\emph{MAD}}$ of 1.48. However, as the time-series also has overlapping values and is very dense, the $\delta_{\emph{MAD}}$ and $\delta_{\emph{IQR}}$ are all within 2 seconds. There is a boolean field type for the store and forward flag which indicates whether the trip record was held in vehicle memory before sending to the vendor because the vehicle did not have a connection to the server. There are 12 floating point field types including latitude and longitude values (high decimal count) and fares (low decimal count). There are integer field types including vendor id, rate code id and drop off timestamps.

\begin{table}%
\caption{Public IoT Datasets used for Experiments}
\label{tab:iot_datasets}
\begin{minipage}{\columnwidth}
\begin{center}
\begin{tabular}{lcrcrrrrrr}
  \toprule
 Dataset & \multicolumn{3}{c}{Metadata} & \multicolumn{3}{c}{Timestamps} & \multicolumn{3}{c}{Field Types}\\
  	\midrule
  {} & {} & Rows & Fields  & Precision & $\delta_{\emph{MAD}}$ & $\delta_{\emph{IQR}}$ & Bool & FP & Int  \\	
  SRBench     	& Weather 	& 647k		& 4.3m		& s				& 0\footnote{As there were 4702 stations, a median of the MAD and IQR of all stations was taken, the means are 538 and 2232k}	& 0 	& \~0\footnote{A mean across the 4702 stations was taken for each field type in SRBench}
  & \~6 		& 0 			\\
  Shelburne    	& Agriculture & 12.4m		& 74.7m		& ms/ns			& 0/0 & 0.29/293k & 0 & 6 & 0 \\
  GreenTaxi    	& Taxi		& 4.4m		& 88.9m		& s				& 1.48		& 2 & 1 & 12 & 7	\\
    \bottomrule
\end{tabular} 
\end{center}
\bigskip\centering
\end{minipage}
\end{table}%

\subsection{Compressing Timestamps and Values}
\label{subsec:compress}

In Section \ref{subsec:spacing}, we saw that there was a mix of unevenly-spaced and evenly-spaced time-series in the IoT. We also saw in Section \ref{subsec:shape} that the majority of IoT data is numerical. These characteristics offer the opportunity to study specialised compression algorithms for timestamps and values of time-series data individually.

\subsubsection{Timestamp compression}
\label{subsubsec:timestamp_compression}

\begin{figure}
\centerline{\includegraphics[width=5.1in]{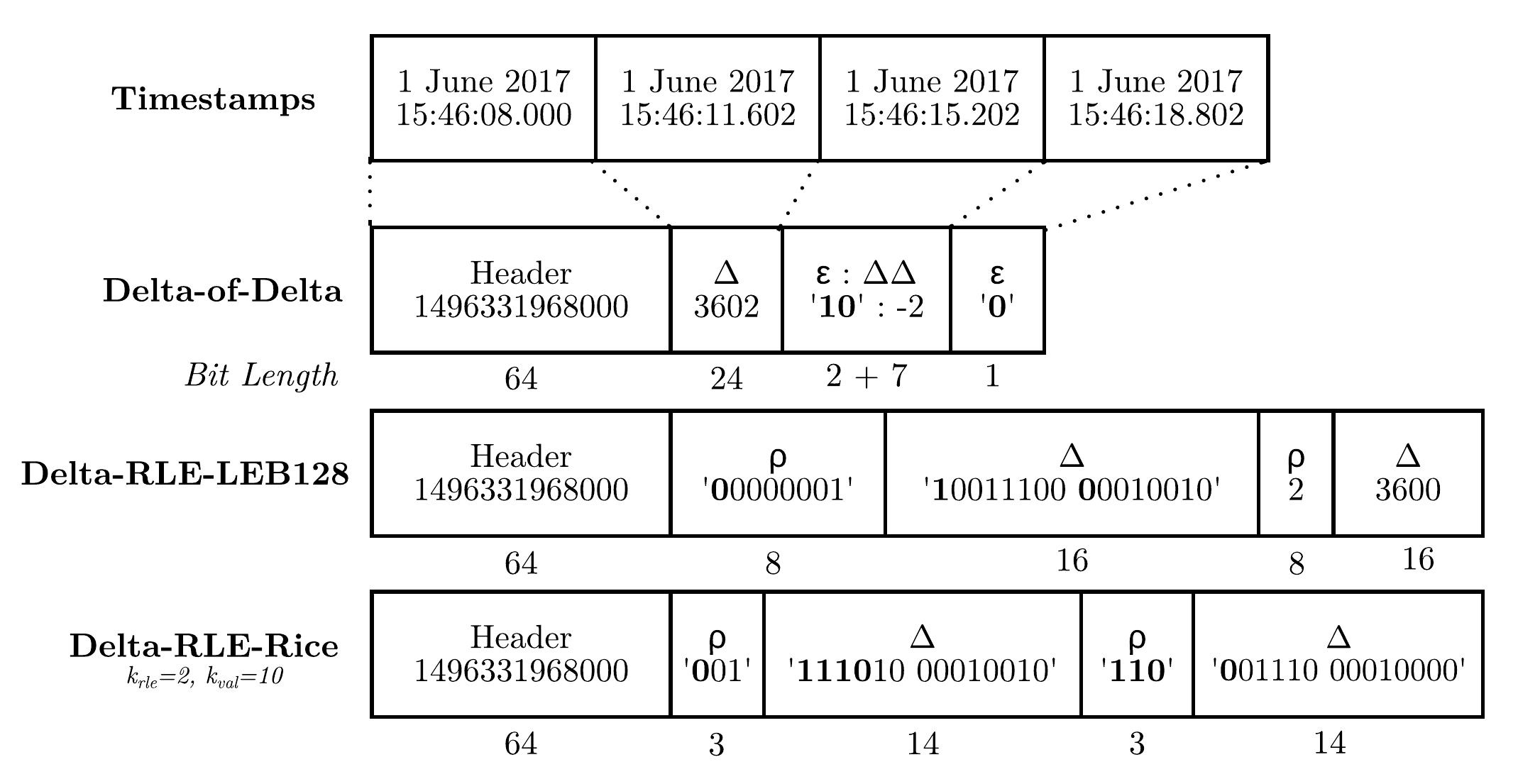}}
\caption{Visualising millisecond timestamp compression with varying delta-based methods}
\label{fig:timestamp}
\end{figure}

Timestamps in a series can be compressed to great effect based on the knowledge that in practice, the delta of a timestamp, the difference between this timestamp and the previous, is a fraction of the length of the timestamp itself and can be combined with variable length encoding to reduce storage size. If the series is somewhat evenly-spaced, run length encoding can be applied to further compress the timestamp deltas. For high precision timestamps (e.g. in nanoseconds), where deltas themselves are large however, delta-of-delta compression that stores the difference between deltas can often be more effective. Fig. \ref{fig:timestamp} depicts various methods of compressing a series of four timestamps with millisecond precision.

\emph{Delta-of-delta compression} builds on the technique for compressing timestamps introduced by Pelkonen \emph{et~al.} \cite{Pelkonen2015} to support effective compression on varying timestamp precision. The header stores a full starting timestamp for the block in 64 bits and the next variable length of bits depends on the timespan of a block and the precision of the timestamps. In this example in Fig. \ref{fig:timestamp}, a 24 bit delta of the value 3602 is stored (for a 4 hour block at millisecond precision with delta assumed to be postive). With knowledge of the timestamp precision during ingestion and a pre-defined block size, a suitable variable length can be determined on the fly (e.g. for a 4 hour block, 14 bits for seconds, 24 bits for milliseconds and 44 bits for nanoseconds precision). $\epsilon$ is a 1 to 4 bit value to indicate the next number of bits to read. `0' means the delta-of-delta ($\Delta\Delta$) is 0, while `10' means read the next 7 bits as the value is between -63 and 64 (range of $2^{7}$), `110' the next 24 bits, `1110' the next 32 bits. Finally, an $\epsilon$ of `1111' means reading 64 bits $\Delta\Delta$. The example follows with $\Delta\Delta$s of -2 and 0 stored in just 10 bits which reflect deltas of 3600 for the next 2 timestamps.

\emph{Delta-RLE-LEB128} The LEB128 encoding format is a variable-length encoding recommended in the DWARF debugging format specification \cite{dwarf2010dwarf} and used in Android's Dalvik Executable format. Numerical values like timestamps can be compressed efficiently along byte boundaries (minimum of 1 byte). In the example in Fig. \ref{fig:timestamp}, the header stores a full starting timestamp for the block in 64 bits followed by a run-length value, $\rho$, of 1 and the actual delta, $\Delta$, of 3602, both compressed with LEB128 to 8 and 16 bits respectively. The first bit in each 8 bits is a control bit that signifies to read another byte for the sequence if `1' or the last byte in the sequence if `0'. The remaining 7 bits are appended with any others in the sequence to form the numerical value. Binary `00001110 00010010' is formed from appending the last 7 bits from each byte of $\Delta$ which translates to the value of 3602 in base 10. This is followed by a run-length, $\rho$, of 2 $\Delta$s of 3600 each in the example.

\emph{Delta-RLE-Rice} We utilise the Rice coding format \cite{rice1971} to build a backward adaptation strategy inspired by Malvar's Run-Length/Golomb-Rice encoding \cite{Malvar2006} for tuning a $k$ parameter which allows us to adapt to timestamps and run-lengths of varying precision and periodicity respectively. Rice coding divides a value, $u$, into two parts based on $k$, giving a quotient $q =\left \lfloor{u/2^{k}}\right \rfloor $ and the remainder, $r = u\%2^{k}$. The quotient, $q$ is stored in unary coding, for example, the $\Delta$ value 3602 with a $k$ of 10 has a quotient of 3 and is stored as `1110'. The remainder, $r$, is binary coded in $k$ bits. Initial $k$ values of 2 and 10 are used in this example and are adaptively tuned based on the previous value in the sequence so this can be reproduced during decoding. 3 rules govern the tuning based on $q$.
\begin{displaymath}
    \text{if } q=
    \begin{cases}
      0, & k \to k-1 \\
      1, & \text{no change in } k \\
      \textgreater1, & k \to k+q 
    \end{cases}
  \end{displaymath}
This adaptive coding adjusts $k$ based on the actual data to be encoded so no other information needs to be retrieved on the side for decoding, has a fast learning rate that chooses good, though not necessarily optimal, $k$ values and does not have the delay of forward adaptation methods. $k$ is adapted from 2 and 10 to 1 and 13 respectively in Fig. \ref{fig:timestamp}.

\begin{table}%
\caption{Compressed size of timestamps and values in datasets with varying methods}
\label{tab:compression}
\begin{minipage}{\columnwidth}
\begin{center}
\begin{tabular}{lrrrrrrrr}
  \toprule
 Dataset & \multicolumn{4}{c}{Timestamps (MB)\footnote{$\delta_{\Delta}$ = Delta-of-delta, $\delta_{\emph{leb}}$ = Delta-RLE-LEB128, $\delta_{\emph{rice}}$ = Delta-RLE-Rice, $\delta_{\varnothing}$ = Delta-Uncompressed}} & \multicolumn{4}{c}{Values (MB)\footnote{$C_{\emph{gor}}$ = Gorilla, $C_{\emph{fpc}}$ = FPC, $C_{\Delta}$ = Delta-of-delta , $C_{\varnothing}$ = Uncompressed}}\\
  	\midrule
  {} & $\delta_{\Delta}$ & $\delta_{\emph{leb}}$ & $\delta_{\emph{rice}}$ & $\delta_{\varnothing}$ 
  & $C_{\emph{gor}}$ & $C_{\emph{fpc}}$ & $C_{\Delta}$ & $C_{\varnothing}$ \\	
  SRBench     	& 0.6		& 0.5		& \textbf{0.4}			& 5.2		 
  & \textbf{8.2}			& 23.8		& 21.9		& 33.9			\\
  Shelburne (ms)& \textbf{8.0}		& 18.3		& 13.6			& 99.5		
  & \multirow{2}{*}{440.8}		& \multirow{2}{*}{\textbf{419.3}}	 & \multirow{2}{*}{426.6} 	& \multirow{2}{*}{597.4} 		\\
  Shelburne (ns)& \textbf{35.9}		& 56.2		& 44.1			& 99.5		
  & 			& 			&  			& 			\\
  GreenTaxi  	& 4.0		& 6.9		& \textbf{1.5}			& 35.5 		
  & 342.1		& \textbf{317.1}		& 318.8		& 710.9		\\
    \bottomrule
\end{tabular}
\end{center}
\bigskip\centering
\end{minipage}
\end{table}%

Table \ref{tab:compression} shows the results of running each of the timestamp compression methods against each dataset. We observe that Delta-RLE-Rice, $\delta_{\emph{rice}}$, performs best for low precision timestamps (to the second) while Delta-of-delta compression, $\delta_{\Delta}$, performs well on high precision, milli and nanosecond timestamps. The adaptive $\delta_{\emph{rice}}$ performed exceptionally well on the GreenTaxi timestamps which were very small due to precision to seconds and small deltas. $\delta_{\Delta}$ performed well on Shelburne due to the somewhat evenly-spaced but large deltas (due to high precision).

\subsubsection{Value compression}
\label{subsec:val_compress}

As can be observed from Table \ref{tab:compression}, even the worse compression method for timestamps occupies  but a fraction of the total space using the best value compression method, $ \delta_{\emph{max}} \div (\delta_{\emph{max}} + C_{\emph{min}}) \times 100\%, $ which results in percentages of 6.8\%, 11.8\% and 2.1\% for SRBench, Shelburne and Green Taxi respectively. Hence, an effective compression method supporting hard-to-compress numerical values (both floating point numbers and long integers) can greatly improve compression ratios. We look at FPC, the fast floating point compression algorithm by Burtscher \emph{et~al.} \cite{Burtscher2009}, the simplified method used in Facebook's Gorilla \cite{Pelkonen2015} and Delta-of-delta in BTrDb \cite{Andersen2016}.

During compression, the FPC algorithm uses the more accurate of an fcm \cite{sazeides1997predictability} or a dfcm \cite{goeman2001differential} value predictor to predict the next value in a double-precision numerical sequence. Accuracy is determined by the number of significant bits shared by the two values. After an XOR operation between the predicted and actual values, the leading zeroes are collapsed into a 3-bit value and appended with a single bit indicating which predictor was used and the remaining non-zero bytes. As XOR is reversible and the predictors are effectively hash tables, lossless decompression can be performed. Gorilla does away with predictors and instead merely compares the current value to the previous value. After an XOR operation between the values, the result, $r$, is stored according to the output from a function \emph{gor()} described below, where $.$ is an operator that appends bits together, $p$ is the previous XOR value, \emph{lead()} and \emph{trail()} return the number of leading and trailing zeroes respectively, \emph{len()} returns the length in bits and $n$ are remaining meaningful bits within the value.
\begin{displaymath}
    \emph{gor}(r)=
    \begin{cases}
      '0', & \text{if } r = 0 \\
      '10'.n, & \text{if } lead(r) >= lead(p) \text{ and } trail(r) = trail(p)\\
      '11'.l.m.n, & \text{else, where } l = lead(r) \text{ and } m = len(n)
    \end{cases}
  \end{displaymath}

Anderson \emph{et~al.} \cite{Andersen2016} suggest the use of a delta-of-delta method for compressing the mantissa and exponent components of floating point numbers within a series separately. The method is not described in the paper but we interpret it as such: a IEEE-754 double precision floating point number \cite{ieee2008standard} can be split into sign, exponent and mantissa components. The 1 bit sign is written, followed by at most 11 bits delta-of-delta of the exponent, $\delta_{\emph{exp}}$, encoded by a function $E_{\emph{exp}}()$, described as follows, and at most 53 bits delta-of-delta of the mantissa, $\delta_{\emph{mantissa}}$, encoded by $E_{\emph{mantissa}}()$.
\begin{displaymath}
    E_{\emph{exp}}(\delta_{\emph{exp}})=
    \begin{cases}
      '0', & \text{if } \delta_{\emph{exp}} = 0 \\
      '1'.e, & \text{else, where } e = \delta_{\emph{exp}} + (2^{11}-1)\\
    \end{cases}
  \end{displaymath}
  \begin{displaymath}
    E_{\emph{mantissa}}(\delta_{\emph{mantissa}})=
    \begin{cases}
      '0', & \text{if } \delta_{\emph{mantissa}} = 0 \\
      '10'.m, & \text{if } -2^{6}+1 <= \delta_{\emph{mantissa}} <= 2^{6}\text{, } m = \delta_{\emph{mantissa}} + (2^{6}-1) \\
      '110'.m, &  \text{if } -2^{31}+1 <= \delta_{\emph{mantissa}} <= 2^{31}\text{, } m = \delta_{\emph{mantissa}} + (2^{31}-1)\\
      '1110'.m, & \text{if } -2^{47}+1 <= \delta_{\emph{mantissa}} <= 2^{47}\text{, } m = \delta_{\emph{mantissa}} + (2^{47}-1)\\
      '1111'.m, & \text{else, where } m = \delta_{\emph{mantissa}} + (2^{53}-1)\\
    \end{cases}
  \end{displaymath}
The operator $.$ appends binary coded values in the above functions. $e$ and $m$ are expressed in binary coding (of base 2). A maximum of 12 and 53 bits are needed for the exponent and mantissa deltas respectively as they could be negative.

Table \ref{tab:compression} shows the results comparing Gorilla, FPC and delta-of-delta value compression against each of the datasets. Each compression method has advantages, however, in terms of compression and decompression times, Gorilla compression consistently performs best as shown in Table \ref{tab:compress_speed} where each dataset is compressed to a file 100 times and the time taken is averaged. Each dataset is then decompressed from the files and time taken is averaged over a 100 tries. A read and write buffer of $2^{12}$ bytes was used. FPC has the best compression ratio on values with high decimal count in Shelburne and is slightly better on a range of field types in GreenTaxi than Delta-of-delta compression, however, even though the hash table prediction has similar speed to the Delta-of-delta technique, it is still up to 25\% slower on encoding than Gorilla. Gorilla though, expectedly trails FPC and delta-of-delta in terms of size for Shelburne and the Taxi datasets with more rows, as this is characteristic of the Gorilla algorithm being optimised for smaller partitioned blocks of data (this is explained in more detail in Section \ref{subsubsec:spaceamp} on Space Amplification).

\begin{table}%
\caption{Average (over 100 attempts) compression/decompression time of datasets}
\label{tab:compress_speed}
\begin{minipage}{\columnwidth}
\begin{center}
\begin{tabular}{lrrrlrrrl}
  \toprule
 Dataset & \multicolumn{4}{c}{Compression (s)\footnote{$C_{\emph{gor}}$ = Gorilla, $C_{\emph{fpc}}$ = FPC, $C_{\Delta}$ = Delta-of-delta}} & \multicolumn{4}{c}{Decompression (s)}\\
  	\midrule
  {} & $C_{\emph{gor}}$ & $C_{\emph{fpc}}$ & $C_{\Delta}$ & Top
  & $C_{\emph{gor}}$ & $C_{\emph{fpc}}$ & $C_{\Delta}$ & Top\\	
  SRBench     	& \textbf{2.10}	& 3.25	& 3.10	& $C_{\emph{gor}}$
  & \textbf{0.97} & 1.61 & 1.36			& $C_{gor}$\\
  Shelburne     & \textbf{30.68} & 42.02 & 40.91	& $C_{\emph{gor}}$
  & \textbf{3.80} & 4.57 &	5.42			& $C_{gor}$\\
  GreenTaxi     & \textbf{28.85} & 32.11 & 32.94	& $C_{\emph{gor}}$
  & \textbf{2.94} & 4.32 &	5.77			& $C_{gor}$\\
    \bottomrule
\end{tabular} 
\end{center}
\bigskip\centering
\end{minipage}
\end{table}%

\subsection{Storage Engine Data Structures and Indexing}

In time-series databases, as we saw in the previous section, Section \ref{subsec:compress}, data can be effectively compressed in time order. A common way of persisting this to disk is to partition each time-series by time to form time-partitioned blocks that can be aligned on page-sized boundaries or within memory-mapped files. In this section, we experiment with generalised implementations of data structures used in state-of-the-art time-series databases to store and retrieve time-partitioned blocks: concurrent B+ trees, Log-structured Merge (LSM) Trees and segmented Hash trees and each is explained in Sections \ref{subsubsec:btree} to \ref{subsubsec:lsmtree}. We also propose a Sorted String Table (SSTable) inspired, Tritan Table (TrTable) data structure for block storage in Section \ref{subsubsec:sstable}. 

Microbenchmarks aim to measure 3 metrics that characterise the performance of each data structure, write performance, read amplification and space amplification. Write performance is measured by the average time taken to ingest each of the datasets over a 100 tries. Borrowing from Kuszmaul's definitions \cite{Kuszmaul2014}, read amplification is `the number of input-output operations required to satisfy a particular query' and we measure this by taking the average of a 100 tries of scanning the whole database and the execution time of range queries over a 100 pairs of deterministic pseudo-random values with a fixed seed from the entire time range of each dataset. Space amplification is the `space required by a data structure that can be inflated by fragmentation or temporary copies of the data' and we measure this by the resulting size of the database after compaction operations. Each time-partitioned block is compressed using $\delta_{\Delta}$ and $C_{\emph{gor}}$ compression. Results for each of the metrics follow in Sections  \ref{subsubsec:spaceamp} to \ref{subsubsec:readamp}.

\subsubsection{B+ Tree-based}
\label{subsubsec:btree}

\begin{figure}
\centerline{\includegraphics[width=5.1in]{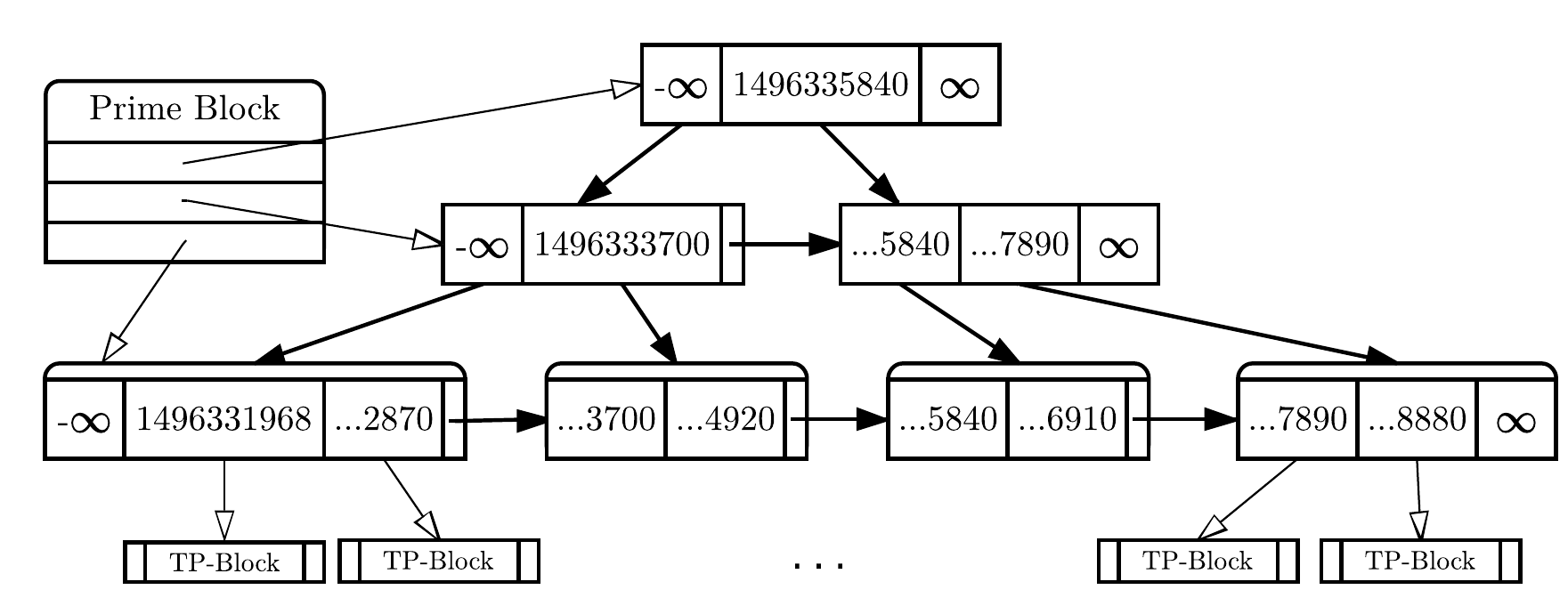}}
\caption{A $B^{\emph{Link}}$ tree with timestamps as keys and time-partitioned (TP) blocks stored outside nodes}
\label{fig:blink}
\end{figure}

A concurrent B+ Tree can be used to store time-partitioned blocks with the keys being block timestamps and the values being the compressed binary blocks. Time-series databases like Akumuli \cite{EugeneLazin2017} (LSM with B+ Trees instead of SSTables) and BTrDb \cite{Andersen2016} (append-only/copy-on-write) use variations of this data structure. We use an implementation of Sagiv’s \cite{Sagiv1986} $B^{\emph{Link}}$ balanced search tree utilising the algorithm verified in the work by Pinto \emph{et~al.} \cite{DaRochaPinto2011} in our experiments. The leaves of the tree are nodes that contain a fixed size list of key and value-pointer pairs stored in order. The value-pointer points to the actual block location so as to minimise the size of nodes that have to be read during traversal. The final pointer in each node’s list, called a link pointer, points to the next node at that level which allows for fast sequential traversal between nodes. A prime block includes pointers to the first node in each level. Fig. \ref{fig:blink} shows a $B^{\emph{Link}}$ tree with timestamps as keys and time-partitioned (TP) blocks stored off node.

\subsubsection{Hash Tree-based}
\label{subsubsec:htree}

Given that hashing is commonly used in building distributed storage systems and various time-series databases like Riak-TS (hash ring) \cite{Basho2016} and OpenTSDB on HBase (hash table) \cite{OpenTSDB} utilise hash-based structures internally, we investigate the generalised concurrent hash map data structure. A central difficulty of hash table implementations is defining an initial size of the root table especially for streaming time-series' of indefinite sizes. Instead of using a fixed sized hash table that suffers from fragmentation and requires rehashing data when it grows, an auto-expanding hash tree of hash indexes is used instead. Leaves of the tree contain expanding nodes with keys and value pointers. Concurrency is supported by implementing a variable (the concurrency factor) segmented Read-Write-Lock approach similar to that implemented in JDK7's \emph{ConcurrentHashMap} data structure \cite{Burnison2013} and 32 bit hashes for block timestamp keys are used.

\subsubsection{LSM Tree-based}
\label{subsubsec:lsmtree}

The Log-Structured Merge Tree \cite{ONeil1996} is a write optimised data structure used in time-series databases like InfluxDb \cite{InfluxData2017} (a variation called time-structured merge tree is used) and Cassandra-based \cite{LakshamAvinash2010} databases. High write throughput is achieved by performing sequential writes instead of dispersed, update-in-place operations that some tree based structures require. This particular implementation of the LSM tree is based on the bLSM design by Sears \emph{et~al.} \cite{Sears2012} and has an in-memory buffer, a \emph{memtable}, that holds block timestamp keys and time-partitioned blocks as values within a red-black tree (to preserve key ordering). When the memtable is full, the sorted data is flushed to a new file on disk requiring only a sequential write. Any new blocks or edits simply create successive files which are traversed in order during reads. The system periodically performs a compaction to merge files together, removing duplicates.

\subsubsection{TrTables}
\label{subsubsec:sstable}

Tritan Tables (TrTables) are our novel IoT time-series optimised storage data structure inspired by Sorted String Tables (SSTables) which consist a persistent, ordered, immutable map from keys to values used in  many big data systems like BigTable \cite{Chang2008}. TrTables include support for out-of-order timestamps within a time window with a \emph{quantum re-ordering buffer}, efficient sequential reads and writes due to maintaining a sorted order in-memory with a \emph{memtable} and on disk with a TrTable. Furthermore, a block index table also boosts range and aggregation queries. Keys in TrTables are block timestamps while values are compressed, time-partitioned blocks. TrTables also inherit other beneficial characteristics from SSTables, which are fitting for storing time-series IoT data, like simple locking semantics for only the \emph{memtable} with no contention on immutable TrTables. Furthermore, there is no need for a background compaction process like in LSM-tree based storage engines using SSTables as the \emph{memtable} for a time-series is always flushed to a single TrTable file. However, TrTables do not support expensive updates and deletions as we argue that there is no established use case for individual points within an IoT time-series in the past to be modified.

\begin{definition}[Quantum Re-ordering Buffer, $Q$, and Quantum, $q$] A quantum re-ordering buffer, $Q$, is a list-like window that contains a number of timestamp-row pairs as elements.  A quantum, $q$, is the amount of elements within $Q$ to cause an \emph{expiration operation} where an insertion sort is performed on the timestamps of $q$ elements and the first $a \times q$ elements are flushed to the \emph{memtable}, where $1 < a < 0$. The remaining $ (1-a) \times q $ elements now form the start of the window.
\end{definition}

\begin{figure}
\centerline{\includegraphics[width=\textwidth]{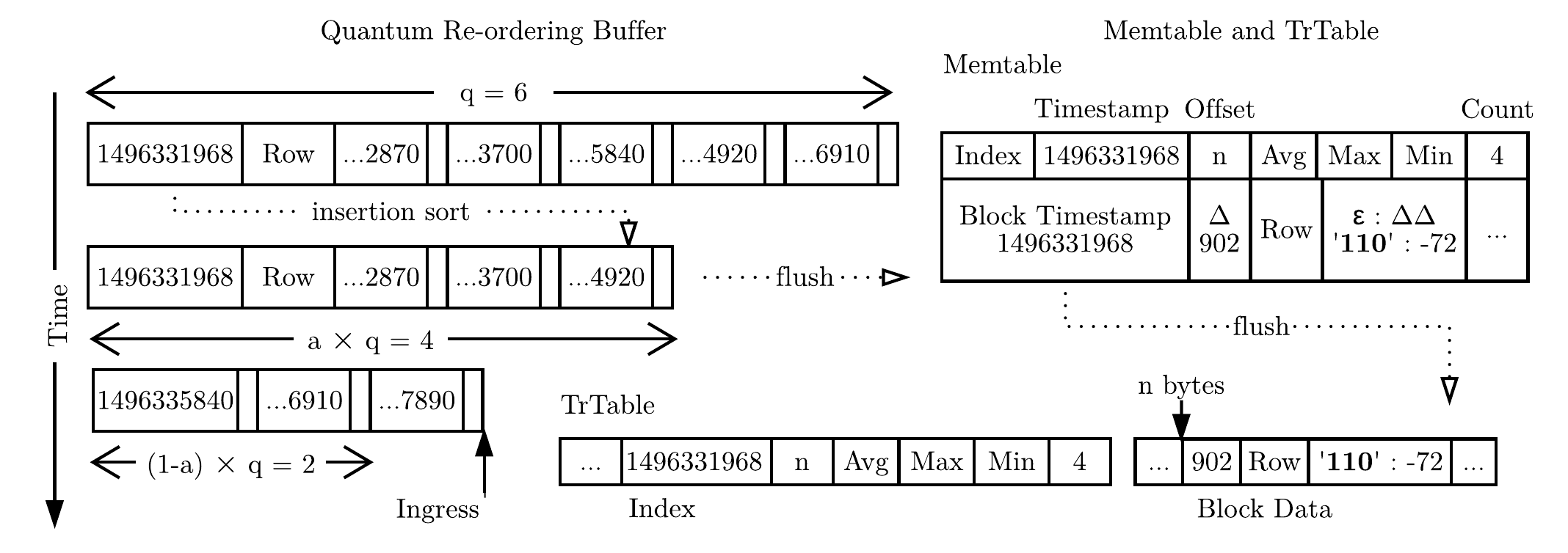}}
\caption{The quantum re-ordering buffer, memtable and TrTable in operation over time}
\label{fig:trtables}
\end{figure}

The insertion sort is efficient as the window is already substantially sorted, so the complexity is $O(nk)$ where $k$, the furthest distance of an element from its final sorted position, is small. Any timestamp now entering the re-ordering buffer less than the minimum allowed timestamp, $t_{\emph{minA}}$ (the first timestamp in the buffer) is rejected, marked as `late' and returned with a warning. Fig. \ref{fig:trtables} shows the operation of $Q$ over time (along the y-axis). When $Q$ has 6 elements and $q = 6$, an expiration operation occurs where an insertion sort is performed and the first 4 sorted elements are flushed to the \emph{memtable}. A new element that enters has timestamp, $t = 1496337890$, which is greater than $t_{\emph{minA}} = 1496335840$ and hence is appended at the end of $Q$. 

The \emph{memtable}, also shown in Fig. \ref{fig:trtables}, consists of an index entry, $i$, that stores values of the block timestamp, current TrTable offset and average, maximum, minimum and counts of the row data which are updated when elements from $Q$ are inserted. It also stores a block entry, $b$, which contains the current compressed time-partitioned block data. The memtable gets flushed to a TrTable on disk once it reaches the time-partitioned block size, $b_{\emph{size}}$. Each time-series has a \emph{memtable} and corresponding TrTable file on disk.

\subsubsection{Space Amplification and the effect of block size, $b_{\emph{size}}$}
\label{subsubsec:spaceamp}

The block size, $b_{\emph{size}}$, refers to the maximum size that each time-partitioned block occupies within a data structure. Base 2 multiples of $2^{12}$, the typical block size on file systems, are used such that $b_{\emph{size}} = 2^{12} \times 2^{x}$ and in these experiments we use $x = \{2..8\}$. Fig. \ref{fig:sizeamp} shows the database size in bytes, which suggests the space amplification, for the Shelburne and Taxi datasets of each data structure at varying $b_{\emph{size}}$. Both TrTables-LSM-tree and B+-tree-Hash-tree pairs have database sizes that are almost identical with the maximum difference only about 0.2\%, hence, they are grouped together in the figure. 

We notice a trend, the database size decreases as $b_{\emph{size}}$ decreases. This is a characteristic of the $C_{\emph{gor}}$ algorithm used for value compression described in Section \ref{subsec:val_compress} as more `localised' compression occurs. Each new time-partitioned block will trigger the else clause in the $gor(r)$ function to encode the longer $`11'.l.m.n$, however, the subsequent $lead(p)$ and $trail(p)$ are likely to be smaller and more `localised' and fewer significant bits will need to be used for values in these datasets.

TrTables and LSM-trees have smaller database sizes than the B+-tree and Hash-tree data structures for both datasets. As sorted keys and time-partitioned blocks in append-only, immutable structures like TrTables and the LSM-trees after compaction are stored in contiguous blocks on disk, they are expectedly more efficiently stored (size-wise). Results from SRBench are omitted as the largest time-partitioned block across all the stations is smaller than the smallest $b_{\emph{size}}$ where $x = 2$, hence, there is no variation across different $x$ values and $b_{\emph{size}}$.

We also avoid key clashing in tree-based stores for the Taxi dataset, where multiple trip records have the same starting timestamp, by using time-partitioned blocks where $b_{\emph{size}} > s_{\emph{size}}$, the longest compressed sequence with the same timestamp.

\input{minibench_datastructure}

\subsubsection{Write Performance}
\label{subsubsec:writeamp}

Fig. \ref{fig:ingestionamp} shows the ingestion time in milliseconds for the Shelburne and Taxi datasets of each data structure while varying $b_{\emph{size}}$. Both TrTables and LSM-tree perform consistently across $b_{\emph{size}}$ due to append-only sequential writes which corresponds to their log-structured nature. TrTables are about 8 and 16 seconds faster on average than LSM-tree for the Taxi and Shelburne datasets respectively due to no overhead of a compaction process. Both the Hash-Tree and B+-Tree perform much slower (up to 10 times slower on Taxi between the B+ tree and TrTables when $x=2$) on smaller $b_{\emph{size}}$ as each of these data structures are comparatively not write-optimised and the trees become expensive to maintain as the amount of keys grow. When $x=8$, the ingestion time for LSM-tree and Hash-trees converge, B+-trees are still slower while TrTables are still about 10s faster for both datasets. At this point, the bottleneck is no longer due to write amplification but rather subject to disk input-output.

For the concurrent B+-tree and and Hash-tree, both parallel and sequential writers were tested and the faster parallel times were used. In the parallel implementation, the write and commit operation for each time-partitioned block (a key-value pair) is handed to worker threads from a common pool using Kotlin's asynchronous coroutines \footnote{https://github.com/Kotlin/kotlinx.coroutines}.

\subsubsection{Read Amplification}
\label{subsubsec:readamp}

Fig. \ref{fig:scan} shows the execution time for a full scan on each data structure while varying $b_{\emph{size}}$ and Fig. \ref{fig:range} show the execution time for range queries. All scans and queries were averaged across a 100 tries and for the range queries, the same pseudo-random ranges with a fixed seed were used. The write-optimised LSM-tree performed the worst for full scans and while B+-trees and Hash-trees performed quite similarly, TrTables recorded the fastest execution times as a full scan on a TrTable file is efficient with almost no read amplification (a straightforward sequential read of the file with no intermediate seeks necessary).

From the results of the range queries in Fig. \ref{fig:range}, we see that LSM-tree has highest read amplification trying to access a sub-range of keys as a scan of keys across levels has to be performed, for both datasets, while the Hash-tree has the second highest read amplification, which is expected as it has to perform random input-output operations to retrieve time-partitioned blocks based on the distribution by the hash function. It is possible to use an order-preserving minimal perfect hashing function \cite{Czech1992} at the expense of hashing performance and space, however, this is out of the scope of our microbenchmarks. TrTables still has better performance on both datasets than the read-optimised B+-tree due to its index that guarantees a maximum of just one seek operation.

From these experiments and these datasets, $2^{12} \times 2^{4}$ bytes is the most suitable $b_{\emph{size}}$ for reads and TrTables has the best performance for both full scans and range queries at this $b_{\emph{size}}$. 

\input{minibench_datastructure2}

\subsubsection{Rounding up performance: TrTables and 64KB}
\label{subsubsec:roundup}

TrTables has the best write performance and storage size due to its simple, immutable, compressed, write-optimised structure that benefits from fast, batched sequential writes. The in-memory quantum re-ordering buffer and memtable support ingestion of out-of-order, unevenly-spaced data within a window, which is a requirement for IoT time-series data from wireless sensors. Furthermore, the memtable allows batched writes and amortises the compression time. A $b_{\emph{size}}$ of 64KB when $x = 4$ with TrTables also provides the best read performance across full scans and range queries of the various datasets.

B+-trees and Hash-trees have higher write amplification, especially for smaller $b_{\emph{size}}$ and LSM-trees have higher read amplification.

The specifications of the experimental setup for microbenchmarks had a $4 \times 3.2$GHz CPU, 8 GB memory and average disk data rate of 146.2 MB/s. 

\section{Query Translation}
\label{sec:translation}

\begin{figure}
\centerline{\includegraphics[width=4.8in]{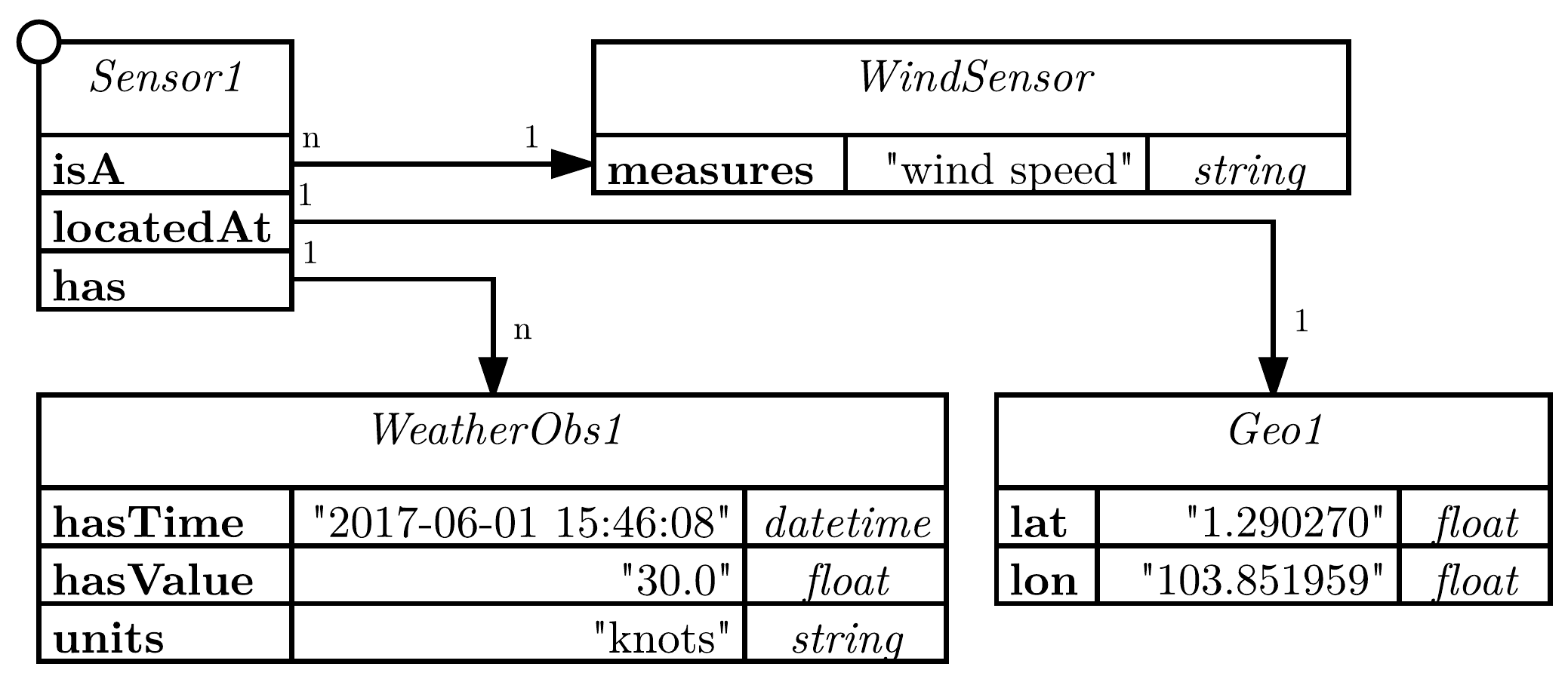}}
\caption{A data model of a sensor and observation measuring the wind speed}
\label{fig:data_model_wind}
\end{figure}

Consider the observation \texttt{`WeatherObs1'} recorded by \texttt{`Sensor1'} in Fig. \ref{fig:data_model_wind} and the data model in which it is represented. Such a rich data model has advantages for IoT data integration in that metadata useful for queries, applications and other Thing's to understand context can be attached to observation data (e.g. the location and type of sensor or the unit of measurement). Having a common \texttt{`WindSensor'} sensor class on top of a common data model for example, also helps Things to interoperate and understand the context of each others observations.

We can represent the above rich data model as a tree, a restricted, hierarchical form of directed graph without cycles and where a child node can have only one parent. JSON and the eXtensible Markup Language (XML) are popular implementations of a tree-based data model. However, if \texttt{`Sensor1'} is the root, the tree model cannot represent the many-to-one multiplicity of the relationship between the class of \texttt{`Sensor1'} and \texttt{`WindSensor'}. Hence, the query to find the wind speed observations across all sensors would require a full scan of all \texttt{`Sensor'} nodes in database terms. Furthermore, if we receive a stream of weather observations, we might like to model \texttt{`WeatherObs1'} as the root of each data point in the stream, an example of which is modelled using JSON Schema in Listing \ref{lst:json_schema_obs} with the actual data in a JSON document in Listing \ref{lst:json_obs}. Listing \ref{lst:json_schema_sensor} and \ref{lst:json_schema_sensortype} show the corresponding \texttt{`Sensor1'} and \texttt{`WindSensor'} schema models. Hence, each observation produces a significant amount of repetitive metadata derived from the sensor and sensor type schemata.

\noindent\begin{minipage}{.46\textwidth}
\begin{lstlisting}[style=wind,caption={sensor.json},label={lst:json_schema_sensor},frame=tlrb,basicstyle=\tiny\ttfamily]
{"$schema":".../draft-04/schema#",
 "description": ..., "type": "object",
 "properties": {
    "name": { "type": "string" },
    "isA": {"$ref":"sensortype.json"},
    "locatedAt": {
    	"$ref":"http://json-schema.org/geo" }}}
\end{lstlisting}
\end{minipage}\hfill
\begin{minipage}{.46\textwidth}
\begin{lstlisting}[style=wind,caption={sensortype.json},label={lst:json_schema_sensortype},frame=tlrb,basicstyle=\tiny\ttfamily]
{"$schema": ..., "description": ...,
 "type": "object",
 "properties": {
    "name": { "type": "string" },
    "measures": { "type": "string" },
   }
}
\end{lstlisting}
\end{minipage}

\begin{lstlisting}[style=wind,caption={JSON Schema of an observation, observation.json},label={lst:json_schema_obs},frame=tlrb,basicstyle=\tiny\ttfamily]
{"$schema": "http://json-schema.org/draft-04/schema#",
 "title": "observation", "description": "A weather observation document", "type": "object",
 "properties": {
    "name": { "type": "string" }, "hasValue": { "type": "number" },
    "hasTime": { "type": "string", "format": "date-time" }, 
    "units": { "type": "string" }, "has": {"$ref":"sensor.json"} }}
\end{lstlisting}

\begin{lstlisting}[style=wind,caption={JSON document of a single weather observation, weatherObs1.json},label={lst:json_obs},frame=tlrb,basicstyle=\tiny\ttfamily]
{"name": "WeatherObs1", "hasValue": 30.0, "units": "knots", "hasTime": "2017-06-01 15:46:08",
 "has" : { "name": "Sensor1", 
 	   "locatedAt" : {"latitude": 1.290270, "longitude":103.851959 }	,
 	   "isA": { "name": "WindSensor", "measures": "wind speed" }}}
\end{lstlisting}

The graph model, is less restrictive and relations can be used to reduce the repetition by referencing sensors and sensor types. The graph can be realised either as a property graph as per Fig. \ref{fig:data_model_wind}, where nodes can also store properties (key-value pairs), or as a general graph like an RDF graph where all properties are first class nodes as well. This means that we have more flexibility to model the multiplicities in the relationship between \texttt{`WeatherObs1'} and \texttt{`Sensor1'} with the former as parent and the similar many-to-one relationship between \texttt{`Sensor1'} and \texttt{`WindSensor'}. However, as studied in Section \ref{subsec:rdf_iot_data}, the characteristics of RDF IoT data show that there is a expansion of metadata in modelling observation metadata.

Hence, although both models are rich and promote interoperability, they also repetitively encode sensor and observation metadata which deviates from the efficient time-series storage structures we benchmarked in Section \ref{sec:microbenchmarks}. Therefore, we present a novel abstraction of a query translation algorithm titled \emph{map-match-operate} that allows us to query rich data models while preserving the efficient underlying time-series storage that exploits the characteristics of IoT data. We use examples of RDF graphs (as a rich data model) and corresponding SPARQL \cite{harris2013sparql} queries building on previous SPARQL-to-SQL work \cite{Siow2016b}. The abstraction can also be applied on other graph models or tree-based models like JSON documents with JSON Schema, which are restricted forms of a graph, but is not the focus of the paper.

\subsection{Map-Match-Operate: An Formal Abstraction for Time-Series Query Translation}
\label{subsec:map_match_operate}

We define \emph{map-match-operate} formally in Definition \ref{def:mmo} and define each step, map, match and operate in the following Sections \ref{subsubsec:map_op} to \ref{subsubsec:operate_op}. This process is meant to act on a rich graph data model abstracting time-series data, so as to translate a graph query to a set of operators that are executed on the underlying time-series database.

\begin{definition}[Map-Match-Operate, $\mu$] Given a time-series database, $T$, which stores a set of time-series, $t \in T$, a graph data model for each time-series, $m \in M$ where $M$ is the union of data models and a query, $q$, whose intention is to extract data from the $T$ through $M$, the Map-Match-Operate function, $\mu(q,M,T)$, returns an appropriate result set, $r$, of the query from set $M \times T$.
\label{def:mmo}
\end{definition}

\subsubsection{Map: Binding $M$ to $T$}
\label{subsubsec:map_op}

A rich graph data model, $m = (V,E)$, consists of a set of vertices, V, and edges, E. A time-series $t$, consists of a set of timestamps, $\tau$ and a set of all columns $C$ where each individual column $c \in C$. Definition \ref{def:map} describes the \emph{map} step on $m$ and $t$, which are elements of $M$ and $T$ respectively.

\begin{definition}[Map, $\mu_{\emph{map}}$] The map function, $\mu_{\emph{map}}(m,t) \rightarrow \mathbb{B}$, produces a binary relation, $\mathbb{B}$, between the set of vertices, $V$, and the set $(\tau \times C)$. Each element, $b \in \mathbb{B}$, is called a binding and $b = (x,y)$, where $x \in V $ and $y \in (\tau \times C) $. A data model mapping, $m_{\emph{map}}$, where $m_{\emph{map}} = m \bigcup \mathbb{B}$, integrates the binary relation consisting of bindings, $\mathbb{B}$, within a data model $m$. 
\label{def:map}
\end{definition}

An RDF graph, $m_{\emph{RDF}}$ is a type of graph data model that consists of a set of triple patterns, $tp = (s,p,o)$, whereby each triple pattern has a subject, $s$, predicate, $p$, and an object, $o$. A triple pattern describes a relationship where a vertex, $s$, is connected to a vertex, $o$, via an edge, $p$. Each $ s = \{I,B\} $ and each $ o = \{I,B,L\} $, where $I$ is a set of Internationalised Resource Identifiers (IRI), $B$ is a set of blank nodes and $L$ is a set of literal values. A binding $b_{\emph{RDF}} = (x_{\emph{RDF}},y)$, where $x_{\emph{RDF}} = (I \times L)$, is an element of $\mathbb{B_{\emph{RDF}}}$. The detailed formalisation of a data model mapping, $m_{\emph{map}}^{\emph{RDF}} = m_{\emph{RDF}} \bigcup \mathbb{B_{\emph{RDF}}}$, that extends the RDF graph can be found in work on S2SML \cite{Siow2016a}.

\subsubsection{Match: Retrieving $\mathbb{B}_{\emph{match}}$ by matching $q_{\emph{graph}}$ to $M_{\emph{map}}$}
\label{subsubsec:match_op}

The union of all data model mappings, $M_{\emph{map}} = \bigcup m_{\emph{map}}$, where each $m_{\emph{map}}$ relates to a subset of time-series in $T$ is used by the \emph{match} step expressed in Definition \ref{def:match}. $q_{\emph{graph}}$ is a subset of query, $q$, which describes variable vertices $V_{\emph{var}}$ and edges $E_{\emph{var}}$ within a graph model, intended to be retrieved from $M$ and subjected to other operators in $q$.

\begin{definition}[Match, $\mu_{\emph{match}}$] The match function, $\mu_{\emph{match}}(q_{\emph{graph}},M_{\emph{map}}) \rightarrow \mathbb{B}_{\emph{match}}$, produces a binary relation, $\mathbb{B}_{\emph{match}}$, between the set of variables from $q_{\emph{graph}}$, $\upsilon$, and the set $(\tau \times C \times V)$ from $T$ and $M_{\emph{map}}$ respectively. This is done by graph matching $q_{\emph{graph}}$ and the relevant $m_{\emph{map}}$ within $M_{\emph{map}}$. Each element, $b_{\emph{match}} \in \mathbb{B}_{\emph{match}}$, is a binding match where $b_{\emph{match}} = (a,b)$, $a \in \upsilon $ and $b \in (\tau \times C \times V) $.
\label{def:match}
\end{definition}

A graph query on an RDF graph can be expressed in the SPARQL Query Language for RDF \cite{harris2013sparql}. A SPARQL query can express multiple Basic Graph Patterns (BGPs), each consisting of a set of Triple Patterns, $tp$ and relating to a specific $m_{\emph{map}}^{\emph{RDF}}$. Any of the $s$, $p$ or $o$ in $tp$ can be a query variable from the set $\upsilon_{\emph{RDF}}$ within a BGP. Hence, $\mu_{\emph{match}}$ for RDF, is the matching of BGPs to the relevant $m_{\emph{map}}^{\emph{RDF}}$ and retrieving a result set, $\mathbb{B}_{\emph{match}}^{\emph{RDF}}$.

\subsubsection{Operate: Executing $q$'s Operators on $T$ and $M$ using the results of $\mathbb{B}_{\emph{match}}$}
\label{subsubsec:operate_op}

\begin{figure}[h!]
	\centering
\Tree[.\textit{Project, $\Pi$_{\emph{station}}} [.\textit{Union, $\cup$} [.\textit{Filter, $\sigma$_{\emph{time,snow}}} $Q_{\emph{graph}}^{\emph{snow}}$ ]
			[.\textit{Union, $\cup$} [.\textit{Filter, $\sigma$_{>30,\emph{time}}} $Q_{\emph{graph}}^{\emph{rain}}$ ]
				[.\textit{Filter, $\sigma$_{>100,\emph{time}}} $Q_{\emph{graph}}^{\emph{windSpeed}}$ ]]]]
\caption{Query Tree of Operators, checking stations where weather conditions are poor}
\label{diag:query_tree}
\end{figure}

A graph query, $q$, can be parsed to form a tree of operators (utilising well-known relational algebra vocabulary \cite{Ozsu2011}) like the one shown in Fig. \ref{diag:query_tree}. The leaf nodes of the tree are made up of specific $Q_{\emph{graph}}$ operators, which when executed, retrieve a set of values from $M \times T$ according to the specific $\mathbb{B}_{\emph{match}}$. For example, $Q_{\emph{graph}}^{\emph{windSpeed}}$ retrieves from $\mathbb{B}_{\emph{match}}^{\emph{windSpeed}}$ with a binding match $b_{\emph{match}}^{\emph{windSpeed}}$, the values from column $c = \emph{windSpeedCol}$, from $t = \emph{weatherTs}$, based on a $m$ like in Fig. \ref{fig:data_model_wind}. By traversing the tree from leaves to root, a sequence of operations, a high-level query execution plan, $s_q$, can be obtained and by executing each operation in $s_q$, a final result set, $R$, can be obtained. Such a sequence of operations to produce $R$ for Fig. \ref{diag:query_tree} can be seen in equations \ref{eq:union} and \ref{eq:seq}.

\begin{equation}\label{eq:union}
\cup_{1} = \cup(\sigma_{\emph{windSpeed} > 100 \land x < \emph{time} < y}(Q_{\emph{graph}}^{\emph{windSpeed}}),\sigma_{\emph{rain} > 100 \land x < \emph{time} < y}(Q_{\emph{graph}}^{\emph{rain}}))
\end{equation}

\begin{equation}\label{eq:seq}
R = \Pi_{\emph{station}}(\cup(\cup_{1},\sigma_{\emph{snow}=\emph{true} \land x < \emph{time} < y}(Q_{\emph{graph}}^{\emph{snow}})))
\end{equation}

A SPARQL query on an RDF graph model produces a tree of operators like in Fig. \ref{diag:query_tree} and the sequence represented in equations \ref{eq:union} and \ref{eq:seq} with each $Q_{\emph{graph}}$ operation working on the relevant $\mathbb{B}_{\emph{match}}^{\emph{RDF}}$ relation of a BGP match. Query 6 in SRBench \cite{Zhang2012a} that returns the stations that have observed extremely low visibility in the last hour has a query tree such as the example. Appendix \ref{Appendix:operate_translation} describes TritanDB operators and their conversion from SPARQL algebra operators.

\subsection{Practical Considerations for IoT data}
\label{subsec:practical_considerations}

In previous work on SPARQL-to-SQL translation by Siow \emph{et~al.} \cite{Siow2016b} for time-series IoT data that is flat and wide, storing row data in relational databases with query translation resulted in performance improvements on Things from 2 times to 3 orders of magnitude as compared to RDF stores. Conceptually, relational databases consists of two-dimensional table structures that can compactly store rows of wide observations. Physically, the interface to storage hardware is a one-dimensional one represented by a seek and retrieval, which native time-series databases seek to optimise. By generalising the solution with the formal \emph{map-match-operate} model, we look also to exploit the fact that there is a high proportion of numeric observation data and that it can be compressed efficiently, that point data in time-series is largely immutable and that there is the possibility of the IoT community converging on any of the various rich graph or tree-based data models for interoperability. As such, Section \ref{sec:design} and \ref{sec:experiments} seek to show the design and evaluation of TritanDB that address concerns of \begin{inparaenum}[1)]
\item the overhead of query translation, 
\item the performance against other state-of-the-art stores for IoT data and queries including relational stores,
\item the generalisability to rich data models and query languages other than RDF and SPARQL, 
\item and the ease of designing rich data models for the IoT with a reduced configuration philosophy and templating.
\end{inparaenum}

\section{Designing a Time-Series Database for rich IoT data models}
\label{sec:design}

To handle the high volume of incoming IoT data for ingestion and querying while balancing this with the Fog Computing use case in mind of deployments on both resource-constrained Things and the Cloud, we design and implement a high performance input stack on top of our TrTables storage engine in TritanDB.

\subsection{The Input Stack: A Non-blocking Req-Rep Broker and the Disruptor Pattern}
\label{subsec:input_stack}

\begin{figure}
\centerline{\includegraphics[width=4.1in]{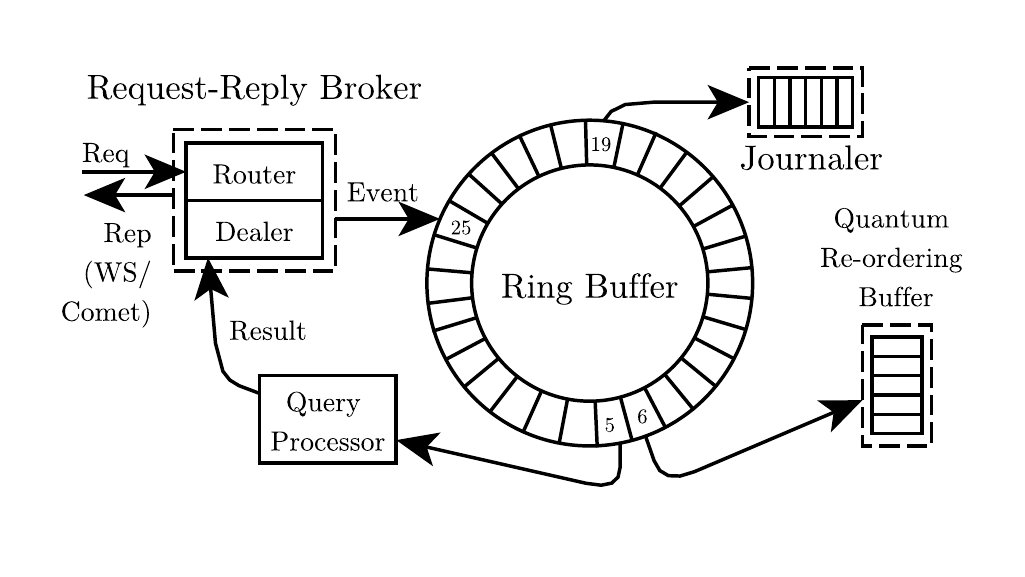}}
\caption{A Request-Reply broker for ingestion and querying utilising the Disruptor Pattern}
\label{fig:ringbuffer}
\end{figure}

The Constrained Application Protocol (CoAP) \footnote{http://coap.technology/}, MQTT \footnote{http://mqtt.org/} and HTTP are just some of many protocols used to communicate between devices in the IoT. Instead of making choices between these protocols, we design a non-blocking Request-Reply broker that works with ZeroMQ \footnote{http://zeromq.org/} sockets and library instead, so any protocol can be implemented on top of it. The broker is divided into a Router frontend component that clients bind to and send requests and a Dealer backend component that binds to a worker to forward requests. Replies are sent through the dealer to the router and then to clients. Fig. \ref{fig:ringbuffer} shows the broker design. All messages are serialised as protocol buffers, which are a small, fast, and simple means of binary transport with minimal overhead for structured data.

The worker that the dealer binds to is a high performance queue drawing inspiration from work on the Disruptor pattern \footnote{https://lmax-exchange.github.io/disruptor/} used in high frequency trading that reduces both cache misses at the CPU-level and locks requiring kernel arbitration by utilising a single thread. Data is referenced, as opposed to memory being copied, within a ring buffer structure. Furthermore, multiple processes can read data from the ring buffer without overtaking the head, ensuring consistency in the queue. Fig. \ref{fig:ringbuffer} shows the ring buffer with the producer, the dealer component of the broker, writing an entry at slot 25, which it has claimed by reading from a \emph{write counter}. Write contention is avoided as data is owned by only one thread for write access. Once done, the producer updates a \emph{read counter} with slot 25, representing the cursor for the latest entry available to consumers. The pre-allocated ring buffer with pointers to objects has a high chance of being laid out contiguously in main memory and thus supporting cache striding. Garbage collection is also avoided with pre-allocation. Consumers wait on the memory barrier and check they never overtake the head with \emph{read counter}.

A journaler at slot 19 records data on the ring buffer for crash recovery. If two journalers are deployed, one could record even slots while the other odd slots for better concurrent performance. The Quantum Re-ordering Buffer reads from slot 6 a row of time-series data to be ingested. Unfortunately, the memory needs to be copied and deserialised in this step. A Query Processor also reads a query request of slot 5, processes it and a reply is sent through the router to the client that contains the result of the query. 

The disruptor pattern describes an event-based asynchronous system. Hence, requests are converted to events when the worker bound to a dealer places them on the ring buffer. Replies are independent of the requests although they do contain the address of the client to respond to. Therefore, in a HTTP implementation on top of the broker, replies are sent chunked via a connection utilising either Comet style programming (long polling) or Websockets to clients.

\subsection{The Storage Engine: TrTables and $M_{\emph{map}}$ models and templates}
\label{subsec:trtables_design}

Tritan Tables (TrTables) form the basis of the storage engine and are a persistent, compressed, ordered, immutable and optimised time-partitioned block data structure. TrTables consist of four major components: a \emph{quantum re-ordering buffer} to support ingestion of out-of-order timestamps within a time quantum, a sorted in-memory time-partitioned block, a \emph{memtable} and persistent on-disk, sorted TrTable files for each time-series, consisting of \emph{time-partition blocks} and a \emph{block and aggregate index}. Section \ref{subsubsec:sstable} covers the design of TrTables in more detail.

Each time-partitioned block is compressed using the adaptive Delta-RLE-Rice encoding for lower precision timestamps and Delta-Delta compression for higher precision timestamps (milliseconds onwards) as explained in Section \ref{subsubsec:timestamp_compression}. Value Compression uses the Gorilla algorithm explained in Section \ref{subsec:val_compress}. Time-partitioned blocks of 64KB are used as analysed in Section \ref{subsubsec:roundup}.

When a time-series is created and a row of data is added to TritanDB, a $m_{\emph{map}}$ for this time-series is automatically generated according to a customisable set of templates based on the Semantic Sensor Network Ontology \cite{Compton2012} that models each column as an observation. The $m_{\emph{map}}$ can subsequently be modified on-the-fly, imported from RDF serialisation formats (XML, JSON, turtle \footnote{https://www.w3.org/TR/turtle/}, etc.) and exported. Internally, TritanDB stores the union of all $m_{\emph{map}}$, $M_{\emph{map}}$, as a fast in-memory model. Changes are persisted to disk using an efficient binary format, RDF Thrift \footnote{https://afs.github.io/rdf-thrift/}. The use of customisable templates helps to realise a reduced configuration philosophy on setup and input of time-series data, but still allows the flexibility of evolving a `schema-less' rich data model (limited only by bindings to time-series columns).

\subsection{The Query Engine: Swappable Interfaces}
\label{subsec:query_engine_design}

\begin{figure}
\centerline{\includegraphics[width=4.6in]{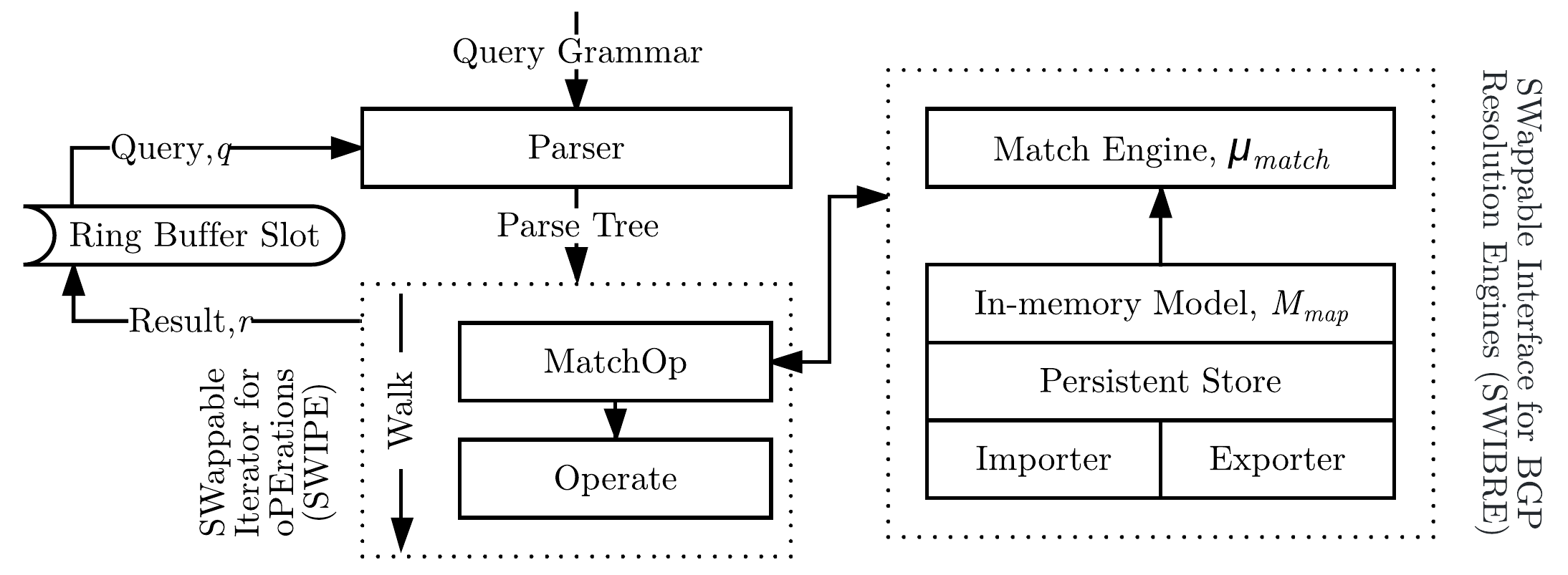}}
\caption{Modular query engine with swappable interfaces for \emph{Match}, \emph{Operate} and query grammar}
\label{fig:query_engine_design}
\end{figure}

Fig. \ref{fig:query_engine_design} shows the modular query engine design in TritanDB that can be extended to support other rich data models and query languages besides RDF and SPARQL. We argue that this is important for the generalisability to other graph and tree data models and any impact on runtime performance is minimised through the use of a modular design connected by pre-compiled interfaces and reflection in Kotlin. There are three main modular components, the \emph{parser}, the \emph{matcher} and the \emph{operator}. The compiled query grammar enables a \emph{parser} to produce a parse tree from an input query, $q$. The query request is accessed from the input ring buffer in Section \ref{subsec:input_stack}. The parse tree is walked by the \emph{operator} component that sends the $q_{\emph{graph}}$ leaves of a parse tree to the \emph{matcher}. The \emph{matcher} performs $\mu_{\emph{match}}$ based on the relevant $m_{\emph{map}}$ model from the in-memory $M_{\emph{map}}$ model described in Section \ref{subsec:trtables_design}. The match engine performing $\mu_{\emph{match}}$ can be overridden and a custom implementation based on a minimal, stripped-down version of Apache Jena's \footnote{https://jena.apache.org/} matcher is included. Alternative full Jena and Eclipse rdf4j \footnote{http://rdf4j.org/} matchers are also included. The $\mathbb{B}_{\emph{match}}$ is returned to the \emph{operator} which continues walking the parse tree and executing operations till a result, $r$ is returned at the root. This result is sent back to the requesting client through the Request-Reply broker. There is an open source implementation of TritanDB on Github \footnote{https://github.com/eugenesiow/tritandb-kt}. Details of the SWappable Iterator for oPerations (SWIPE) and the SWappable Interface for BGP Resolution (SWIBRE) build on previous work \footnote{https://eugenesiow.gitbooks.io/tritandb/}.

\subsection{Designing for Concurrency}
\label{subsec:concurrency}

Immutable TrTable files simplify the locking semantics to only the quantum re-ordering buffer (QRB) and memtable in TritanDB. Furthermore, reads on time-series data can always be associated with a range of time (if a range is unspecified, then the whole range of time) which simplifies the look up via a block index across the QRB, memtable and TrTable files. The QRB has the additional characteristic of minimising any blocking on the memtable writes as it flushes and writes to disk a TrTable as long as $t_{q}$, the time taken for the QRB to reach the next quantum expiration and flush to memtable, is more than $t_{\emph{write}}$, the  time taken to write the current memtable to disk.

The following listings describe some functions within TritanDB that elaborate on maintaining concurrency during ingestion and queries. The QRB is backed by a concurrent ArrayBlockingQueue in this implementation and inserting is shown in Listing \ref{lst:insert_code} where the flush to memtable has to be synchronised. The insertion sort needs to synchronise on the QRB as the remainder $(1-a) \times q$ values are put back in. The \texttt{`QRB.min'} is now the maximum of the flushed times.  Listing \ref{lst:flush_code} shows the synchronised code on the memtable and index to flush to disk and add to the BlockIndex. A synchronisation lock is necessary as time-series data need not be idempotent (i.e. same data in the memtable and TrTable at the same time is incorrect on reads). The memtable stores compressed data to amortise write cost, hence flushing to disk, $t_{write}$, is kept minimal and the time blocking is reduced as well. Listing \ref{lst:query_code} shows that a range query checks the index to obtain the blocks it needs to read, which can be from the QRB, memtable or TrTable, before it actually retrieves each of these blocks for the relevant time ranges. Listing \ref{lst:qrb_code} shows the internal get function in the QRB for iterating across rows to retrieve a range.

\begin{listing}
\refstepcounter{listing}
\noindent\begin{minipage}[b]{.48\textwidth}
    \begin{lstlisting} [language=Kotlin,frame=tlrb,basicstyle=\tiny\ttfamily]
fun insert(row) {
	if(QRB.length >= q) { 
		synchronized(QRB) { 
			arr = insertionSort(QRB.drain())
			QRB.put(remainder = arr.split(a*q,arr.length)) 
		} synchronized(memtable) { 
			memtable.addAll(flushed = arr.split(0,a*q-1)) 
			memidx.update()
		}} 
	row.time > QRB.min ? QRB.put(row, row.time) }
    \end{lstlisting}
    \captionof{sublisting}{Quantum Re-ordering Buffer (QRB) insert}
    \label{lst:insert_code}            
    \end{minipage}%
    \hfill
\begin{minipage}[b]{.48\textwidth}
    \begin{lstlisting}[language=Kotlin,frame=tlrb,basicstyle=\tiny\ttfamily]
fun flushMemTable() {
	synchronized(memtable) {
		TrTableWriter.flushToDisk(memtable,memidx)
		BlockIndex.add(memidx)
		memidx.clear() 
		memtable.clear()
	} }
    \end{lstlisting}
    \captionof{sublisting}{Flush memtable and index to disk}
    \label{lst:flush_code}            
\end{minipage}\\
\noindent\begin{minipage}[b]{.48\textwidth}
    \begin{lstlisting} [language=Kotlin,frame=tlrb,basicstyle=\tiny\ttfamily]
fun query(start,end):Result {
	blocks = BlockIndex.get(start,end) 
	for((btype,s,e,o) in blocks) { //relevant blocks
		when(btype) {
			`QRB' -> r += QRB.get(s,e)
			`memtable'-> r += memtable.get(s,e) 
			`trtable' -> r += trReader.get(s,e,o) //offset 
		}}
	return r }
    \end{lstlisting}
    \captionof{sublisting}{Query a range across memory and disk}
    \label{lst:query_code}            
    \end{minipage}%
    \hfill
\begin{minipage}[b]{.48\textwidth}
    \begin{lstlisting}[language=Kotlin,frame=tlrb,basicstyle=\tiny\ttfamily]
fun QRB.get(start,end):Result {
	for(row in this.iterator()) {
		if(row.time in start...end) r.add(row)
		else if(row.time > end) break }
	return r }
    \end{lstlisting}
    \captionof{sublisting}{Internal QRB functions get and put}
    \label{lst:qrb_code}            
\end{minipage}
\caption{Functions in TritanDB supporting concurrency for buffer, memtable and disk}
\label{lst:concurrency_code}
\end{listing}

\section{Experiments, Results and Discussion}
\label{sec:experiments}

The following section covers an experimental evaluation of TritanDB with other time-series, relational and NoSQL databases commonly used to store time-series data. Results are presented and discussed across a range of experimental setups, datasets and metrics for each database.

\subsection{Experimental Setup and Experiment Design}
\label{subsec:experimental_setup}

Due to the emergence of large volumes of streaming IoT data and a trend towards Fog Computing networks that Chiang \emph{et~al.} \cite{Chiang2017} describe as an `end-to-end horizontal architecture that distributes computing, storage, control, and networking functions closer to users along the cloud-to-thing continuum', there is a case for experimenting on cloud and Thing setups with varying specifications.

Table \ref{tab:experiment_spec} summarises the CPU, memory, disk data rate and Operating System (OS) specifications of each experimental setup. The disk data rate is measured by copying a file with random chunks and syncing the filesystem to remove the effect of caching. \emph{Server1} is a high memory setup with high disk data rate but lower compute (less cores). \emph{Server2} on the other hand is a lower memory setup with more CPU cores and a similarly high disk data rate. Both of these setups represent cloud-tier specifications in a Fog Computing network. The \emph{Pi2 B+} and \emph{Gizmo2} setups represent the Things-tier as compact, lightweight computers with low memory and CPU, an ARM and x86 processors respectively and a Class 10 SD card and mSATA SSD drives respectively with relatively lower disk data rates. The Things in these setups perform the role of low-powered, portable base stations or embedded sensor platforms within a Fog Computing network.

Databases tested, as we looked at in Related Work in Section \ref{sec:related_work}, include state-of-the-art time-series databases InfluxDB and Akumuli with innovative LSM-tree and B+-tree inspired storage engine designs respectively. We also benchmark against two popular NoSQL, schema-less databases that underly many emerging time-series databases: MongoDb and Cassandra. OpenTSDB, an established open-source time-series database that works on HBase, a distributed key-value store, is also tested. Other databases tested against include the lightweight but fully-featured relational database, H2 SQL and the search-index-based ElasticSearch which was shown to perform well for time-series monitoring by Mathe \emph{et~al.} \cite{Mathe2015}.

\begin{table}%
\caption{Specifications of each experimental setup}
\label{tab:experiment_spec}
\begin{minipage}{\columnwidth}
\begin{center}
\begin{tabular}{l|cccc}
  \toprule
  Specification & Server1 & Server2 & Gizmo2 & Pi2 B+\\
  	\midrule
CPU	     		& $2 \times 2.6$ GHz			& $4 \times 2.6$ GHz		& $2 \times 1$ GHz	& $4 \times 0.9$ GHz				\\
Memory 			& 32 GB 					& 4 GB					& 1 GB				& 1 GB\\
Disk Data Rate 	& 380.7 MB/s 				& 372.9 MB/s   			&  154 MB/s 			& 15.6 MB/s\\
OS 				& \multicolumn{3}{c}{Ubuntu 14.04 64-bit}   								& Raspbian Jessie 32-bit \\
\bottomrule
\end{tabular}
\end{center}
\end{minipage}
\end{table}%

We perform experiments on each setup described in Sections \ref{subsubsec:experiment_design_ingest} and \ref{subsubsec:experiment_design_query} to test the ingestion performance and query performance respectively for IoT data. Results and discussion for ingestion and storage performance are presented in Section \ref{subsec:storage} and for query performance in Section \ref{subsec:query_performance}.

\subsubsection{Ingestion Experimentation Design}
\label{subsubsec:experiment_design_ingest}

\begin{figure}
\centerline{\includegraphics[width=5.4in]{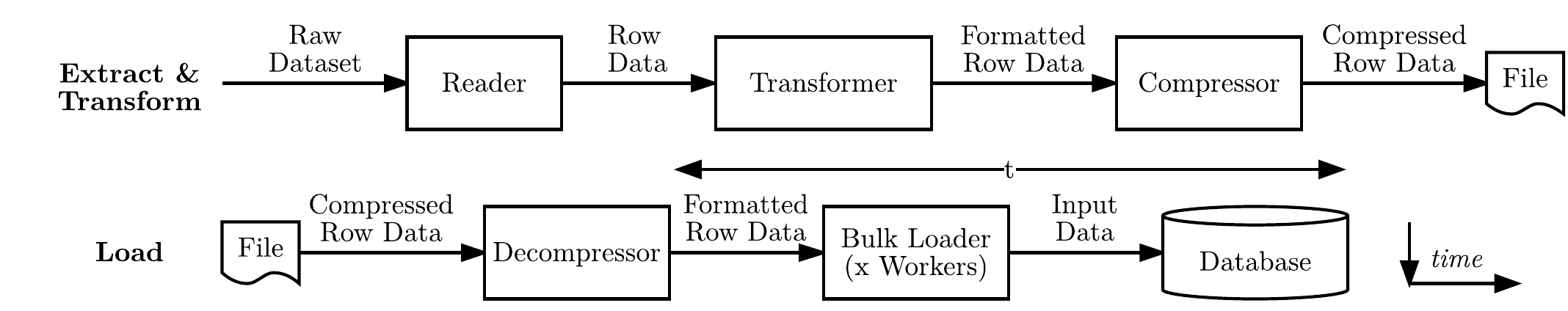}}
\caption{Ingestion experiment design described in terms of Extract, Transform and Load by time}
\label{fig:ingestion_design}
\end{figure}

Fig. \ref{fig:ingestion_design} summarises the ingestion experiment process in well-defined Extract, Transform and Load stages. A reader sends the raw dataset as rows to a transformer in the Extract stage. In the Transform stage, the transformer formats the data according to the intended database's bulk write protocol format and compressed using Gzip to a file. In the Load stage, the file is decompressed and the formatted data is sent to the database by a bulk loader which employs $x$ workers, where $x$ corresponds to the number of cores on an experimental setup. The average ingestion time, $t$, is measured by averaging across 5 runs for each setup, dataset and database. The average rate of ingestion for each setup, $s^1$, $s^2$, $p$ and $g$ is calculated by dividing the number of rows of each dataset by the average ingestion time. The storage space required for the database is measured 5 minutes after ingestion. Each database is deployed in a Docker container.

The schema design for MongoDB, Cassandra and OpenTSDB are optimised for reads in ad-hoc querying and follow the recommendations of Persen \emph{et~al.} in their series of technical papers on performantly mapping the time-series use case to each of these databases \cite{Persen2016,Persen2016a,Persen2016b}. This approach models each row by their individual fields in documents, columns or key-value pairs respectively with the tradeoff of storage space for query performance.

\subsubsection{Query Experimentation Design}
\label{subsubsec:experiment_design_query}

The aim of the query experimentation is to determine the overhead of query translation and the performance of TritanDB against other state-of-the-art stores for IoT data and queries. Particularly, we look at the following types of queries advised by literature for measuring the characteristics of time-series financial databases \cite{Jacob2000}, each is averaged across 100 fixed seed pseudo-random time ranges:
\begin{enumerate}
	\item Cross-sectional range queries that access all columns of a dataset.
	\item Deep-history range queries that access a random single column of a dataset.
	\item Aggregating a subset of columns of a dataset by arithmetic mean (average).
\end{enumerate}

The execution time of each query is measured as the time from sending the query request to when the query results have been completely written to a file on disk. The above queries are measured on the Shelburne and GreenTaxi datasets. 


Database-specific formats for ingestion and query experiments build on time-series database comparisons from InfluxDb and Akumuli \footnote{https://github.com/Lazin/influxdb-comparisons}.

\subsection{Discussing the storage and ingestion results}
\label{subsec:storage}

\begin{table}%
\caption{Storage space (in GB) required for each dataset on different databases}
\label{tab:storage}
\begin{minipage}{\columnwidth}
\begin{center}
\begin{tabular}{l|ccc}
  \toprule
  Database & Shelburne & Taxi & SRBench \\
  	\midrule
  TritanDB     		& \textbf{0.350}		& \textbf{0.294}		& 0.009		\\
  InfluxDb  		& 0.595		& \textbf{0.226}\footnote{InfluxDb points with the same timestamp are silently overwritten (due to its log-structured-merge-tree-based design), hence, database size is smaller as there are only $3.2 \times 10^6$ unique timestamps of $4.4 \times 10^6$ rows.}		& 0.015		\\
  Akumuli  			& 0.666		& 0.637		& \textbf{0.005}		\\
  MongoDb  			& 5.162		& 6.828		& 0.581		\\
  OpenTSDB  		& 0.742		& 1.958		& 0.248		\\
  H2 SQL	 		& 1.109/2.839\footnote{(size without indexes, size with an index on the timestamp column)}			& 0.579/1.387		& 0.033		\\
  Cassandra  		& 1.088		& 0.838		& 0.064		\\
  ElasticSearch (ES)	& 2.225			& 1.134		& - \footnote{As each station is an index, ES on even the high RAM \emph{Server1} setup failed when trying to create 4702 stations.}	\\
    \bottomrule
\end{tabular}
\end{center}
\end{minipage}
\end{table}%

Table \ref{tab:storage} shows the storage space, in gigabytes (GB), required for each dataset with each database. TritanDB that makes use of time-series compression, time-partitioning blocks and TrTables that have minimal space amplification has the best storage performance for the Shelburne and GreenTaxi datasets. It comes in second to Akumuli for the SRBench dataset. InfluxDb and Akumuli that also utilise time-series compression produce significantly smaller database sizes than the other relational and NoSQL stores. 

MongoDb needs the most storage space amongst the databases for the read-optimised schema design chosen while search index based ElasticSearch (ES) also requires more storage. ES also struggles with the SRBench dataset where creating many time-series as separate indexes fails even on the high RAM \emph{Server1} configuration. In this design, each of the 4702 stations is an index on its own to be consistent with the other database schema designs. 

As InfluxDb silently overwrites rows with the same timestamp, it shows a smaller database size for the GreenTaxi dataset of trips as trips for different taxis that start at the same timestamp are overwritten. Only $3.2 \times 10^6$ of $4.4 \times 10^6$ are stored eventually. It is possible to use tags to differentiate taxis in InfluxDb but this is limited by a fixed maximum tag cardinality of 100k.

TritanDB has from 1.7 times to an order of magnitude better storage efficiency than other databases for the larger Shelburne and Taxi datasets. It has a similar 1.7 to an order of magnitude advantage over all other databases except Akumuli for SRBench.

\begin{table}%
\caption{Average rate of ingestion for each dataset on different databases}
\label{tab:insertion}
\begin{minipage}{\columnwidth}
\begin{center}
\begin{tabular}{l|rrrrrr}
  \toprule
  Database & \multicolumn{3}{c}{Server1 ($10^3$ rows/s)} & \multicolumn{3}{c}{Server2 ($10^3$ rows/s)}  \\
  	\midrule
  {} 			&  $s_{\emph{shelburne}}^1$ & $s_{\emph{taxi}}^1$ & $s_{\emph{srbench}}^1$ &  $s_{\emph{shelburne}}^2$ & $s_{\emph{taxi}}^2$ & $s_{\emph{srbench}}^2$  \\
  TritanDB     	& \textbf{173.59}	& \textbf{68.28}	& \textbf{94.01}	& \textbf{252.82}	& \textbf{110.07}	& \textbf{180.19}	\\
  InfluxDb  	& 1.08	& 1.05	& 1.88	& 1.39		& 1.34		& 1.09			\\
  Akumuli  		& 49.63 & 18.96	& 61.78	& 46.44	& 17.71		& 59.23			\\
  MongoDb  		& 1.35	& 0.39	& 1.23	& 1.96	& 0.58		& 1.81			\\
  OpenTSDB  	& 0.26	& 0.08	& 0.24	& 0.25	& 0.07		& 0.22		\\
  H2 SQL	 	& 80.22	& 45.23	& 51.89	& 84.42	& 52.67 		& 77.12 \\
  Cassandra  	& 0.90	& 0.25	& 0.78	& 1.47	& 0.45 		& 1.66		\\
  ES			& 0.10	& 0.09	& - 		& 0.11	& 0.04		& -\\
    \toprule
    & \multicolumn{3}{c}{Pi2 B+ ($10^2$ rows/s)\footnote{Note the difference in order of magnitude of $10^2$ rather than $10^3$}} & \multicolumn{3}{c}{Gizmo2 ($10^3$ rows/s)} \\
    \midrule
   {} 			& $p_{\emph{shelburne}}$ & $p_{\emph{taxi}}$ & $p_{\emph{srbench}}$ &  $g_{\emph{shelburne}}$ & $g_{\emph{taxi}}$ & $g_{\emph{srbench}}$ \\   
 TritanDB 		& \textbf{73.68}	& \textbf{26.58}			& \textbf{48.42}		& \textbf{32.62}	 & \textbf{12.77}	& \textbf{14.05}	\\		
 InfluxDb 		& 1.33	& 1.28	 	& 1.43 		& 0.26		& 0.25	& 0.28	\\
 Akumuli 		& -\footnote{At the time of writing, Akumuli does not support ARM systems.} 	& -	& - & 9.79	& 3.84	& 10.48 \\
 MongoDb 		& -\footnote{Ingestion on MongoDb on the 32-bit Pi2 for larger datasets fails due to memory limitations.}	& -			& 1.78	& 0.22	 & 0.06		& 0.21	\\
 OpenTSDB 		& 0.10		& 0.03 		& 0.08	 & 0.05 		& 0.02 	& 0.05 \\
 H2 SQL 		& 32.11		& 18.80	 	& 34.26	& 15.13		& 8.30	& 10.42 \\
 Cassandra  	& 0.67	 	& 0.27		& 0.85	& 0.16		& 0.05	& 0.15	\\
 ES 			& -\footnote{Ingestion on ElasticSearch fails due to memory limitations (Java heap space).} 		& - 		& -		& 0.03 		& 0.01 	& - \\
 \bottomrule  
\end{tabular}
\end{center}
\end{minipage}
\end{table}%

Table \ref{tab:insertion} shows the average rate of ingestion, in rows per second, for each dataset with each database, across setups. From \emph{Server1} and \emph{Server2} setups, we notice that TritanDB, InfluxDb, MongoDb, H2 SQL and Cassandra all perform better with more processor cores rather than more memory while Akumuli and OpenTSDB perform slightly better on the high memory \emph{Server2} setup with slightly better disk data rate. For both setups and all datasets, TritanDB has the highest rate of ingestion from 1.5 times to 3 orders of magnitude higher on \emph{Server1} and from 2 times to 3 orders of magnitude higher on \emph{Server2} due to the ring buffer and sequential write out to TrTables.

The Class 10 SD card, a Sandisk Extreme with best-of-class advertised write speeds, of the \emph{Pi2 B+} setup is an order of magnitude slower than the mSATA SSD of the \emph{Gizmo2} setup. Certain databases like Akumuli did not support the 32-bit ARM \emph{Pi2 B+} setup at the time of writing so some experiments could not be carried out. On the \emph{Gizmo2}, TritanDB ingestion rates were about 8 to 12 times slower than on \emph{Server2} due to a slower CPU with less cores, however, it still performed the best amongst the databases and was at least 1.3 times faster than its nearest competitor, H2 SQL.

\subsection{Evaluating Query Performance and Translation Overhead}
\label{subsec:query_performance}

\subsubsection{Query Translation Overhead}
\label{subsubsec:query_translation_overhead}

The translation overhead is the time taken to parse the input query, perform the \emph{match} and \emph{operate} steps and produce a query plan for execution. The JVM is shutdown after each run and a gradle compile and execute task starts the next run to minimise the impact of previous runs on run time. Time for loading the models in the \emph{map} step is not included as this occurs on startup of TritanDB rather than at query time. Table \ref{tab:query_overhead} shows the query translation overhead, averaged across a 100 different queries of each type (e.g. cross-sectional, deep-history) and then averaged amongst datasets, across different setups. 

The mean query overhead for all three types of queries are similar with deep-history queries the simplest in terms of query tree complexity followed by aggregation and then cross-sectional queries which involve unions between graphs. The results reflect this order. Queries on the Pi2 B+ and Gizmo2 are an order of magnitude slower than those running on the server setups, however, still execute in sub-second times and can be improved with caching of query trees. When executed in a sequence without restarting the JVM, subsequent query overhead is under 10ms on the Pi2 B+ and Gizmo2 and under 2ms on the server setups. The Gizmo2 is faster than the Pi2 B+ in processing queries and Server2 is slightly faster than Server1.

\begin{table}%
\caption{Average query overhead (in ms) for various queries across different setups}
\label{tab:query_overhead}
\begin{minipage}{\columnwidth}
\begin{center}
\begin{tabular}{l|ccc}
  \toprule
  Setup & Cross-sectional & Deep-history & Aggregation \\
  	\midrule
  Server1     		& 53.99		& 52.16		& 53.72		\\
  Server2     		& 58.54		& 53.31		& 53.99		\\
  Pi2 B+  			& 581.99	& 531.95		& 537.80		\\
  Gizmo2  			& 449.33	& 380.18 	& 410.58 	\\
    \bottomrule
\end{tabular}
\end{center}
\end{minipage}
\end{table}%

\subsubsection{Cross-sectional, Deep-history and Aggregation Queries}
\label{subsubsec:range_queries}

Fig. \ref{fig:cross_sec_server_range} shows the results of a cross-sectional range query on the server setups $s^1$ and $s^2$. As cross-sectional queries are wide and involve retrieving many fields/columns from each row of data, the columnar schema design in MongoDb (each document as a field of a row) has the slowest  average execution time. Furthermore, the wider Taxi dataset (20 columns) has longer execution times than the narrower Shelburne dataset (6 columns). This disparity between datasets is also true for Cassandra, where a similar schema design is used. Row-based H2 SQL and ElasticSearch (where each row is a document), show the inverse phenomena between datasets. Purpose-built time-series databases TritanDB, OpenTSDB and Akumuli perform the best for this type of query. TritanDB has the fastest average query execution time of about 2.4 times better than the next best OpenTSDB running on HBase (which does not support the Taxi dataset due to multiple duplicate timestamps in the dataset) and 4.7 times faster than third best Akumuli for cross-sectional range queries on server setups.

\input{bench_query}

\begin{table}%
\caption{Average query execution time for various queries with TritanDB on the Pi2 B+ and Gizmo2}
\label{tab:pi_gizmo_query}
\begin{minipage}{\columnwidth}
\begin{center}
\begin{tabular}{l|rrrrrr}
  \toprule
  TritanDB & \multicolumn{2}{c}{Pi2 B+ (s)} & \multicolumn{2}{c}{Gizmo2 (s)} & \multicolumn{2}{c}{Ratio ($s$:$p$:$g$)}  \\
  	\midrule
  {} 			& $p_{\emph{shelburne}}$ & $p_{\emph{taxi}}$ &  $g_{\emph{shelburne}}$ & $g_{\emph{taxi}}$  &  $r_{\emph{shelburne}}$ & $r_{\emph{taxi}}$ \\
  Cross-Sectional   & 388.18	& 375.97		& 54.44 	& 62.19 	& 1:35:5 	& 1:36:6\\
  Deep-History  	& 111.10 	& 47.95		& 19.03 	& 21.37 & 1:22:4 	& 1:30:13\\
  Aggregation  		& 0.13		& 0.16		& 0.07 	& 0.06	& 1:6:3		& 1:15:6 \\
 \bottomrule  
\end{tabular}
\end{center}
\end{minipage}
\end{table}%

 Fig. \ref{fig:deep_hist_server_range} shows the average execution time for each database on a mean of $s^1$ and $s^2$ setups for deep-history range queries. We see that all databases and not only those that utilise columnar storage design perform better on the Taxi dataset than on Shelburne when retrieving deep-history with a single column due to there being less rows of data in Taxi. TritanDB has the fastest query execution times for deep-history queries as well and is 1.1 times faster than OpenTSDB and 3 times faster than the third best Cassandra. Both OpenTSDB and Cassandra have columnar schema design optimised for retrieving deep history queries which explains the narrower performance gap than for cross-sectional queries. ElasticSearch which stores rows as documents and requires a filter to retrieve a field from each document performs poorly for deep-history queries.
 
 Table \ref{tab:pi_gizmo_query} shows the average execution time for various queries on TritanDB on both Things setups. The Gizmo2 is faster than the Pi2 B+ and is from 3 to 13 times slower than the mean of the server setups execution times across various queries. The Pi2 B+ setup is 6 to 36 times slower than the servers. We observe an inversion of results between the narrow Shelburne and wide Taxi datasets on the Gizmo2 for both the cross-sectional and deep-history queries where the bottleneck is the CPU for reading and decompressing  time-partitioned blocks. However, the bottleneck shifts to the slow write speed of the Pi2 B+ to SD card and so the more rows of Shelburne take precedence in performance metrics.

Fig. \ref{fig:aggr_server_range} shows the average execution time for each database on a mean of $s^1$ and $s^2$ setups for aggregation range queries. An average aggregator is used in the queries on a subset of columns and a $10^{1} log_{10}$ scale is used to fit the range of execution times in the graph. TritanDB,  Akumuli have the fastest execution times (within about 10-100ms) as they both store aggregates for blocks (e.g. sum, max, min, count) within the block index in memory and B+ tree structure respectively. TritanDB performs a fast lookup of the index in-memory and scans the first and last blocks and is 3.3 and 1.2 times faster than Akumuli for the Shelburne and Taxi datasets respectively. Native time-series databases like InfluxDB, TritanDB and Akumuli perform the best for aggregation queries as this is a key optimisation for time-series rollup and resampling operations. ElasticSearch also performs well for aggregation queries with indexing tuned specifically for time-series metrics, agreeing with independent benchmark results \cite{Mathe2015}. Additional results are presented in Appendix \ref{Appendix:evaluation_extra}.


\section{Time-series Analytics on TritanDB}
\label{sec:analytics}

\subsection{Resampling and converting unevenly-spaced to evenly spaced time-series}
\label{subsec:resampling}

It was discovered that IoT data collected from across a range of domains in the IoT consisted of both evenly-spaced and non-evenly spaced time-series. While there exists an extensive body of literature on the analysis of evenly-spaced time-series data \cite{Lutkepohl2010}, few methods exist specifically for unevenly-spaced time series data. As Eckner \cite{Eckner2014} explains, this was because the basic theory for time-series analysis was developed ``when limitations in computing resources favoured the analysis of equally spaced data'', where efficient linear algebra routines could be used to provide explicit solutions. TritanDB that works across both resource-constrained fog computing platforms and the cloud, provides two methods for dealing with unevenly-spaced data.

One method of transforming unevenly-spaced to evenly-spaced time-series in TritanDB is resampling. This is achieved by splitting the time series into time buckets and applying an aggregation function, such as an \texttt{`AVG'} function, to perform linear interpolation on the values in that series. Listing \ref{lst:sparql_resample} shows a SPARQL 1.1 query that converts an unevenly-spaced time-series to an hourly aggregated time-series of average wind speed values per hour. Unfortunately, as time-partitioned blocks in TritanDB are based on a fixed block size, the index is created without knowledge of hourly boundaries. In the worse case, a full scan will have to be performed on each block for such a query.

\begin{lstlisting}[language=sql,style=sparql,caption={SPARQL query on TritanDB to resample the wind speed time-series by hours},label={lst:sparql_resample},basicstyle=\scriptsize\ttfamily]
SELECT (AVG(?wsVal) AS ?val) WHERE {
	?sensor isA windSensor;
		has ?obs.
	?obs hasValue ?wsVal;
		hasTime ?time.
	FILTER (?time>"2003-04-01T00:00:00" && ?time<"2003-04-01T01:00:00"^^xsd:dateTime)
} GROUP BY hours(?time)
\end{lstlisting}

As Eckner \cite{Eckner2014} summarises from a series of examples, performing the conversion from unevenly-spaced to evenly-spaced time-series results in data loss with dense points and dilution with sparse points, the loss of time information like the frequency of observations, and affects causality. The linear interpolation used in resampling also ``ignores the stochasticity around the conditional mean'' which leads to a difficult-to-quantify but significant bias when various methods of analysing evenly-spaced time-series are applied, as shown in experiments comparing correlation analysis techniques by Rehfeld \emph{et~al.} \cite{Rehfeld2011}.

Hence, a more graceful approach to working with unevenly-spaced time-series is to use Simple Moving Averages (SMA). Each successive average is calculated from a moving window of a certain time horizon, $\tau$, over the time-series. An efficient algorithm to do so is from an SMA function defined in Definition \ref{def:sma} as proposed by Eckner \cite{Eckner2017}. 

\begin{definition}[Simple Moving Average, $\emph{SMA}(X,\tau)_t$] Given an unevenly-spaced time-series, $X$, the simple moving average, for a time horizon of $\tau$ where $\tau>0$, for $t \in T(X)$ is $\emph{SMA}(X,\tau)_t = \frac{1}{\tau} \int_{0}^{\tau} X[t-s] ds$. $T(X)$ is the vector of the observation times within the time-series $X$, $X[t]$ is the sampled value of time series $X$ at time $t$ and $s$ is the spacing of observation times.
\label{def:sma}
\end{definition}

\begin{figure}[ht]
\centering
\includegraphics[width=0.7\textwidth]{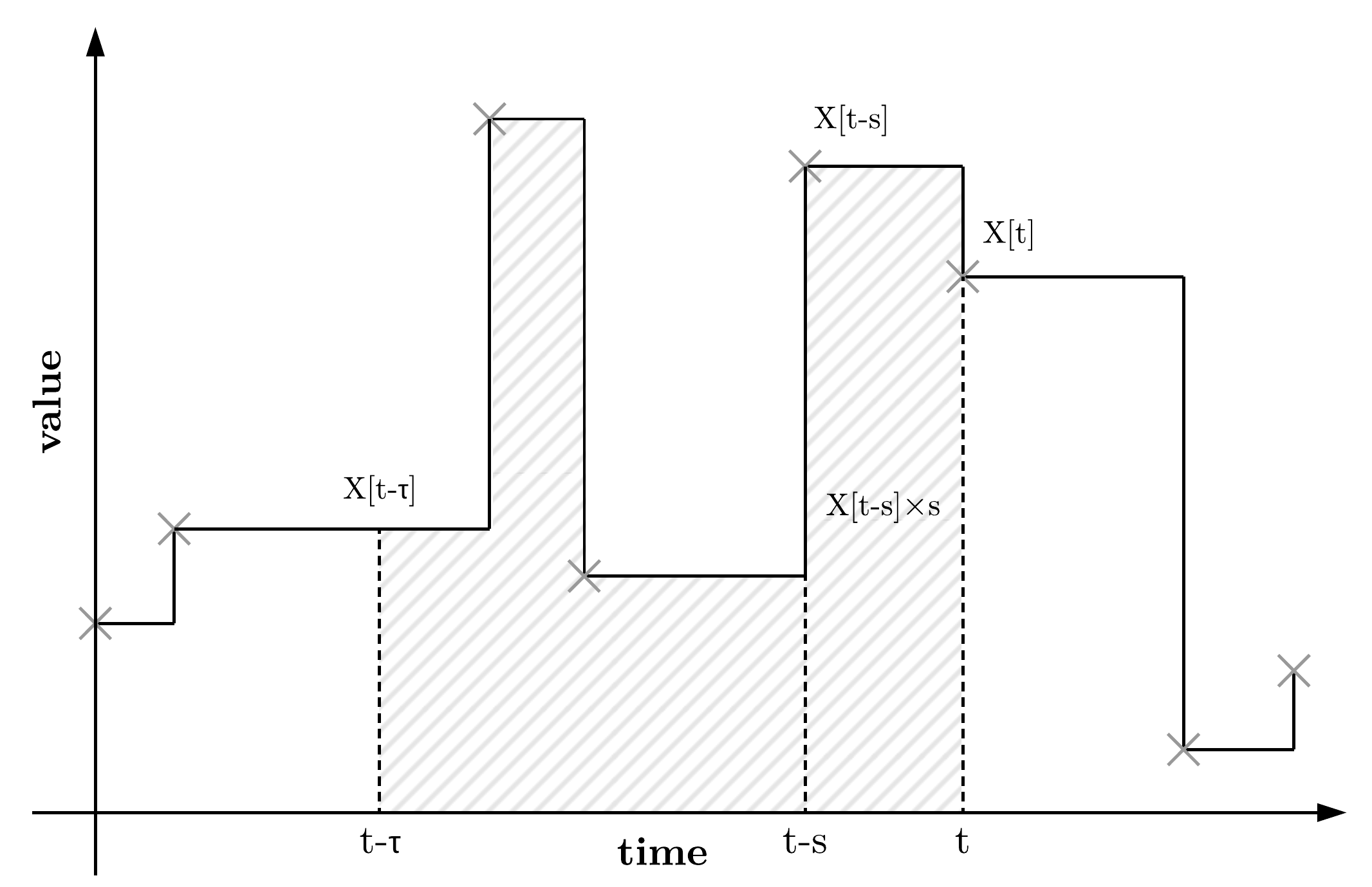}
\caption{Visualisation of Simple Moving Average Calculation}
\label{fig:sma_diagram}
\end{figure}

Figure \ref{fig:sma_diagram} shows a visualisation of how SMA is calculated. Each observation is marked by a cross in the figure and this particular time horizon is from $t-\tau$ to $t$. The area under the graph averaged over $\tau$ gives the SMA value for this window. In this case, $s$ is the time interval between the rightmost observation at $t$ and the previous observation.

The \texttt{`hours()'} function in the query in Listing \ref{lst:sparql_resample} can be changed to a \texttt{`sma(?time,tau)'} function from an extension to SPARQL implemented in TritanDB. This produces an SMA time-series using the efficient algorithm implementing the SMA function by Eckner \cite{Eckner2017} and shown in Listing \ref{lst:moving_average}.

\bigskip
\begin{lstlisting}[language=Kotlin,caption={Algorithm to Calculate Simple Moving Average},label={lst:moving_average},basicstyle=\scriptsize\ttfamily]
left = 1; area = left_area = X[1] * tau; SMA[1] = X[1];
for (right in 2:N(X)) {
	// Expand interval on right end
	area = area + X[right-1] * (T[right] - T[right-1]);
	// Remove truncated area on left end
	area = area - left_area;
	// Shrink interval on left end
	t_left_new = T[right] - tau;
	while (T[left] <= t_left_new) {
		area = area - X[left] * (T[left+1] - T[left]);
		left = left + 1;
	}
	// Add truncated area on left end
	left_area = X[max(1, left-1)] * (T[left] - t_left_new)
	area = area + left_area;
	// Save SMA value for current time window
	SMA[right] = area / tau;
}   	
\end{lstlisting}

This algorithm incrementally calculates the SMA for a N(X)-sized moving window, where N(X) is the number of observations in time-series X, reusing the previous calculations. There are four main areas (under the graph) involved for each SMA value calculated, the right area, the central area, the new left area and the left area. The central area is unchanged in each calculation. The algorithm first expands and adds the new right area from T[right] - T[right-1], where \texttt{`right'} is the new rightmost observation. The leftmost area from the previous iteration is removed and any additional area to the left less than $T[\emph{right}] - \tau$, the time horizon, is also removed. The removed area is the left area. A new left area from $T[\emph{right}] - \tau$ till the next observation is then calculated and added to the total area. This new total area is then divided by the time horizon value, $\tau$, to obtain the SMA for this particular window.

\subsection{Models: Seasonal ARIMA and forecasting}
\label{subsec:forecasting}

TritanDB includes an extendable model operator that is added to queries as function extensions to SPARQL. An example is shown in Listing \ref{lst:sparql_model} which shows how a forecasting of the next months points of evenly-spaced time-series can be made using a moving average function, a seasonal ARIMA model function with a 4-week seasonal cycle and a years worth of time-series data in a SPARQL query.

\begin{lstlisting}[language=sql,style=sparql,caption={Forecasting a month with a 4-week-cycle Seasonal ARIMA Model on a year of time-series},label={lst:sparql_model},basicstyle=\scriptsize\ttfamily]
SELECT (FORECAST(?tVal,30) AS ?val) WHERE {
	?sensor isA tempSensor;
		has ?obs.
	?obs hasValue ?tVal;
		hasTime ?time.
	FILTER (?time>"2011-04-01T00:00:00" && ?time<"2012-04-01T00:00:00"^^xsd:dateTime)
} GROUP BY ARIMA_S(sma(?time,1d),4w)
\end{lstlisting}

The result set of the query includes a forecast of 30 points representing values of the temperature sensor in the next month. ARIMA and random walk models are also included from an open source time-series analysis library \footnote{https://github.com/jrachiele/java-timeseries}.

\section{Conclusions and Future Work}
\label{sec:conclusion}

In this paper, we tackled the requirements of performance and interoperability when handling the increasing amount of streaming data from the Internet of Things (IoT), building on advances in time-series databases for telemetry data and efficient query translation on rich data models. The investigation of the structure of public IoT data provides a basis to design database systems according to the characteristics of flat, wide, numerical and a mix of both evenly and unevenly-spaced time-series data. The microbenchmarks and benchmarks also provide strong arguments for the effectiveness of time-partitioned blocks, timestamp and value compression algorithms and immutable data structures with in-memory tables for time-series IoT storage and processing. Furthermore, benchmarks on both cloud servers and resource-constrained Things, comparing across native time-series databases, relational databases and NoSQL storage provides a foundation for understanding performance within the IoT and Fog Computing networks. In terms of performance, there is still a disparity between cloud and Things performance which provides a case for resampling and aggregations for real-time analysis.

The included generalised map-match-operate method for query translation encourages the development of rich data models for data integration and interoperability in the IoT and we develop one possible actualisation with the Resource Description Framework (RDF) and SPARQL query language. Simple analytical features like models for forecasting with time-series data are also explored. The possibilities for future research that we are  pursuing are the specific optimisation of query plans for time-series data and workloads and understanding the challenges of scaling and partitioning time-series data especially across Fog Computing networks.
\appendix
\section{Additional Evaluation Results}
\label{Appendix:evaluation_extra}

In this appendix, we present the additional evaluation results that were omitted in the main paper for brevity. 
Table \ref{tab:cross_sec_pi_gizmo} shows the average cross-sectional query execution time on the Pi2 B+ and the Gizmo2 while Table \ref{tab:deep_history_pi_gizmo} shows the results for deep-history queries and Table \ref{tab:aggregation_pi_gizmo} shows the results for aggregation. TritanDB was faster than other databases on each type of query on the Thing setups of the Pi2 B+ and Gizmo2 as well.

\begin{table}[h!]%
\caption{Average query execution time for cross-sectional queries on the Pi2 B+ and Gizmo2}
\label{tab:cross_sec_pi_gizmo}
\begin{minipage}{\columnwidth}
\begin{center}
\begin{tabular}{l|rrrr}
  \toprule
  Database & \multicolumn{2}{c}{Pi2 B+ ($10^2$ s)} & \multicolumn{2}{c}{Gizmo2 ($10^2$ s)} \\
  	\midrule
  {} 			& $p_{\emph{shelburne}}$ & $p_{\emph{taxi}}$ &  $g_{\emph{shelburne}}$ & $g_{\emph{taxi}}$  \\
  InfluxDb  	& -	\footnote{InfluxDB encounters out of memory errors for cross-sectional and deep-history queries on both setups.}		& -			& -	 	& -		  \\
  Akumuli	  	& -			& -			& 2.46 	& 2.04	  \\
  MongoDb  		& -			& -			& 28.35 	& 41.22	  \\
  OpenTSDB 		& -	\footnote{OpenTSDB on both setups runs out of memory incurring Java Heap Space errors on all 3 types of queries.}		& -			& -	 	& -		  \\
  H2 SQL  		& 21.65		& 11.49		& 10.45 	& 3.75	  \\
  Cassandra 	& 26.44		& 25.28		& 6.82 	& 6.15	  \\
  ElasticSearch	& -			& -			& 8.74	& 5.54	  \\
 \bottomrule  
\end{tabular}
\end{center}
\end{minipage}
\end{table}%

\begin{table}[h!]%
\caption{Average query execution time for deep-history queries on the Pi2 B+ and Gizmo2}
\label{tab:deep_history_pi_gizmo}
\begin{minipage}{\columnwidth}
\begin{center}
\begin{tabular}{l|rrrr}
  \toprule
  Database & \multicolumn{2}{c}{Pi2 B+ ($10^2$ s)} & \multicolumn{2}{c}{Gizmo2 ($10^2$ s)}  \\
  	\midrule
  {} 			& $p_{\emph{shelburne}}$ & $p_{\emph{taxi}}$ &  $g_{\emph{shelburne}}$ & $g_{\emph{taxi}}$ \\
  InfluxDb  	& -			& -			& -	 		& -			  \\
  Akumuli	  	& -			& -			& 1.14 		& 0.40		  \\
  MongoDb  		& -			& -			& 4.44		& 1.88		  \\
  OpenTSDB 		& -			& -			& -	 		& -			  \\
  H2 SQL  		& 13.52		& 5.03		& 3.04	 	& 1.25		  \\
  Cassandra 	& 4.77		& 1.30		& 1.10 		& 0.32		  \\
  ElasticSearch	& -			& -			& 8.64 		& 3.53		 \\
 \bottomrule  
\end{tabular}
\end{center}
\end{minipage}
\end{table}%

\begin{table}[h!]%
\caption{Average query execution time for aggregation queries on the Pi2 B+ and Gizmo2}
\label{tab:aggregation_pi_gizmo}
\begin{minipage}{\columnwidth}
\begin{center}
\begin{tabular}{l|rrrr}
  \toprule
  Database & \multicolumn{2}{c}{Pi2 B+ (s)} & \multicolumn{2}{c}{Gizmo2 (s)}  \\
  	\midrule
  {} 			& $p_{\emph{shelburne}}$ & $p_{\emph{taxi}}$ &  $g_{\emph{shelburne}}$ & $g_{\emph{taxi}}$ \\
  InfluxDb  	& 7.10		& 1.76		& 2.75 		& 0.62		  \\
  Akumuli	  	& -			& -			& 0.37 		& 0.33		  \\
  MongoDb  		& -			& -			& 96.45 		& 57.90		  \\
  OpenTSDB 		& -			& -			& -	 		& -			  \\
  H2 SQL  		& 400.40	& 192.54		& 126.49		& 54.04		  \\
  Cassandra 	& 102.67	& 56.75		& 35.32 		& 19.89		  \\
  ElasticSearch	& -			& -			& 0.45 		& 0.24		  \\
 \bottomrule  
\end{tabular}
\end{center}
\end{minipage}
\end{table}%

\section{Operate: SPARQL to TritanDB Operators}
\label{Appendix:operate_translation}

In this appendix, we present the set of SPARQL algebra operators (excluding property path operations which are not supported) and their corresponding translation to TritanDB operators in the operate step of Map-Match-Operate. The list of SPARQL algebra is obtained from the SPARQL 1.1 specification under the `Translation to SPARQL algebra' section \footnote{https://www.w3.org/TR/sparql11-query/\#sparqlQuery} and follows the OpVisitor \footnote{https://jena.apache.org/documentation/javadoc/arq/org/apache/jena/sparql/algebra/OpVisitor.html} implementation from Apache Jena. The implementation of the set of TritanDB operators was inspired by the relational algebra Application Programming Interface (API) specification of Apache Calcite \footnote{https://calcite.apache.org/docs/algebra.html}. Table \ref{tab:operators} shows the conversion from SPARQL algebra to TritanDB operator.

\begin{table}[h!]%
\caption{SPARQL Algebra and the corresponding TritanDB Operator in the Operate step}
\label{tab:operators}
\begin{minipage}{\columnwidth}
\begin{center}
\begin{tabular}{l|r}
  \toprule
SPARQL Algebra & TritanDB Operator  \\
  	\midrule
\multicolumn{2}{c}{Graph Pattern}\\
  	\midrule
BGP, $Q_{\emph{graph}}$	& match(BGP,map), scan(TS) \\
Join, $\bowtie$			& join(expr...) \\
LeftJoin, $\ltimes$		& semiJoin(expr) \\
Filter,$\sigma$			& filter(expr...) \\
Union, $\cup$			& union() \\
Graph					& setMap(map) \\
Extend					& extend(expr,var) \\
Minus					& minus() \\
Group/Aggregation		& aggregate(groupKey, aggr) \\
  	\midrule
\multicolumn{2}{c}{Solution Modifiers}\\
\midrule
OrderBy					& sort(fieldOrdinal...) \\
Project, $\Pi$			& project(exprList [, fieldNames]) \\
Distinct				& distinct() \\
Reduced					& distinct() \\
Slice					& limit(offset, fetch) \\
 \bottomrule  
\end{tabular}
\end{center}
\end{minipage}
\end{table}%

\texttt{match} is described in Definition \ref{def:match} which matches a Basic Graph Pattern (BGP) from a query with a mapping to produce a binding $\mathbb{B}$. A set of time-series are referenced within $\mathbb{B}$. \texttt{scan} is an operator that returns an iterator over a time-series \texttt{TS}.

\texttt{join} combines two time-series according to conditions specified as \texttt{expr} while \texttt{semiJoin} joins two time-series according to some condition, but outputs only columns from the left input.

\texttt{filter} modifies the input to return an iterator over points for which the conditions specified in \texttt{expr} evaluate to true. A common filter condition would be one over time for a time-series.

\texttt{union} returns the union of the input time-series and bindings $\mathbb{B}$. If the same time-series is referenced within inputs, only the bindings need to be merged. If two different time-series are merged, the iterator is formed in linear time by a comparison-based sorting algorithm, the merge step within a merge sort, as the time-series are retrieved in sorted time order.

\texttt{setMap} is used to apply the specified mapping to its algebra tree leaf nodes for \texttt{match}.

\texttt{extend} allows the evaluation of an expression \texttt{expr} to be bound to a new variable \texttt{var}. This evaluation is performed only if \texttt{var} is projected. There are three means in SPARQL to produce the algebra: using \texttt{bind}, expressions in the \texttt{select} clause or expressions in the \texttt{group by} clause.

\texttt{minus} returns the iterator of first input excluding points from the second input.

\texttt{aggregate} produces an iteration over a set of aggregated results from an input. To calculate aggregate values for an input, the input is first divided into one or more groups by the \texttt{groupKey} field and the aggregate value is calculated for the particular \texttt{aggr} function for each group. The functions supported are \texttt{count}, \texttt{sum}, \texttt{avg}, \texttt{min}, \texttt{max}, \texttt{sample} and \texttt{groupconcat}.

\texttt{sort} imposes a particular sort order on its input based on a sequence consisting of \texttt{fieldOrdinals}, each defining the time-series field index (zero-based) and specifying a positive ordinal for ascending and negative for descending order.

\texttt{project} computes the set of chosen variables to 'select' from its input, as specified by \texttt{exprList}, and returns an iterator to the result containing only the selected variables. The default name of variables provided can be renamed by specifying the new name within the \texttt{fieldNames} argument.

\texttt{distinct} eliminates all duplicate records while \texttt{reduced}, in the TritanDB implementation, performs the same function. The SPARQL specification defines the difference being that \texttt{distinct} ensures duplicate elimination while \texttt{reduced} simply permitting duplicate elimination. Given that time-series are retrieved in sorted order of time, the \texttt{distinct} function works the same for both and eliminates immediately repeated duplicate result rows.

\texttt{limit} computes a window over the input returning an iterator over results that are of a maximum size (in rows) of \texttt{fetch} and are a distance of \texttt{offset} from the start of the results.

\bibliographystyle{ACM-Reference-Format}
\bibliography{tritandb}

%% file: minibench_datastructure.tex
\begin{figure}
\centering

\begin{minipage}{.5\textwidth}
  \centering
  \begin{tikzpicture}
\begin{axis}[
    mark list fill={.!75!white},
    title={},
    xlabel={Block Size, $b_{\emph{size}}$ ($2^{12} \times 2^{x}$  bytes)},
    ylabel={Database Size (bytes)},
    ylabel near ticks,
    xmin=2, xmax=8,
    ymin=280000000, ymax=420000000,
    legend pos=north west,
    ymajorgrids=true,
    grid style=dashed,
    legend style={font=\tiny},
    height=2.45in,
    cycle list name=mycolorlist
]

\addplot
    coordinates {
(2,337193353)
(3,343118952)
(4,350423715)
(5,359187669)
(6,369108421)
(7,378393344)
(8,386287757)
};

\addplot
    coordinates {
(2,283692640)	
(3,287916828)	
(4,294397455)
(5,302702911)	
(6,310169149)	
(7,316288764)	
(8,320270263)
};

\addplot
    coordinates {
(2,343932928)
(3,355467264)	
(4,375390208)	
(5,384827392)	
(6,395313152)	
(7,401604608)	
(8,412090368)
};

\addplot
    coordinates {
(2,290455552)	
(3,298844160)	
(4,314572800)	
(5,324009984)	
(6,331350016)	
(7,333447168)	
(8,341835776)
};

\legend{$S_{\emph{Tr}}$/$S_{\emph{lsm}}$,
		$T_{\emph{Tr}}$/$T_{\emph{lsm}}$,
		$S_{B+}$/$S_{H}$,
		$T_{B+}$/$T_{H}$}
    
\end{axis}
\end{tikzpicture}
  \captionof{figure}{Database size at varying $b_{\emph{size}}$}
  \label{fig:sizeamp}
\end{minipage}%
\begin{minipage}{.5\textwidth}
  \centering
    \begin{tikzpicture}
\begin{axis}[
    mark list fill={.!75!white},
    title={},
    xlabel={Block Size, $b_{\emph{size}}$ ($2^{12} \times 2^{x}$  bytes)},
    ylabel={Ingestion Time (ms)},
    xmin=2, xmax=8,
    ymin=28000, ymax=520000,
    ylabel near ticks, yticklabel pos=right,
    legend pos=north east,
    ymajorgrids=true,
    grid style=dashed,
    legend style={font=\tiny},
    height=2.45in
]

\addplot+[Dark2-6-1,mark options={Dark2-6-1}]
    coordinates {
(2,76373)
(3,70545)
(4,75976)
(5,76435)
(6,72321)
(7,70325)
(8,70344)
};

\addplot+[Dark2-6-A,mark options={Dark2-6-A}]
    coordinates {
(2,35942)
(3,32442)
(4,29531)
(5,28387)
(6,27537)
(7,27677)
(8,28716)
};

\addplot+[Dark2-6-2,mark options={Dark2-6-2},mark=otimes]
    coordinates {
(2,487600)
(3,279659)
(4,202063)
(5,125418)
(6,98304)
(7,87222)
(8,78288)
};

\addplot+[Dark2-6-B,mark options={Dark2-6-B}]
    coordinates {
(2,306578)
(3,179522)
(4,104566)
(5,67182)
(6,56972)
(7,43121)
(8,38053)
};

\addplot+[Dark2-6-3,mark options={Dark2-6-3}]
    coordinates {
(2,518202)
(3,381662)
(4,256121)
(5,161424)
(6,135963)
(7,113940)
(8,96008)
};

\addplot+[Dark2-6-C,mark options={Dark2-6-C}]
    coordinates {
(2,324727)
(3,173834)
(4,107820)
(5,69779)
(6,58893)
(7,47063)
(8,38571)
};

\addplot+[Dark2-6-4,mark options={Dark2-6-4}]
    coordinates {
(2,91270)
(3,93607)
(4,87385)
(5,90303)
(6,89416)
(7,89972)
(8,86876)
};

\addplot+[Dark2-6-D,mark options={Dark2-6-D}]
    coordinates {
(2,37774)
(3,40913)
(4,37923)
(5,38911)
(6,37865)
(7,37021)
(8,38110)
};

\legend{$S_{\emph{Tr}}$,
		$T_{\emph{Tr}}$,
		$S_{H}$,
		$T_{H}$,
		$S_{B+}$,
		$T_{B+}$,
		$S_{\emph{lsm}}$,
		$T_{\emph{lsm}}$,
		}
    
\end{axis}
\end{tikzpicture}
  \captionof{figure}{Ingestion times at varying $b_{\emph{size}}$}
  \label{fig:ingestionamp}
\end{minipage}
\hfill \break
{\tiny \textbf{Datasets:} Shelburne = $S$, Taxi = $T$, 
\textbf{Data Structures:} TrTables = Tr, LSM-Tree = lsm, B+ Tree = B+, Hash Tree = H}
\end{figure}

%% file: minibench_datastructure2.tex
\begin{figure}
\centering
\begin{subfigure}{.5\textwidth}
  \centering
  \begin{tikzpicture}
\begin{axis}[
    mark list fill={.!75!white},
    title={},
    xlabel={Block Size, $b_{\emph{size}}$ ($2^{12} \times 2^{x}$  bytes)},
    ylabel={Execution Time (s)},
    ylabel near ticks,
    xmin=2, xmax=8,
    ymin=3.6, ymax=4.2,
    legend pos=north west,
    ymajorgrids=true,
    grid style=dashed,
    legend style={font=\tiny},
    legend columns=4,
    height=2.4in,
    cycle list name=mycolorlist
]

\addplot
    coordinates {
(2,3.775555556)
(3,3.632111111)
(4,3.605888889)
(5,3.615777778)
(6,3.638)
(7,3.699777778)
(8,3.803111111)
};

\addplot
    coordinates {
(2,3.889333333)
(3,3.776666667)
(4,3.824555556)
(5,3.864666667)
(6,3.941666667)
(7,3.974444444)
(8,4.033777778)
};

\addplot
    coordinates {
(2,3.888888889)
(3,3.752555556)
(4,3.891222222)
(5,3.949222222)
(6,3.923111111)
(7,4.008555556)
(8,4.076222222)
};

\addplot
    coordinates {
(2,4.006888889)
(3,3.933666667)
(4,3.901)
(5,3.915)
(6,3.965)
(7,4.042)
(8,4.111)
};

\legend{$\emph{Tr}$,
		$B+$,
		$H$,
		$\emph{lsm}$,}
    
\end{axis}
\end{tikzpicture}
  \caption{Shelburne Dataset}
  \label{fig:scan_s}
\end{subfigure}%
\begin{subfigure}{.5\textwidth}
  \centering
\begin{tikzpicture}
\begin{axis}[
    mark list fill={.!75!white},
    title={},
    xlabel={Block Size, $b_{\emph{size}}$ ($2^{12} \times 2^{x}$  bytes)},
    ylabel={Execution Time (s)},
    ylabel near ticks, yticklabel pos=right,
    xmin=2, xmax=8,
    ymin=3.0, ymax=3.8,
    legend pos=north west,
    ymajorgrids=true,
    grid style=dashed,
    legend style={font=\tiny},
    legend columns=4,
    height=2.4in,
    cycle list name=mycolorlist
]

\addplot
    coordinates {
(2,3.145444444)
(3,3.065222222)
(4,3.045555556)
(5,3.055666667)
(6,3.106777778)
(7,3.209888889)
(8,3.216888889)
};

\addplot
    coordinates {
(2,3.390333333)
(3,3.298888889)	
(4,3.403666667)
(5,3.445444444)
(6,3.459777778)
(7,3.484777778)
(8,3.556666667)
};

\addplot
    coordinates {
(2,3.493555556)
(3,3.381222222)
(4,3.470555556)
(5,3.550222222)
(6,3.536666667)
(7,3.575555556)
(8,3.719666667)
};

\addplot
    coordinates {
(2,3.706666667)
(3,3.605888889)
(4,3.575)
(5,3.593)
(6,3.633)
(7,3.657)
(8,3.755)
};

\legend{$\emph{Tr}$,
		$B+$,
		$H$,
		$\emph{lsm}$,}
    
\end{axis}
\end{tikzpicture}
  \caption{Taxi Dataset}
  \label{fig:scan_t}
\end{subfigure}
\caption{Full scan execution time per $b_{\emph{size}}$}
\label{fig:scan}
{\tiny \textbf{Data Structures:} TrTables = Tr, LSM-Tree = lsm, B+ Tree = B+, Hash Tree = H}
\end{figure}

\begin{figure}
\centering
\begin{subfigure}{.5\textwidth}
  \centering
  \begin{tikzpicture}
\begin{axis}[
    mark list fill={.!75!white},
    title={},
    xlabel={Block Size, $b_{\emph{size}}$ ($2^{12} \times 2^{x}$  bytes)},
    ylabel={Execution Time (s)},
    ylabel near ticks,
    xmin=2, xmax=8,
    ymin=1, ymax=1.3,
    legend pos=north east,
    ymajorgrids=true,
    grid style=dashed,
    legend style={font=\tiny},
    legend columns=4,
    height=2.4in,
    cycle list name=mycolorlist
]

\addplot
    coordinates {
(2,1.07291)
(3,1.03748)
(4,1.01436)
(5,1.02252)
(6,1.04581)
(7,1.02191)
(8,1.04254)
};

\addplot
    coordinates {
(2,1.06047)
(3,1.03854)
(4,1.08913)
(5,1.08867)
(6,1.09569)
(7,1.10028)
(8,1.12503)
};

\addplot
    coordinates {
(2,1.2084)
(3,1.13258)
(4,1.12868)
(5,1.13873)
(6,1.17416)
(7,1.17773)
(8,1.18152)
};

\addplot
    coordinates {
(2,1.25357)
(3,1.2331)
(4,1.225)
(5,1.232)
(6,1.244)
(7,1.242)
(8,1.249)
};

\legend{$\emph{Tr}$,
		$B+$,
		$H$,
		$\emph{lsm}$,}
    
\end{axis}
\end{tikzpicture}
  \caption{Shelburne Dataset}
  \label{fig:range_s}
\end{subfigure}%
\begin{subfigure}{.5\textwidth}
  \centering
\begin{tikzpicture}
\begin{axis}[
    mark list fill={.!75!white},
    title={},
    xlabel={Block Size, $b_{\emph{size}}$ ($2^{12} \times 2^{x}$  bytes)},
    ylabel={Execution Time (s)},
    ylabel near ticks, yticklabel pos=right,
    xmin=2, xmax=8,
    ymin=0.8, ymax=1.2,
    legend pos=north west,
    ymajorgrids=true,
    grid style=dashed,
    legend style={font=\tiny},
    legend columns=4,
    height=2.4in,
    cycle list name=mycolorlist
]

\addplot
    coordinates {
(2,0.88743)
(3,0.86973)
(4,0.85109)
(5,0.85846)
(6,0.88307)
(7,0.88343)
(8,0.87557)
};

\addplot
    coordinates {
(2,0.8699)
(3,0.88401)
(4,0.88277)
(5,0.90728)
(6,0.92426)
(7,0.96565)
(8,0.96522)
};

\addplot
    coordinates {
(2,1.03409)
(3,1.00701)
(4,0.99208)
(5,1.00485)
(6,1.0253)
(7,1.03974)
(8,1.04473)
};

\addplot
    coordinates {
(2,1.17008)
(3,1.14126)
(4,1.120)
(5,1.129)
(6,1.144)
(7,1.146)
(8,1.150)
};

\legend{$\emph{Tr}$,
		$B+$,
		$H$,
		$\emph{lsm}$,}
    
\end{axis}
\end{tikzpicture}
  \caption{Taxi Dataset}
  \label{fig:range_t}
\end{subfigure}
\caption{Range query execution time per $b_{\emph{size}}$}
\label{fig:range}
{\tiny \textbf{Data Structures:} TrTables = Tr, LSM-Tree = lsm, B+ Tree = B+, Hash Tree = H}
\end{figure}

%% file: bench_query.tex
\begin{figure}
\centering
\begin{minipage}{\columnwidth}
\begin{center}
  \begin{tikzpicture}
\begin{axis}[
	xticklabels={TritanDB,InfluxDb, Akumuli, MongoDb, OpenTSDB\footnote{OpenTSDB queries cannot be executed on the Taxi dataset because multiple duplicate timestamps are not supported}, H2 SQL, Cassandra, ES},xtick={0,1,2,3,4,5,6,7,8},
	x tick label style={rotate=90,anchor=east},
	ylabel=Average Execution Time ($log_{10}$) (s),
	ymode=log,
	enlargelimits=0.05,
	legend style={at={(0.5, 1.15)},
	anchor=north,legend columns=-1},
	ybar interval=0.7
]
\addplot[fill=gray2]
[error bars/.cd,y dir=both, y explicit]
	coordinates {
	(0,11.18) +- (0.34,0.34)
	(1,106.26) +- (2.52,2.52)
	(2,52.41) +- (7.08,7.08) 
	(3,625.97) +- (5.45,5.45) 
	(4,26.51) +- (0.56,0.56) 
	(5,84.56) +- (0.57,0.57) 
	(6,83.64) +- (1.12,1.12) 
	(7,84.38) +- (20.39,20.39) 
	(8,1) };
\addplot[fill=gray1]
[error bars/.cd,y dir=both, y explicit]
	coordinates {
	(0,10.53) +- (0.51,0.51) 
	(1,53.94) +- (4.21,4.21)
	(2,50.01) +- (5.42,5.42)
	(3,870.80) +- (6.48,6.48) 
	(4,1) 
	(5,44.27) +- (2.54,2.54)
	(6,114.16) +- (1.68,1.68) 
	(7,46.57)  +- (12.73,12.73) 
	(8,1) };
\legend{Shelburne,Taxi}
\end{axis}
\end{tikzpicture}
\end{center}
\end{minipage}
\caption{Cross-sectional range query average execution time for each database for $s^1$ and $s^2$}
\label{fig:cross_sec_server_range}
{\tiny The execution times are the mean of $s^1$ and $s^2$ while the confidence interval to the left of each bar indicates the range of execution time.}
\end{figure}

\begin{figure}
\begin{minipage}{\columnwidth}
\begin{center}
  \begin{tikzpicture}
\begin{axis}[
	xticklabels={TritanDB,InfluxDb, Akumuli, MongoDb, OpenTSDB\footnote{OpenTSDB queries cannot be executed on the Taxi dataset because multiple duplicate timestamps are not supported}, H2 SQL, Cassandra, ES},xtick={0,1,2,3,4,5,6,7,8},
	x tick label style={rotate=90,anchor=east},
	ylabel=Average Execution Time ($log_{10}$) (s),
	ymode=log,
	enlargelimits=0.05,
	legend style={at={(0.5, 1.15)},
	anchor=north,legend columns=-1},
	ybar interval=0.7
]
\addplot[fill=gray2]
	coordinates {
	(0,5.08)
	(1,37.81)
	(2,22.30)
	(3,96.11)
	(4,5.55) 
	(5,45.16)
	(6,15.40) 
	(7,154.32) (8,1) };
\addplot[fill=gray1]
[error bars/.cd,y dir=both, y explicit]
	coordinates {
	(0,1.59)
	(1,8.72) 
	(2,7.32) 
	(3,39.17)
	(4,1) 
	(5,18.31) 
	(6,7.54) 
	(7,63.19) (8,1) };
\legend{Shelburne,Taxi}
\end{axis}
\end{tikzpicture}
\end{center}
\end{minipage}
\caption{Deep-history range query average execution time mean of $s^1$ and $s^2$ on each database}
\label{fig:deep_hist_server_range}
\end{figure}

\begin{figure}
\begin{minipage}{\columnwidth}
\begin{center}
  \begin{tikzpicture}
\begin{axis}[
	xticklabels={TritanDB\footnote{As the 100 queries are executed in sequence, the query translation overhead decreases to 0 to 2ms after the initial query},InfluxDb, Akumuli, MongoDb, OpenTSDB\footnote{OpenTSDB queries cannot be executed on the Taxi dataset because multiple duplicate timestamps are not supported}, H2 SQL, Cassandra, ES},xtick={0,1,2,3,4,5,6,7,8},
	x tick label style={rotate=90,anchor=east},
	ylabel=Average Execution Time ($10^{1} log_{10}$) (ms),
	ymode=log,
	enlargelimits=0.05,
	legend style={at={(0.5, 1.15)},
	anchor=north,legend columns=-1},
	ybar interval=0.7
]
\addplot[fill=gray2]
	coordinates {
	(0,2.24)
	(1,79.45)
	(2,7.3)
	(3,1179)
	(4,445) 
	(5,1306)
	(6,1677) 
	(7,25.43) (8,1) };
\addplot[fill=gray1]
[error bars/.cd,y dir=both, y explicit]
	coordinates {
	(0,1.11)
	(1,29.49) 
	(2,1.3) 
	(3,675)
	(4,1) 
	(5,759) 
	(6,532) 
	(7,13.79) (8,1) };
\legend{Shelburne,Taxi}
\end{axis}
\end{tikzpicture}
\end{center}
\end{minipage}
\caption{Aggregation range query average execution time mean of $s^1$ and $s^2$ on each database}
\label{fig:aggr_server_range}
\end{figure}

%% file: tritandb.bbl

\begin{thebibliography}{61}


\ifx \showCODEN    \undefined \def \showCODEN     #1{\unskip}     \fi
\ifx \showDOI      \undefined \def \showDOI       #1{#1}\fi
\ifx \showISBNx    \undefined \def \showISBNx     #1{\unskip}     \fi
\ifx \showISBNxiii \undefined \def \showISBNxiii  #1{\unskip}     \fi
\ifx \showISSN     \undefined \def \showISSN      #1{\unskip}     \fi
\ifx \showLCCN     \undefined \def \showLCCN      #1{\unskip}     \fi
\ifx \shownote     \undefined \def \shownote      #1{#1}          \fi
\ifx \showarticletitle \undefined \def \showarticletitle #1{#1}   \fi
\ifx \showURL      \undefined \def \showURL       {\relax}        \fi
\providecommand\bibfield[2]{#2}
\providecommand\bibinfo[2]{#2}
\providecommand\natexlab[1]{#1}
\providecommand\showeprint[2][]{arXiv:#2}

\bibitem[\protect\citeauthoryear{Abadi, Marcus, Madden, and Hollenbach}{Abadi
  et~al\mbox{.}}{2009}]%
        {Abadi2009}
\bibfield{author}{\bibinfo{person}{Daniel~J. Abadi}, \bibinfo{person}{Adam
  Marcus}, \bibinfo{person}{Samuel~R. Madden}, {and} \bibinfo{person}{Kate
  Hollenbach}.} \bibinfo{year}{2009}\natexlab{}.
\newblock \showarticletitle{{SW-Store: a vertically partitioned DBMS for
  Semantic Web data management}}.
\newblock \bibinfo{journal}{{\em The VLDB Journal\/}} \bibinfo{volume}{18},
  \bibinfo{number}{2} (\bibinfo{date}{feb} \bibinfo{year}{2009}),
  \bibinfo{pages}{385--406}.
\newblock
\showISSN{1066-8888}


\bibitem[\protect\citeauthoryear{Andersen and Culler}{Andersen and
  Culler}{2016}]%
        {Andersen2016}
\bibfield{author}{\bibinfo{person}{Michael~P Andersen} {and}
  \bibinfo{person}{David~E Culler}.} \bibinfo{year}{2016}\natexlab{}.
\newblock \showarticletitle{{BTrDB : Optimizing Storage System Design for
  Timeseries Processing}}. In \bibinfo{booktitle}{{\em Proceedings of the 14th
  USENIX Conference on File and Storage Technologies}}.
  \bibinfo{pages}{39--52}.
\newblock
\showISBNx{9781931971287}


\bibitem[\protect\citeauthoryear{Banker}{Banker}{2011}]%
        {banker2011mongodb}
\bibfield{author}{\bibinfo{person}{Kyle Banker}.}
  \bibinfo{year}{2011}\natexlab{}.
\newblock \bibinfo{booktitle}{{\em {MongoDB in action}}}.
\newblock \bibinfo{publisher}{Manning Publications Co.}
\newblock


\bibitem[\protect\citeauthoryear{Barnaghi, Wang, Henson, and Taylor}{Barnaghi
  et~al\mbox{.}}{2012}]%
        {Barnaghi2012}
\bibfield{author}{\bibinfo{person}{Payam Barnaghi}, \bibinfo{person}{Wei Wang},
  \bibinfo{person}{Cory Henson}, {and} \bibinfo{person}{Kerry Taylor}.}
  \bibinfo{year}{2012}\natexlab{}.
\newblock \showarticletitle{{Semantics for the Internet of Things: early
  progress and back to the future}}.
\newblock \bibinfo{journal}{{\em International Journal on Semantic Web and
  Information Systems\/}} \bibinfo{volume}{8}, \bibinfo{number}{1}
  (\bibinfo{year}{2012}), \bibinfo{pages}{1--21}.
\newblock


\bibitem[\protect\citeauthoryear{Basho}{Basho}{2016}]%
        {Basho2016}
\bibfield{author}{\bibinfo{person}{Basho}.} \bibinfo{year}{2016}\natexlab{}.
\newblock \bibinfo{title}{{RiakTS: NoSQL Time-series Database}}.
\newblock   (\bibinfo{year}{2016}).
\newblock
\showURL{%
\url{http://basho.com/products/riak-ts/}}


\bibitem[\protect\citeauthoryear{Bishop, Kiryakov, and Ognyanoff}{Bishop
  et~al\mbox{.}}{2011}]%
        {Bishop2011}
\bibfield{author}{\bibinfo{person}{Barry Bishop}, \bibinfo{person}{Atanas
  Kiryakov}, {and} \bibinfo{person}{Damyan Ognyanoff}.}
  \bibinfo{year}{2011}\natexlab{}.
\newblock \showarticletitle{{OWLIM: A family of scalable semantic
  repositories}}.
\newblock \bibinfo{journal}{{\em Semantic Web\/}} \bibinfo{volume}{2},
  \bibinfo{number}{1} (\bibinfo{year}{2011}), \bibinfo{pages}{33--42}.
\newblock
\showISSN{15700844}


\bibitem[\protect\citeauthoryear{Bizer, Heath, and Berners-Lee}{Bizer
  et~al\mbox{.}}{2009}]%
        {Bizer2009}
\bibfield{author}{\bibinfo{person}{Chris Bizer}, \bibinfo{person}{Tom Heath},
  {and} \bibinfo{person}{Tim Berners-Lee}.} \bibinfo{year}{2009}\natexlab{}.
\newblock \showarticletitle{{Linked Data - The Story So Far}}.
\newblock \bibinfo{journal}{{\em International Journal on Semantic Web and
  Information Systems\/}}  \bibinfo{volume}{5} (\bibinfo{year}{2009}),
  \bibinfo{pages}{1--22}.
\newblock
\showISBNx{1552-6283}


\bibitem[\protect\citeauthoryear{Bormann and Hoffman}{Bormann and
  Hoffman}{2013}]%
        {bormann2013concise}
\bibfield{author}{\bibinfo{person}{Carsten Bormann} {and} \bibinfo{person}{Paul
  Hoffman}.} \bibinfo{year}{2013}\natexlab{}.
\newblock \bibinfo{booktitle}{{\em {Concise Binary Object Representation
  (CBOR)}}}.
\newblock \bibinfo{type}{{T}echnical {R}eport}. \bibinfo{institution}{Internet
  Engineering Task Force}.
\newblock
\showURL{%
\url{https://tools.ietf.org/html/rfc7049}}


\bibitem[\protect\citeauthoryear{Buil-Aranda and Hogan}{Buil-Aranda and
  Hogan}{2013}]%
        {Buil-Aranda2013}
\bibfield{author}{\bibinfo{person}{C Buil-Aranda} {and} \bibinfo{person}{Aidan
  Hogan}.} \bibinfo{year}{2013}\natexlab{}.
\newblock \showarticletitle{{SPARQL Web-Querying Infrastructure: Ready for
  Action?}}. In \bibinfo{booktitle}{{\em Proceedings of the International
  Semantic Web Conference}}.
\newblock


\bibitem[\protect\citeauthoryear{Burnison}{Burnison}{2013}]%
        {Burnison2013}
\bibfield{author}{\bibinfo{person}{Richard Burnison}.}
  \bibinfo{year}{2013}\natexlab{}.
\newblock \bibinfo{title}{{The Concurrency Of ConcurrentHashMap}}.
\newblock   (\bibinfo{year}{2013}).
\newblock
\showURL{%
\url{https://www.burnison.ca/articles/the-concurrency-of-concurrenthashmap}}


\bibitem[\protect\citeauthoryear{Burtscher and Ratanaworabhan}{Burtscher and
  Ratanaworabhan}{2009}]%
        {Burtscher2009}
\bibfield{author}{\bibinfo{person}{Martin Burtscher} {and}
  \bibinfo{person}{Paruj Ratanaworabhan}.} \bibinfo{year}{2009}\natexlab{}.
\newblock \showarticletitle{{FPC: A High-Speed Compressor for Double-Precision
  Floating-Point Data}}.
\newblock \bibinfo{journal}{{\it IEEE Trans. Comput.}} \bibinfo{volume}{58},
  \bibinfo{number}{1} (\bibinfo{year}{2009}), \bibinfo{pages}{18--31}.
\newblock


\bibitem[\protect\citeauthoryear{Chang, Dean, Ghemawat, Hsieh, Wallach,
  Burrows, Chandra, Fikes, and Gruber}{Chang et~al\mbox{.}}{2008}]%
        {Chang2008}
\bibfield{author}{\bibinfo{person}{Fay Chang}, \bibinfo{person}{Jeffrey Dean},
  \bibinfo{person}{Sanjay Ghemawat}, \bibinfo{person}{Wilson~C Hsieh},
  \bibinfo{person}{Deborah~A Wallach}, \bibinfo{person}{Mike Burrows},
  \bibinfo{person}{Tushar Chandra}, \bibinfo{person}{Andrew Fikes}, {and}
  \bibinfo{person}{Robert~E Gruber}.} \bibinfo{year}{2008}\natexlab{}.
\newblock \showarticletitle{{Bigtable: A Distributed Storage System for
  Structured Data}}.
\newblock \bibinfo{journal}{{\em ACM Transactions on Computer Systems\/}}
  \bibinfo{volume}{26}, \bibinfo{number}{2} (\bibinfo{year}{2008}),
  \bibinfo{pages}{1--26}.
\newblock


\bibitem[\protect\citeauthoryear{Chiang, Ha, I, Risso, and Zhang}{Chiang
  et~al\mbox{.}}{2017}]%
        {Chiang2017}
\bibfield{author}{\bibinfo{person}{Mung Chiang}, \bibinfo{person}{Sangtae Ha},
  \bibinfo{person}{Chih-Lin I}, \bibinfo{person}{Fulvio Risso}, {and}
  \bibinfo{person}{Tao Zhang}.} \bibinfo{year}{2017}\natexlab{}.
\newblock \showarticletitle{{Clarifying Fog Computing and Networking: 10
  Questions and Answers}}.
\newblock \bibinfo{journal}{{\em IEEE Communications Magazine\/}}
  \bibinfo{number}{April} (\bibinfo{year}{2017}), \bibinfo{pages}{1--15}.
\newblock


\bibitem[\protect\citeauthoryear{Compton, Barnaghi, Bermudez,
  Garc{\'{i}}a-Castro, Corcho, Cox, Graybeal, Hauswirth, Henson, Herzog, Huang,
  Janowicz, Kelsey, {Le Phuoc}, Lefort, Leggieri, Neuhaus, Nikolov, Page,
  Assant, and Sheth}{Compton et~al\mbox{.}}{2012}]%
        {Compton2012}
\bibfield{author}{\bibinfo{person}{Michael Compton}, \bibinfo{person}{Payam
  Barnaghi}, \bibinfo{person}{Luis Bermudez}, \bibinfo{person}{Ra{\'{u}}l
  Garc{\'{i}}a-Castro}, \bibinfo{person}{Oscar Corcho}, \bibinfo{person}{Simon
  Cox}, \bibinfo{person}{John Graybeal}, \bibinfo{person}{Manfred Hauswirth},
  \bibinfo{person}{Cory Henson}, \bibinfo{person}{Arthur Herzog},
  \bibinfo{person}{Vincent Huang}, \bibinfo{person}{Krzysztof Janowicz},
  \bibinfo{person}{W.~David Kelsey}, \bibinfo{person}{Danh {Le Phuoc}},
  \bibinfo{person}{Laurent Lefort}, \bibinfo{person}{Myriam Leggieri},
  \bibinfo{person}{Holger Neuhaus}, \bibinfo{person}{Andriy Nikolov},
  \bibinfo{person}{Kevin Page}, \bibinfo{person}{Alexandre Assant}, {and}
  \bibinfo{person}{Amit Sheth}.} \bibinfo{year}{2012}\natexlab{}.
\newblock \showarticletitle{{The SSN ontology of the W3C semantic sensor
  network incubator group}}.
\newblock \bibinfo{journal}{{\em Journal of Web Semantics\/}}
  \bibinfo{volume}{17} (\bibinfo{year}{2012}), \bibinfo{pages}{25--32}.
\newblock
\showISBNx{1570-8268}
\showISSN{15708268}


\bibitem[\protect\citeauthoryear{Cordell and Newton}{Cordell and
  Newton}{2016}]%
        {cordell2016language}
\bibfield{author}{\bibinfo{person}{Pete Cordell} {and} \bibinfo{person}{Andrew
  Newton}.} \bibinfo{year}{2016}\natexlab{}.
\newblock \showarticletitle{{A Language for Rules Describing JSON Content}}.
\newblock  (\bibinfo{year}{2016}).
\newblock
\showURL{%
\url{https://www.ietf.org/id/draft-newton-json-content-rules-08.txt}}


\bibitem[\protect\citeauthoryear{Czech, Havas, and Majewski}{Czech
  et~al\mbox{.}}{1992}]%
        {Czech1992}
\bibfield{author}{\bibinfo{person}{Zbigniew~J. Czech}, \bibinfo{person}{George
  Havas}, {and} \bibinfo{person}{Bohdan~S. Majewski}.}
  \bibinfo{year}{1992}\natexlab{}.
\newblock \showarticletitle{{An optimal algorithm for generating minimal
  perfect hash functions}}.
\newblock \bibinfo{journal}{{\it Inform. Process. Lett.}} \bibinfo{volume}{43},
  \bibinfo{number}{5} (\bibinfo{year}{1992}), \bibinfo{pages}{257--264}.
\newblock
\showISSN{00200190}
\showDOI{%
\url{https://doi.org/10.1016/0020-0190(92)90220-P}}


\bibitem[\protect\citeauthoryear{{da Rocha Pinto}, Dinsdale-Young, Dodds,
  Gardner, and Wheelhouse}{{da Rocha Pinto} et~al\mbox{.}}{2011}]%
        {DaRochaPinto2011}
\bibfield{author}{\bibinfo{person}{Pedro {da Rocha Pinto}},
  \bibinfo{person}{Thomas Dinsdale-Young}, \bibinfo{person}{Mike Dodds},
  \bibinfo{person}{Philippa Gardner}, {and} \bibinfo{person}{Mark Wheelhouse}.}
  \bibinfo{year}{2011}\natexlab{}.
\newblock \showarticletitle{{A simple abstraction for complex concurrent
  indexes}}.
\newblock \bibinfo{journal}{{\em ACM SIGPLAN Notices\/}} \bibinfo{volume}{46},
  \bibinfo{number}{10} (\bibinfo{year}{2011}), \bibinfo{pages}{845}.
\newblock
\showISBNx{9781450309400}
\showISSN{03621340}
\showDOI{%
\url{https://doi.org/10.1145/2076021.2048131}}


\bibitem[\protect\citeauthoryear{{DWARF Debugging Information Format
  Committee}}{{DWARF Debugging Information Format Committee}}{2010}]%
        {dwarf2010dwarf}
\bibfield{author}{\bibinfo{person}{{DWARF Debugging Information Format
  Committee}}.} \bibinfo{year}{2010}\natexlab{}.
\newblock \showarticletitle{{DWARF debugging information format, version 5}}.
\newblock \bibinfo{journal}{{\em Free Standards Group\/}}
  (\bibinfo{year}{2010}).
\newblock
\showURL{%
\url{http://www.dwarfstd.org/doc/DWARF5.pdf}}


\bibitem[\protect\citeauthoryear{Elastic}{Elastic}{2017}]%
        {Elastic}
\bibfield{author}{\bibinfo{person}{Elastic}.} \bibinfo{year}{2017}\natexlab{}.
\newblock \bibinfo{title}{{Elasticsearch: RESTful, Distributed Search {\&}
  Analytics}}.
\newblock   (\bibinfo{date}{jun} \bibinfo{year}{2017}).
\newblock


\bibitem[\protect\citeauthoryear{Galiegue, Zyp, and Others}{Galiegue
  et~al\mbox{.}}{2013}]%
        {galiegue2013json}
\bibfield{author}{\bibinfo{person}{Francis Galiegue}, \bibinfo{person}{Kris
  Zyp}, {and} \bibinfo{person}{Others}.} \bibinfo{year}{2013}\natexlab{}.
\newblock \showarticletitle{{JSON Schema: Core definitions and terminology}}.
\newblock \bibinfo{journal}{{\em Internet Engineering Task Force (IETF)\/}}
  (\bibinfo{year}{2013}).
\newblock
\showURL{%
\url{http://json-schema.org/latest/json-schema-core.html}}


\bibitem[\protect\citeauthoryear{Goeman, Vandierendonck, and {De
  Bosschere}}{Goeman et~al\mbox{.}}{2001}]%
        {goeman2001differential}
\bibfield{author}{\bibinfo{person}{Bart Goeman}, \bibinfo{person}{Hans
  Vandierendonck}, {and} \bibinfo{person}{Koenraad {De Bosschere}}.}
  \bibinfo{year}{2001}\natexlab{}.
\newblock \showarticletitle{{Differential FCM: Increasing value prediction
  accuracy by improving table usage efficiency}}. In \bibinfo{booktitle}{{\em
  Proceedings of the Seventh International Symposium on High-Performance
  Computer Architecture}}. IEEE, \bibinfo{pages}{207--216}.
\newblock


\bibitem[\protect\citeauthoryear{Harris, Seaborne, and Prud'hommeaux}{Harris
  et~al\mbox{.}}{2013}]%
        {harris2013sparql}
\bibfield{author}{\bibinfo{person}{Steve Harris}, \bibinfo{person}{Andy
  Seaborne}, {and} \bibinfo{person}{Eric Prud'hommeaux}.}
  \bibinfo{year}{2013}\natexlab{}.
\newblock \showarticletitle{{SPARQL 1.1 query language}}.
\newblock \bibinfo{journal}{{\em W3C recommendation\/}} \bibinfo{volume}{21},
  \bibinfo{number}{10} (\bibinfo{year}{2013}).
\newblock


\bibitem[\protect\citeauthoryear{Heath and Bizer}{Heath and Bizer}{2011}]%
        {Heath2011}
\bibfield{author}{\bibinfo{person}{Tom Heath} {and} \bibinfo{person}{Christian
  Bizer}.} \bibinfo{year}{2011}\natexlab{}.
\newblock \showarticletitle{{Linked Data Evolving the Web into a Global Data
  Space}}.
\newblock In \bibinfo{booktitle}{{\em Synthesis Lectures on the Semantic Web:
  Theory and Technology}}.
\newblock
\showISBNx{9781608454303}


\bibitem[\protect\citeauthoryear{{IEEE Standards Association}}{{IEEE Standards
  Association}}{2008}]%
        {ieee2008standard}
\bibfield{author}{\bibinfo{person}{{IEEE Standards Association}}.}
  \bibinfo{year}{2008}\natexlab{}.
\newblock \showarticletitle{{Standard for Floating-Point Arithmetic}}.
\newblock \bibinfo{journal}{{\em IEEE 754-2008\/}} (\bibinfo{year}{2008}).
\newblock


\bibitem[\protect\citeauthoryear{{Influx Data}}{{Influx Data}}{2017}]%
        {InfluxData2017}
\bibfield{author}{\bibinfo{person}{{Influx Data}}.}
  \bibinfo{year}{2017}\natexlab{}.
\newblock \bibinfo{title}{{InfluxDB Documentation}}.
\newblock   (\bibinfo{date}{jun} \bibinfo{year}{2017}).
\newblock
\showURL{%
\url{https://docs.influxdata.com/influxdb/v1.2/}}


\bibitem[\protect\citeauthoryear{Jacob, Stanley, Witter, York, and
  Shasha}{Jacob et~al\mbox{.}}{2000}]%
        {Jacob2000}
\bibfield{author}{\bibinfo{person}{Kaippallimalil~J Jacob},
  \bibinfo{person}{Morgan Stanley}, \bibinfo{person}{Dean Witter},
  \bibinfo{person}{New York}, {and} \bibinfo{person}{Dennis Shasha}.}
  \bibinfo{year}{2000}\natexlab{}.
\newblock \showarticletitle{{FinTime - a financial time series benchmark}}.
\newblock  \bibinfo{volume}{28}, \bibinfo{number}{4} (\bibinfo{year}{2000}),
  \bibinfo{pages}{42--48}.
\newblock
\showISBNx{0163-5808}
\showISSN{01635808}


\bibitem[\protect\citeauthoryear{KairosDB}{KairosDB}{2015}]%
        {kairosdb}
\bibfield{author}{\bibinfo{person}{KairosDB}.} \bibinfo{year}{2015}\natexlab{}.
\newblock \bibinfo{title}{{KairosDB}}.
\newblock   (\bibinfo{year}{2015}).
\newblock
\showURL{%
\url{https://kairosdb.github.io/}}


\bibitem[\protect\citeauthoryear{Kuszmaul}{Kuszmaul}{2014}]%
        {Kuszmaul2014}
\bibfield{author}{\bibinfo{person}{Bradley~C Kuszmaul}.}
  \bibinfo{year}{2014}\natexlab{}.
\newblock \showarticletitle{{A Comparison of Fractal Trees to Log-Structured
  Merge (LSM) Trees}}.
\newblock \bibinfo{journal}{{\em DZone Refcardz\/}} (\bibinfo{year}{2014}),
  \bibinfo{pages}{1--15}.
\newblock
\showURL{%
\url{https://yadi.sk/i/hlwakkCpkL9yR}}


\bibitem[\protect\citeauthoryear{{Laksham Avinash} and {Prashant
  Malik}}{{Laksham Avinash} and {Prashant Malik}}{2010}]%
        {LakshamAvinash2010}
\bibfield{author}{\bibinfo{person}{{Laksham Avinash}} {and}
  \bibinfo{person}{{Prashant Malik}}.} \bibinfo{year}{2010}\natexlab{}.
\newblock \showarticletitle{{Cassandra: a decentralized structured storage
  system}}.
\newblock \bibinfo{journal}{{\em ACM SIGOPS Operating Systems Review\/}}
  (\bibinfo{year}{2010}), \bibinfo{pages}{1--6}.
\newblock
\showISBNx{9781605583969}
\showISSN{01635980}


\bibitem[\protect\citeauthoryear{Lazin}{Lazin}{2017}]%
        {EugeneLazin2017}
\bibfield{author}{\bibinfo{person}{Eugene Lazin}.}
  \bibinfo{year}{2017}\natexlab{}.
\newblock \bibinfo{title}{{Akumuli Time-series Database}}.
\newblock   (\bibinfo{date}{jun} \bibinfo{year}{2017}).
\newblock
\showURL{%
\url{http://akumuli.org/}}


\bibitem[\protect\citeauthoryear{Malvar}{Malvar}{2006}]%
        {Malvar2006}
\bibfield{author}{\bibinfo{person}{Henrique~S. Malvar}.}
  \bibinfo{year}{2006}\natexlab{}.
\newblock \showarticletitle{{Adaptive run-length/golomb-rice encoding of
  quantized generalized gaussian sources with unknown statistics}}.
\newblock \bibinfo{journal}{{\em Proceedings of the Data Compression
  Conference\/}} \bibinfo{number}{June} (\bibinfo{year}{2006}),
  \bibinfo{pages}{23--32}.
\newblock
\showISBNx{0-7695-2545-8}
\showISSN{10680314}


\bibitem[\protect\citeauthoryear{Mathe, Ramo, Stagni, Tomassetti, {Casajus A},
  M, A, {Casajus A}, R, J, Closier, {Casajus A}, SPA, Opentsdb, Elasticsearch,
  Influxdb, Hdfs, Hbase, lucene Apache, Grafana, Kibana, Rabbitmq,
  {Tsaregorodtsev A}, G, and Hadoop}{Mathe et~al\mbox{.}}{2015}]%
        {Mathe2015}
\bibfield{author}{\bibinfo{person}{Z Mathe}, \bibinfo{person}{A~Casajus Ramo},
  \bibinfo{person}{F. Stagni}, \bibinfo{person}{L. Tomassetti},
  \bibinfo{person}{Fernandez V Graciani R Hamar V Mendez V Poss S Sapunov M
  Stagni F Tsaregorodtsev~A {Casajus A}, Ciba~K}, \bibinfo{person}{Ubeda M},
  \bibinfo{person}{Tsaregorodtsev A}, \bibinfo{person}{Puig~A {Casajus A}, Diaz
  R~G}, \bibinfo{person}{Vazquez R}, \bibinfo{person}{Paterson~S J},
  \bibinfo{person}{Closier}, \bibinfo{person}{Graciani~R {Casajus A}},
  \bibinfo{person}{Tsaregorodtsev SPA}, \bibinfo{person}{Opentsdb},
  \bibinfo{person}{Elasticsearch}, \bibinfo{person}{Influxdb},
  \bibinfo{person}{Hdfs}, \bibinfo{person}{Hbase}, \bibinfo{person}{lucene
  Apache}, \bibinfo{person}{Grafana}, \bibinfo{person}{Kibana},
  \bibinfo{person}{Rabbitmq}, \bibinfo{person}{Brook N Ramo A C Castellani G
  Charpentier P Cioffi C Closier J Diaz R G Kuznetsov G Li Y Y Nandakumar R
  Parerson S Santinelli R Smith A C Miguelez M~S {Tsaregorodtsev A},
  Bergiotti~M}, \bibinfo{person}{Jumenez~S G}, {and} \bibinfo{person}{Hadoop}.}
  \bibinfo{year}{2015}\natexlab{}.
\newblock \showarticletitle{{Evaluation of NoSQL databases for DIRAC monitoring
  and beyond}}.
\newblock \bibinfo{journal}{{\em Journal of Physics: Conference Series\/}}
  \bibinfo{volume}{664}, \bibinfo{number}{4} (\bibinfo{year}{2015}),
  \bibinfo{pages}{042036}.
\newblock
\showISSN{1742-6588}
\showDOI{%
\url{https://doi.org/10.1088/1742-6596/664/4/042036}}


\bibitem[\protect\citeauthoryear{Neumann and Weikum}{Neumann and
  Weikum}{2009}]%
        {Neumann2009}
\bibfield{author}{\bibinfo{person}{Thomas Neumann} {and}
  \bibinfo{person}{Gerhard Weikum}.} \bibinfo{year}{2009}\natexlab{}.
\newblock \showarticletitle{{The RDF-3X engine for scalable management of RDF
  data}}.
\newblock \bibinfo{journal}{{\em The VLDB Journal\/}} \bibinfo{volume}{19},
  \bibinfo{number}{1} (\bibinfo{date}{sep} \bibinfo{year}{2009}),
  \bibinfo{pages}{91--113}.
\newblock
\showISSN{1066-8888}


\bibitem[\protect\citeauthoryear{Oetiker}{Oetiker}{2005}]%
        {oetiker2005rrdtool}
\bibfield{author}{\bibinfo{person}{Tobias Oetiker}.}
  \bibinfo{year}{2005}\natexlab{}.
\newblock \bibinfo{title}{{RRDtool}}.
\newblock   (\bibinfo{year}{2005}).
\newblock


\bibitem[\protect\citeauthoryear{O'Neil, Cheng, Gawlick, and O'Neil}{O'Neil
  et~al\mbox{.}}{1996}]%
        {ONeil1996}
\bibfield{author}{\bibinfo{person}{Patrick O'Neil}, \bibinfo{person}{Edward
  Cheng}, \bibinfo{person}{Dieter Gawlick}, {and} \bibinfo{person}{Elizabeth
  O'Neil}.} \bibinfo{year}{1996}\natexlab{}.
\newblock \showarticletitle{{The log-structured merge-tree (LSM-tree)}}.
\newblock \bibinfo{journal}{{\em Acta Informatica\/}} \bibinfo{volume}{33},
  \bibinfo{number}{4} (\bibinfo{year}{1996}), \bibinfo{pages}{351--385}.
\newblock
\showISSN{0001-5903}


\bibitem[\protect\citeauthoryear{{\"{O}}zsu and Valduriez}{{\"{O}}zsu and
  Valduriez}{2011}]%
        {Ozsu2011}
\bibfield{author}{\bibinfo{person}{M~Tamer {\"{O}}zsu} {and}
  \bibinfo{person}{Patrick Valduriez}.} \bibinfo{year}{2011}\natexlab{}.
\newblock \bibinfo{booktitle}{{\em {Principles of distributed database
  systems}}}.
\newblock \bibinfo{publisher}{Springer Science {\&} Business Media}.
\newblock
\showISBNx{1441988343}


\bibitem[\protect\citeauthoryear{Patni, Henson, and Sheth}{Patni
  et~al\mbox{.}}{2010}]%
        {Patni2010}
\bibfield{author}{\bibinfo{person}{Harshal Patni}, \bibinfo{person}{Cory
  Henson}, {and} \bibinfo{person}{Amit Sheth}.}
  \bibinfo{year}{2010}\natexlab{}.
\newblock \showarticletitle{{Linked Sensor Data}}. In \bibinfo{booktitle}{{\em
  Proceedings of the International Symposium on Collaborative Technologies and
  Systems}}.
\newblock
\showISBNx{9781424466191}
\showISSN{16130073}
\showDOI{%
\url{https://doi.org/10.1109/CTS.2010.5478492}}


\bibitem[\protect\citeauthoryear{Pelkonen, Franklin, Teller, Cavallaro, Huang,
  Meza, and Veeraraghavan}{Pelkonen et~al\mbox{.}}{2015}]%
        {Pelkonen2015}
\bibfield{author}{\bibinfo{person}{Tuomas Pelkonen}, \bibinfo{person}{Scott
  Franklin}, \bibinfo{person}{Justin Teller}, \bibinfo{person}{Paul Cavallaro},
  \bibinfo{person}{Qi Huang}, \bibinfo{person}{Justin Meza}, {and}
  \bibinfo{person}{Kaushik Veeraraghavan}.} \bibinfo{year}{2015}\natexlab{}.
\newblock \showarticletitle{{Gorilla: A Fast, Scalable, In-Memory Time Series
  Database}}.
\newblock \bibinfo{journal}{{\em Proceedings of the 41st International
  Conference on Very Large Data Bases\/}} (\bibinfo{year}{2015}),
  \bibinfo{pages}{1816--1827}.
\newblock
\showISSN{21508097}


\bibitem[\protect\citeauthoryear{Persen and Winslow}{Persen and
  Winslow}{2016}]%
        {Persen2016}
\bibfield{author}{\bibinfo{person}{Todd Persen} {and} \bibinfo{person}{Robert
  Winslow}.} \bibinfo{year}{2016}\natexlab{}.
\newblock \bibinfo{booktitle}{{\em {Benchmarking InfluxDB vs. MongoDB for
  Time-Series Data, Metrics {\&} Management}}}.
\newblock \bibinfo{type}{{T}echnical {R}eport}.
\newblock
\showURL{%
\url{https://goo.gl/7eVLZv}}


\bibitem[\protect\citeauthoryear{Priyatna, Corcho, and Sequeda}{Priyatna
  et~al\mbox{.}}{2014}]%
        {Priyatna2014}
\bibfield{author}{\bibinfo{person}{Freddy Priyatna}, \bibinfo{person}{O
  Corcho}, {and} \bibinfo{person}{Juan Sequeda}.}
  \bibinfo{year}{2014}\natexlab{}.
\newblock \showarticletitle{{Formalisation and Experiences of R2RML-based
  SPARQL to SQL Query Translation using Morph}}. In \bibinfo{booktitle}{{\em
  Proceedings of the 23rd International Conference on World Wide Web}}.
  \bibinfo{pages}{479--489}.
\newblock
\showISBNx{9781450327442}
\showDOI{%
\url{https://doi.org/10.1145/2566486.2567981}}


\bibitem[\protect\citeauthoryear{{Project FiFo}}{{Project FiFo}}{2014}]%
        {FiFo2014}
\bibfield{author}{\bibinfo{person}{{Project FiFo}}.}
  \bibinfo{year}{2014}\natexlab{}.
\newblock \bibinfo{title}{{DalmatinerDb: A fast, distributed metric store}}.
\newblock   (\bibinfo{year}{2014}).
\newblock
\showURL{%
\url{https://dalmatiner.io/}}


\bibitem[\protect\citeauthoryear{{Prometheus Authors}}{{Prometheus
  Authors}}{2016}]%
        {PrometheusAuthors2016}
\bibfield{author}{\bibinfo{person}{{Prometheus Authors}}.}
  \bibinfo{year}{2016}\natexlab{}.
\newblock \bibinfo{title}{{Prometheus Monitoring System and Time-series
  database}}.
\newblock   (\bibinfo{year}{2016}).
\newblock
\showURL{%
\url{https://prometheus.io/}}


\bibitem[\protect\citeauthoryear{Rackspace}{Rackspace}{2013}]%
        {Rackspace2013}
\bibfield{author}{\bibinfo{person}{Rackspace}.}
  \bibinfo{year}{2013}\natexlab{}.
\newblock \bibinfo{title}{{Blueflood}}.
\newblock   (\bibinfo{year}{2013}).
\newblock
\showURL{%
\url{http://blueflood.io/}}


\bibitem[\protect\citeauthoryear{Redhat}{Redhat}{2017}]%
        {Redhat}
\bibfield{author}{\bibinfo{person}{Redhat}.} \bibinfo{year}{2017}\natexlab{}.
\newblock \bibinfo{title}{{Hawkular}}.
\newblock   (\bibinfo{date}{jun} \bibinfo{year}{2017}).
\newblock
\showURL{%
\url{http://www.hawkular.org/}}


\bibitem[\protect\citeauthoryear{Rice and Plaunt}{Rice and Plaunt}{1971}]%
        {rice1971}
\bibfield{author}{\bibinfo{person}{R Rice} {and} \bibinfo{person}{J Plaunt}.}
  \bibinfo{year}{1971}\natexlab{}.
\newblock \showarticletitle{{Adaptive Variable-Length Coding for Efficient
  Compression of Spacecraft Television Data}}.
\newblock \bibinfo{journal}{{\em IEEE Transactions on Communication
  Technology\/}} \bibinfo{volume}{19}, \bibinfo{number}{6} (\bibinfo{date}{dec}
  \bibinfo{year}{1971}), \bibinfo{pages}{889--897}.
\newblock
\showISSN{0018-9332}


\bibitem[\protect\citeauthoryear{Rodriguez-Muro and Rezk}{Rodriguez-Muro and
  Rezk}{2014}]%
        {Rodriguez-muro2014}
\bibfield{author}{\bibinfo{person}{Mariano Rodriguez-Muro} {and}
  \bibinfo{person}{Martin Rezk}.} \bibinfo{year}{2014}\natexlab{}.
\newblock \showarticletitle{{Efficient SPARQL-to-SQL with R2RML mappings}}.
\newblock \bibinfo{journal}{{\em Web Semantics: Science, Services and Agents on
  the WWW\/}}  \bibinfo{volume}{33} (\bibinfo{year}{2014}),
  \bibinfo{pages}{141--169}.
\newblock
\showISSN{1570-8268}
\showDOI{%
\url{https://doi.org/10.1016/j.websem.2015.03.001}}


\bibitem[\protect\citeauthoryear{Sagiv}{Sagiv}{1986}]%
        {Sagiv1986}
\bibfield{author}{\bibinfo{person}{Yehoshua Sagiv}.}
  \bibinfo{year}{1986}\natexlab{}.
\newblock \showarticletitle{{Concurrent Operations On B -trees With
  Overtaking}}.
\newblock \bibinfo{journal}{{\it J. Comput. System Sci.}} \bibinfo{volume}{33},
  \bibinfo{number}{2} (\bibinfo{year}{1986}), \bibinfo{pages}{275--296}.
\newblock


\bibitem[\protect\citeauthoryear{Sazeides and Smith}{Sazeides and
  Smith}{1997}]%
        {sazeides1997predictability}
\bibfield{author}{\bibinfo{person}{Yiannakis Sazeides} {and}
  \bibinfo{person}{James~E Smith}.} \bibinfo{year}{1997}\natexlab{}.
\newblock \showarticletitle{{The predictability of data values}}. In
  \bibinfo{booktitle}{{\em Proceedings of the 30th Annual ACM/IEEE
  International Symposium on Microarchitecture}}. IEEE Computer Society,
  \bibinfo{pages}{248--258}.
\newblock


\bibitem[\protect\citeauthoryear{Schonwalder, Bjorklund, and
  Shafer}{Schonwalder et~al\mbox{.}}{2010}]%
        {schonwalder2010network}
\bibfield{author}{\bibinfo{person}{Jurgen Schonwalder}, \bibinfo{person}{Martin
  Bjorklund}, {and} \bibinfo{person}{Phil Shafer}.}
  \bibinfo{year}{2010}\natexlab{}.
\newblock \showarticletitle{{Network Configuration Management Using NETCONF and
  YANG}}.
\newblock \bibinfo{journal}{{\em IEEE Communications Magazine\/}}
  \bibinfo{volume}{48}, \bibinfo{number}{9} (\bibinfo{year}{2010}),
  \bibinfo{pages}{166--173}.
\newblock
\showISSN{0163-6804}
\showDOI{%
\url{https://doi.org/10.1109/MCOM.2010.5560601}}


\bibitem[\protect\citeauthoryear{Sears and Ramakrishnan}{Sears and
  Ramakrishnan}{2012}]%
        {Sears2012}
\bibfield{author}{\bibinfo{person}{Russell Sears} {and} \bibinfo{person}{Raghu
  Ramakrishnan}.} \bibinfo{year}{2012}\natexlab{}.
\newblock \showarticletitle{{bLSM: A General Purpose Log Structured Merge
  Tree}}. In \bibinfo{booktitle}{{\em Proceedings of the 2012 ACM SIGMOD
  International Conference on Management of Data}}. \bibinfo{pages}{217--228}.
\newblock
\showISBNx{9781450312479}


\bibitem[\protect\citeauthoryear{Sharma}{Sharma}{2016}]%
        {Sharma2016}
\bibfield{author}{\bibinfo{person}{Vishal Sharma}.}
  \bibinfo{year}{2016}\natexlab{}.
\newblock \bibinfo{booktitle}{{\em {Graphite Monitoring and Graphs}}}.
\newblock \bibinfo{publisher}{Apress}, \bibinfo{address}{Berkeley, CA},
  \bibinfo{pages}{73--94}.
\newblock
\showISBNx{978-1-4842-1694-1}
\showDOI{%
\url{https://doi.org/10.1007/978-1-4842-1694-1_6}}


\bibitem[\protect\citeauthoryear{Siow, Tiropanis, and Hall}{Siow
  et~al\mbox{.}}{2016a}]%
        {Siow2016a}
\bibfield{author}{\bibinfo{person}{Eugene Siow}, \bibinfo{person}{Thanassis
  Tiropanis}, {and} \bibinfo{person}{Wendy Hall}.}
  \bibinfo{year}{2016}\natexlab{a}.
\newblock \showarticletitle{{Interoperable {\&} Efficient : Linked Data for the
  Internet of Things}}. In \bibinfo{booktitle}{{\em Proceedings of the 3rd
  International Conference on Internet Science}}.
\newblock
\showDOI{%
\url{https://doi.org/10.1007/978-3-319-45982-0_15}}


\bibitem[\protect\citeauthoryear{Siow, Tiropanis, and Hall}{Siow
  et~al\mbox{.}}{2016b}]%
        {Siow2016b}
\bibfield{author}{\bibinfo{person}{Eugene Siow}, \bibinfo{person}{Thanassis
  Tiropanis}, {and} \bibinfo{person}{Wendy Hall}.}
  \bibinfo{year}{2016}\natexlab{b}.
\newblock \showarticletitle{{SPARQL-to-SQL on Internet of Things Databases and
  Streams}}. In \bibinfo{booktitle}{{\em Proceedings of 15th International
  Semantic Web Conference}}.
\newblock
\showDOI{%
\url{https://doi.org/10.1007/978-3-319-46523-4_31}}


\bibitem[\protect\citeauthoryear{Spotify}{Spotify}{2017}]%
        {Spotify}
\bibfield{author}{\bibinfo{person}{Spotify}.} \bibinfo{year}{2017}\natexlab{}.
\newblock \bibinfo{title}{{Heroic Documentation}}.
\newblock   (\bibinfo{date}{jun} \bibinfo{year}{2017}).
\newblock
\showURL{%
\url{https://spotify.github.io/heroic/}}


\bibitem[\protect\citeauthoryear{Square}{Square}{2012}]%
        {Square2012}
\bibfield{author}{\bibinfo{person}{Square}.} \bibinfo{year}{2012}\natexlab{}.
\newblock \bibinfo{title}{{Cube Time-series Data Collection and Analysis}}.
\newblock   (\bibinfo{year}{2012}).
\newblock
\showURL{%
\url{http://square.github.io/cube/}}


\bibitem[\protect\citeauthoryear{{The OpenTSDB Authors}}{{The OpenTSDB
  Authors}}{2017}]%
        {OpenTSDB}
\bibfield{author}{\bibinfo{person}{{The OpenTSDB Authors}}.}
  \bibinfo{year}{2017}\natexlab{}.
\newblock \bibinfo{title}{{OpenTSDB - A scalable, distributed monitoring
  system}}.
\newblock   (\bibinfo{date}{jun} \bibinfo{year}{2017}).
\newblock
\showURL{%
\url{http://opentsdb.net/}}


\bibitem[\protect\citeauthoryear{Timescale}{Timescale}{2017}]%
        {Timescale2017}
\bibfield{author}{\bibinfo{person}{Timescale}.}
  \bibinfo{year}{2017}\natexlab{}.
\newblock \bibinfo{title}{{Timescale: SQL made scalable for time-series data}}.
\newblock   (\bibinfo{year}{2017}).
\newblock


\bibitem[\protect\citeauthoryear{Trubetskoy}{Trubetskoy}{2017}]%
        {Trubetskoy2017}
\bibfield{author}{\bibinfo{person}{Grisha Trubetskoy}.}
  \bibinfo{year}{2017}\natexlab{}.
\newblock \bibinfo{title}{{Tgres}}.
\newblock   (\bibinfo{year}{2017}).
\newblock
\showURL{%
\url{https://github.com/tgres/tgres}}


\bibitem[\protect\citeauthoryear{{W3C Web of Things WG}}{{W3C Web of Things
  WG}}{2016}]%
        {Things2016}
\bibfield{author}{\bibinfo{person}{{W3C Web of Things WG}}.}
  \bibinfo{year}{2016}\natexlab{}.
\newblock \bibinfo{title}{{White Paper for the Web of Things}}.
\newblock   (\bibinfo{year}{2016}).
\newblock
\showURL{%
\url{http://w3c.github.io/wot/charters/wot-white-paper-2016.html}}


\bibitem[\protect\citeauthoryear{Wlodarczyk}{Wlodarczyk}{2012}]%
        {Wlodarczyk2012}
\bibfield{author}{\bibinfo{person}{Tomasz~Wiktor Wlodarczyk}.}
  \bibinfo{year}{2012}\natexlab{}.
\newblock \showarticletitle{{Overview of Time Series Storage and Processing in
  a Cloud Environment}}. In \bibinfo{booktitle}{{\em Proceedings of the 2012
  IEEE 4th International Conference on Cloud Computing Technology and Science}}
  {\em (\bibinfo{series}{CLOUDCOM '12})}. \bibinfo{publisher}{IEEE Computer
  Society}, \bibinfo{address}{Washington, DC, USA}, \bibinfo{pages}{625--628}.
\newblock
\showISBNx{978-1-4673-4511-8}


\bibitem[\protect\citeauthoryear{Zhang, Duc, Corcho, and Calbimonte}{Zhang
  et~al\mbox{.}}{2012}]%
        {Zhang2012a}
\bibfield{author}{\bibinfo{person}{Ying Zhang}, \bibinfo{person}{Pham~Minh
  Duc}, \bibinfo{person}{Oscar Corcho}, {and} \bibinfo{person}{Jean~Paul
  Calbimonte}.} \bibinfo{year}{2012}\natexlab{}.
\newblock \showarticletitle{{SRBench: A streaming RDF/SPARQL benchmark}}. In
  \bibinfo{booktitle}{{\em Proceedings of the International Semantic Web
  Conference}} {\em (\bibinfo{series}{Lecture Notes in Computer Science})}.
\newblock
\showISBNx{9783642351754}
\showISSN{03029743}
\showDOI{%
\url{https://doi.org/10.1007/978-3-642-35176-1-40}}


\end{thebibliography}
